\documentclass[twocolumn,showpacs,amsmath,amssymb,prd,nofootinbib,floatfix,preprintnumbers]{revtex4-2}

\usepackage[dvipdfmx]{graphicx}
\usepackage{dcolumn}
\usepackage{bm}
\usepackage{comment}
\usepackage{color}
\usepackage[rgb]{xcolor}
\usepackage[normalem]{ulem}
\usepackage{booktabs}
\usepackage{multirow}
\usepackage{hyperref}
\hypersetup{colorlinks=true}

\newcommand{\mnras}{Mon.~Not.~Roy.~Astron.~Soc.}
\newcommand{\physrep}{Phys.~Rep.}
\newcommand{\jcap}{JCAP}
\newcommand{\pasj}{Publ.~Astron.~Soc.~Japan}
\newcommand{\apjl}{\apj~Lett.}
\newcommand{\apjs}{\apj~Suppl.}
\newcommand{\aap}{Astronomy~\&~Astrophysics}
\newcommand{\aj}{The~Astronomical~J.}

\newcommand{\ssr}{Space~Science~Rev.}
\newcommand{\bx}{{\mathbf{x}}}

\newcommand{\bk}{{\mathbf{k}}}
\newcommand{\bd}{{\mathbf{d}}}

\newcommand{\bq}{{\mathbf{q}}}
\newcommand{\br}{{\mathbf{r}}}

\newcommand{\bs}{{\mathbf{s}}}
\newcommand{\bp}{{\mathbf{p}}}
\newcommand{\hx}{\hat{x}}

\newcommand{\hk}{\hat{k}}

\newcommand{\hbx}{\hat{\mathbf{x}}}

\newcommand{\hbk}{\hat{\mathbf{k}}}
\newcommand{\hbd}{\hat{\mathbf{d}}}

\newcommand{\hbr}{\hat{\mathbf{r}}}

\newcommand{\hbs}{\hat{\mathbf{s}}}
\newcommand{\Mpc}{\mathrm{Mpc}}
\newcommand{\hMpci}{h\mathrm{Mpc}^{-1}}
\newcommand{\hiMpc}{h^{-1}\mathrm{Mpc}}

\newcommand{\avrg}[1]{\left\langle #1 \right\rangle}
\newcommand{\bC}{{\mathbf{C}}}
\newcommand{\Om}{\Omega_{\mathrm{m}}}
\newcommand{\rmd}{{\mathrm{d}}}
\newcommand{\rmg}{\mathrm{g}}
\newcommand{\rmr}{\mathrm{r}}
\newcommand{\rms}{\mathrm{s}}
\newcommand{\rmI}{\mathrm{I}}

\newcommand{\rmb}{\mathrm{b}}
\newcommand{\rmc}{\mathrm{c}}
\newcommand{\rmm}{\mathrm{m}}
\newcommand{\delg}{\delta_{\rmg}}
\newcommand{\delm}{\delta_{\rmm}}
\newcommand{\hdelg}{\hat{\delta}_{\rmg}}
\newcommand{\hFg}{\hat{F}_{\rmg}}

\newcommand{\hFgam}{\hat{F}_\gamma}
\newcommand{\hFgamij}{\hat{F}_{\gamma,ij}}
\newcommand{\hFgamkl}{\hat{F}_{\gamma,kl}}
\newcommand{\chead}{\multicolumn{1}{c}}

\definecolor{olivegreen}{rgb}{0,0.6,0}
\definecolor{mycyan}{rgb}{0.13, 0.62, 0.8}
\definecolor{deepcarminepink}{rgb}{0.94, 0.19, 0.22}

\begin{document}


\title{
Constraints on anisotropic primordial non-Gaussianity \\
from intrinsic alignments of SDSS-III BOSS galaxies
}

\author{Toshiki Kurita$^{1,2}$}\email{toshiki.kurita@ipmu.jp}
\author{Masahiro Takada$^{1}$}
\affiliation{%
$^1$Kavli Institute for the Physics and Mathematics of the Universe (WPI),\\
 The University of Tokyo Institutes for Advanced Study (UTIAS),\\
 The University of Tokyo, Chiba 277-8583, Japan
}%
\affiliation{%
$^2$Department of Physics, Graduate School of Science,\\
 The University of Tokyo, 7-3-1 Hongo, Bunkyo-ku, Tokyo 113-0033, Japan
}%
\date{\today}

\begin{abstract}
We measure the three-dimensional cross-power spectrum of galaxy density and intrinsic alignment (IA) fields for the first time from the spectroscopic and imaging data of SDSS-III BOSS galaxies, for each of the four samples in the redshift range $0.2 < z < 0.75$.
In the measurement we use the power spectrum estimator, developed in our previous work, to take into account the line-of-sight dependent projection of galaxy shapes onto the sky coordinate and the $E/B$-mode decomposition of the spin-2 shape field. 
Our method achieves a significant detection of the $E$-mode power spectrum with the total signal-to-noise ratio comparable with that of the quadrupole moment of the galaxy density power spectrum, while the measured $B$-mode power spectra are consistent with a null signal to within the statistical errors for all the galaxy samples. 
We also show that, compared to the previous results based on the two-dimensional projected correlation function, our method improves the precision of the linear shape bias parameter estimation by up to a factor of two thanks to the three-dimensional information.
By performing a joint analysis of the galaxy density and IA power spectra in the linear regime, we constrain the isotropic and anisotropic local primordial non-Gaussianities (PNGs) parameters, $f_\mathrm{NL}^{s=0}$ and $f_\mathrm{NL}^{s=2}$, simultaneously, where the two types of PNGs induce characteristic scale-dependent biases at very large scales in the density and IA power spectra, respectively. 
We do not find any significant detection for both PNGs: the constraints
$f^{s=0}_\mathrm{NL}=57^{+30}_{-29}$ and 
$f^{s=2}_\mathrm{NL} = -67_{-269}^{+285}$ ($68\%$ C.L.), respectively.
Our method paves the way for using the IA power spectrum as a cosmological probe for current and future galaxy surveys. 
\end{abstract}
\maketitle

\section{Introduction} 
\label{sec:introduction} 

The $\Lambda$ Cold Dark Matter ($\Lambda$CDM) model has established as the standard cosmological model to describe various cosmological datasets such as cosmic microwave background radiation (CMB) \citep[e.g.][]{Komatsu+2011:WMAPY7,Planck2015_cosmo,Planck2018_cosmo}, type-Ia supernovae \citep[e.g.][]{2018ApJ...859..101S}, and large-scale structure (LSS) probes \citep[e.g.][]{Cole+2005:2dF_cosmo,Eisenstein+2005:BAO_detection,Alam+2017:BOSSDR12_cosmo,d'Amico+2020:EFTofLSS_PS_W,Ivanov+2020:EFTofLSS_PS_E,Kobayashi+2022:FullShape}.
In the standard $\Lambda$CDM scenario, the primordial perturbations, which seeded cosmic structure formation, are assumed to follow an adiabatic, Gaussian and nearly scale-invariant perturbations as predicted by standard (single-field, slow-roll) inflationary cosmology \cite{Maldacena2003:PNG,2003NuPhB.667..119A,2003JCAP...10..003C}
 \citep[also see][]{baumann_2022}. 
Statistical properties of such a Gaussian field are completely described by its power spectrum (or two-point correlation function).

Hence an exploration of primordial non-Gaussianity (PNG), which refers to any deviation from Gaussianity of the primordial perturbations, is a crucial test of the standard cosmological model. 
If any PNG is detected at a significant level, it would give a transformative advance in our understanding of the nature of physical processes involved in the generation of primordial perturbations in the early universe \citep[see e.g.][]{Maldacena2003:PNG,Chen+2007:PNG_single-field_inflation,Meerburg+2019:PNG_overview}. 
In particular, the so-called \textit{local} PNG, which has large amplitudes in the \textit{squeezed} configuration of the bispectrum, has been well studied in the literature \cite{Komatsu&Spergel2001:local-type_PNG}. 
Any detection of $f_\mathrm{NL}^\mathrm{local}$, a parameter to characterize local PNG, would rule out single-field inflation \citep[e.g.][]{Maldacena2003:PNG,Creminelli&Zaldarriaga2004:PNG_single-field_inflation,Alishahiha+2004:PNG_DBI,Chen+2007:PNG_single-field_inflation,Senatore+2010:PNG_single-field_inflation} and thus 
detection or improved limits on $f_\mathrm{NL}^\mathrm{local}$ would give crucial information on the nature of multi-field inflation \citep[e.g.][]{Linde&Mukhanov1997:PNG_multi-field_inflation,Moroi&Takahashi2001:multi-field_inflation,Enqvist&Sloth2002:multi-field_inflation,Lyth&2002Wands:multi-field_inflation,Lyth+2003:multi-field_inflation,Zaldarriaga2004:PNG_multi-field_inflation,Bartolo+2004:PNG_multi-field_inflation,Lyth2005:multi-field_inflation,Byrnes&Choi2010:PNG_multi-field_inflation_review}. 
The CMB bispectrum has been used to obtain tight constraints on local PNG \citep[e.g.][]{Komatsu+2003:Constraints_PNG_WMAPY1,Spergel+2007:WMAPY3,Komatsu+2011:WMAPY7,Planck2015:PNG,Planck2018:PNG}. 
After the pioneer work by Ref.~\cite{Dalal+2008:PNG_halo}, which found that local PNG induces characteristic \textit{scale-dependent} modulation in the linear bias of LSS tracers such as galaxies and quasars, the LSS datasets have also been used to constrain $f_\mathrm{NL}^\mathrm{local}$ \citep[e.g.][]{Slosar+2008:Constraints_PNG,Xia+2011:PNG_constraints_NVSS_SDSS_MegaZ,Ross+2013:PNG_constraints_BOSS_DR9,Giannantonio+2014:PNG_constraints_BOSS_ISW,Ho+2015:PNG_constraints_photo_quasar,Leistedt+2014:Constraints_PNG_SDSS_quasar_photo,Castorina+2019:Constraints_PNG_eBOSSDR14_quasar,Mueller+2021:Constraints_PNG_eBOSSDR16_quasar,Barreira2022:PNGbias} \citep[also see][for the recent constraint further using the galaxy biaspectrum]{Cabass+2022:Constraints_PNG_EFTofLSS_multi,D'Amico+2022:Constraints_PNG_EFTofLSS}. 

As a generalization of local PNG, one can consider \textit{anisotropic} or \textit{directional-dependent} local PNG with additional angular dependence in the primordial bispectrum expanded in terms of the Legendre polynomials \citep{Shiraishi+2013:fNL_aniso}. 
The usual local PNG corresponds to isotropic or monopole component of this generalized bispectrum. 
Several inflationary scenarios predict generation of the anisotropic local PNG: 
the solid inflation \citep[e.g.][]{Endlich+2013:solid_inflation,Bartolo+2013:solid_inflation,Endlich+2014:solid_inflation,Sitwell&Sigurdson2014:solid_inflation,Bartolo+2014:solid_inflation}, 
the existence of gauge vector fields \citep[e.g.][]{Barnaby+2012:PNG_vector_field,Bartolo+2013:PNG_fphiF2,Bartolo+2015:PNG_fphiF2}, 
primordial magnetic field \citep[e.g.][]{Shiraishi2012:PNG_primordial_magnetic,Shiraishi+2012:PNG_primordial_magnetic} \citep[also see][for a review]{Shiraishi+2013:fNL_aniso}, 
and higher-spin fields \citep[e.g.][]{Arkani-Hamed&Maldacena2015:Cosmo_Collider,Lee+2016:PNG_vector,Franciolini+2018:PNG_higher_spin}.
Such dipolar and quadrupolar PNGs have been constrained by the \textit{Planck} CMB bispectrum \cite{Planck2015:PNG,Planck2018:PNG}. 

As predicted by Ref.~\cite{Schmidt+2015:IA_PNG}, in analogy with isotropic local PNG, anisotropic local PNG induces a quadrupolar modulation in the local power of short-mode matter fluctuations, i.e. induces a coupling between the local tidal field and the long-wavelength tidal field. 
The LSS tidal field can be probed via ``intrinsic'' galaxy shapes, more precisely by measuring large-scale correlations of galaxy shapes with the surrounding tidal field of LSS -- the so-called intrinsic alignments (IA) \citep[e.g.][]{Croft&Metzler2000:IA_dawn,Catelan+2001:IA_dawn,Crittenden+2002:IA_dawn}. 
Hence one can realize an importance consequence of such anisotropic local PNG: in a very similar way to the effect of isotropic local PNG on galaxy density field, anisotropic local PNG induces a scale-dependent bias of the large-scale tidal field traced by 
intrinsic galaxy shapes on very large scales \citep[see also][]{Schmidt+2015:IA_PNG,Chisari&Dvorkin2013:IA_PNG,Kogai+2018:IA_PNG,Kogai+2021:PNG_spin_detector,Akitsu+2021:IA_PNG}. 

The IA effect has been mainly considered as one of the most important systematic effects in weak lensing cosmology \citep{Hirata&Seljak2004:IA_LA} \cite[also see][for reviews]{Joachimi+2015:IA_review,Kiessling+2015:IA_review,Kirk+2015:IA_review,Troxel&Ishak2015:IA_review}. In contrast there has been increasing interest in the use of the IA effect as a cosmological probe \citep[e.g.][]{Schmidt&Jeong2012:LSSwithGW_2,Chisari&Dvorkin2013:IA_PNG,Schmidt+2015:IA_PNG,Kogai+2018:IA_PNG,Okumura+2019:IA_density&velocity,Okumura&Taruya2020:IA_improvement,Taruya&Okumura2020:IA_improvement,Kurita+2020:IA_nbody,Vlah+2020:IA_EFT,Vlah+2021:IA_EFT,Akitsu+2021:IA_PNG,Shi+2020:IA_hydro,Shi+2021:IA_hydro,Akitsu+2021:IA_SU,Okumura&Taruya2022:IA_improvement,Akitsu+2022:SU_sim_GW,vanDompseler+2023:IA_BAO_toy_model}
\citep[see][for the recent, actual cosmological application]{Okumura&Taruya2023:IA_constraint_fs8}. 
While standard cosmology analysis of galaxy clustering is done treating galaxies as ``point'' distribution, where the galaxy density field is a scalar field, the galaxy shape field carries information on vector and tensor perturbations of LSS in addition to scalar perturbations
\citep{Vlah+2020:IA_EFT,Kurita&Takada2022:AnalysisIAPS}.
Hence the IA cosmology can open up a new direction, or at least play a complementary role to the standard density analysis, for cosmology. 

Hence the purpose of this paper is to constrain the anisotropic local PNG from measurements of the IA power spectrum from the spectroscopic and imaging SDSS galaxy catalogs.
To do this, we use the power spectrum measurement method, developed in our previous work \citep{Kurita&Takada2022:AnalysisIAPS}, to take into account the line-of-sight dependent projection of galaxy shapes onto the sky coordinate and the $E/B$-mode decomposition of the spin-2 galaxy shape field. 
Compared to the two-dimensional (projected) correlation function that has been commonly used in previous works \citep[][]{Mandelbaum+2006:IA_measurement,Hirata+2007:IA_measurement,Okumura&Jing2009:IA_measurement_letter,Joachimi+2011:AIA_C1,Li+2013:IA_measurement_CMASS,Singh+2015:IA_measurement,Johnston+2019:IA_measurement_GAMA,Samuroff+2019:DESY1_IA,Fortuna+2021:IA_measurement_KiDS-1000,Samuroff+2022:IA_measurement_DES&eBOSS}, 
our power spectrum analysis enables one to extract the full information of IA effects at a two-point statistics level \citep[however, see][for the use of 3D IA correlation functions]{Singh&Mandelbaum2016:IA_measurement,Okumura&Taruya2023:IA_constraint_fs8}. 
For the model template used in parameter inference, we employ the linear alignment model \citep{Hirata&Seljak2004:IA_LA}, including the survey window convolution \cite{Kurita&Takada2022:AnalysisIAPS}, integral constraint, and weak lensing contamination.
For the covariance matrix that describes statistical errors of the IA power spectrum, we use an analytic method by extending the method for the covariance matrix of galaxy density power spectrum \cite{Wadekar&Scoccimarro2020:CovPT}. 
By performing joint likelihood analyses of the measured galaxy clustering and IA power spectra, we will estimate the linear shape bias ($A_\mathrm{IA}$) and obtain constraints on the amplitudes of the isotropic and anisotropic (quadrupolar) local PNGs. 
Our work using the IA effect as a PNG probe is the first of its kind to be performed for the actual galaxy survey dataset. 

The structure of this paper is as follows. 
In Section~\ref{sec:data}, we describe the galaxy samples constructed from the SDSS-III BOSS catalog. 
In Section~\ref{sec:estimator}, we describe the method to measure the galaxy and IA power spectra. 
In Section~\ref{sec:analysis_method}, we first describe the theoretical template based on the linear alignment model with local PNGs including observational effects such as the window convolution, integral constraint, and weak lensing contamination.
Next we describe an analytic method to compute the covariance of the IA power spectrum derived in this work, and then describe the parameters and priors used in the likelihood analysis. 
In Section~\ref{sec:results}, we show the measured IA power spectrum and constraints on the local PNG parameters.
We will give our conclusions in Section~\ref{sec:conclusion}. 

Throughout this paper, we use the following abbreviations:
\begin{align*}
    \int_\bx \equiv \int \mathrm{d}\bx,~
    \int_\bk \equiv \int \frac{\mathrm{d}\bk}{(2\pi)^3}. 
\end{align*}
We also use notations for the Fourier and inverse Fourier transforms as
\begin{align*}
    f(\bk) \equiv \int_\bx f(\bx) e^{-i\bk\cdot\bx},~
    f(\bx) \equiv \int_\bk f(\bk) e^{i\bk\cdot\bx}.
\end{align*}
We quote the mode of 1D posterior for the central value of a parameter and the 68\% credible interval for the parameter uncertainties, unless otherwise stated. 

\section{Data} 
\label{sec:data}

\subsection{Density Sample} 
\label{subsec:density_sample}
We use the publicly available large-scale structure catalog of SDSS-III BOSS data release 12 (DR12), named \texttt{CMASSLOWZTOT} galaxy sample\footnote{\url{https://data.sdss.org/sas/dr12/boss/lss/}}, provided by Ref.~\cite{Reid+2016:LSScatalog}. 
We call this sample as the \textit{density sample} throughout this work. 
In our analysis, we divide the full sample into two redshift bins, ``low-z'' ($0.2<z<0.5$) and ``high-z'' ($0.5<z<0.75$) for each disjoint footprint, Northern Galactic Cap (NGC) and Southern Galactic Cap (SGC), following previous galaxy power spectrum analyses \citep[e.g.][]{Beutler+2017,d'Amico+2020:EFTofLSS_PS_W,Ivanov+2020:EFTofLSS_PS_E,Kobayashi+2022:FullShape}. 
Thus we simultaneously analyze the four no-overlapping data chunks in this work. 
To remove observational, apparent fluctuations and obtain unbiased estimates of the galaxy density field, each galaxy in the BOSS catalog is assigned the total incompleteness weight: 
\begin{align}
    w_{\mathrm{c},i} \equiv w_{\mathrm{sys},i}(w_{\mathrm{fc},i} + w_{\mathrm{rf},i} - 1), 
    \label{eq:w_c}
\end{align}
where $w_\mathrm{sys} \equiv w_\mathrm{star} w_\mathrm{see}$ is the angular systematic weight defined as the product of the stellar-density and seeing weights, 
and 
$w_\mathrm{fc}$ and $w_\mathrm{rf}$ are the nearest neighbor weights responsible for fiber collision and redshift failure, respectively \cite{Ross+2012:wsys,Anderson+2014:wsys_updated,Reid+2016:LSScatalog}. 
Using this weight, we define the weighted number of galaxies as $N'_{\rmg} \equiv \sum_{i=1}^{N_{\rmg}} w_{\mathrm{c},i}$ where $N_{\rmg}$ is the unweighted number of galaxies. 
In addition, we adopt the so-called FKP weight \cite{Feldman+1994:FKP}:
\begin{align}
    w_{\mathrm{FKP},\rmg}(z) \equiv \frac{1}{1+\bar{n}'_{\rmg}(z) P_0}, 
    \label{eq:fkp_weight}
\end{align}
where $\bar{n}'_{\rmg} = w_\mathrm{c} \bar{n}_{\rmg}$ is the weighted number density of the density sample with $P_0 = 10^4 ~(h^{-1}\Mpc)^3$, which has been commonly used in the standard cosmological analysis of the galaxy spectrum \citep[e.g.][]{Beutler+2017}. 
Using these weights, we define the effective redshift for each sample as 
\begin{align}
    z_\mathrm{eff} \equiv \frac{\sum_{i=1}^{N_{\rmg}} w_{\mathrm{c},i} w_{\mathrm{FKP},i} z_i}{\sum_{i=1}^{N_{\rmg}} w_{\mathrm{c},i} w_{\mathrm{FKP},i}}, 
\end{align}
and use this value to compute the model prediction of power spectrum in our analysis. 

For the random particles, we use the random catalog file, named \texttt{random0}, corresponding to the \texttt{CMASSLOWZTOT} sample, which includes $50$ times larger number of particles than that of galaxies in order to represent the redshift and angular distributions of the data.
We call it as the \textit{density randoms}. 

In Table~\ref{tab:sample_property}, we summarize the properties of our samples. 
\begin{table*}
    \centering
    \begin{tabular}{crrrrrrrr}
    \toprule\midrule
    Sample & & \multicolumn{3}{c}{Density ($\rmg$)} & \multicolumn{4}{c}{Shape ($\gamma$)}\\ 
    & \chead{$z_\mathrm{eff}$} & \chead{$N_{\rmg}$} & \chead{$N'_{\rmg}$} & \chead{$\alpha_{\rmg}$} & \chead{$N_\gamma$} & \chead{$N'_\gamma$} & \chead{$\alpha_\gamma$} & \chead{$\sigma_\gamma$}\\
    \midrule
    NGC low-z & \multirow{2}{*}{0.38} & 429182 & 445261 & 0.0206 & 290328 & 299697 & 0.0241 & 0.1626 \\
    SGC low-z & & 174819 & 182677 & 0.0212 & 56301 & 58358 & 0.0330 & 0.1666 \\
    \midrule
    NGC high-z & \multirow{2}{*}{0.61} & 435741 & 467502 & 0.0205 & 273573 & 291536 & 0.0241 & 0.1780 \\
    SGC high-z & & 158262 & 169907 & 0.0212 & 49744 & 52021 & 0.0323 & 0.1806\\
    \midrule\bottomrule
    \end{tabular}
    \caption{Basic characteristics of our density (as denoted by subscript ``g'') and shape (``$\gamma$'') samples in each disjoint region. 
    We list the effective redshift ($z_\mathrm{eff}$), the unweighted and weighted numbers of galaxies
    ($N_{\beta}$ and $N'_{\beta}$), and the ratio of the weighted galaxy number to random particle number ($\alpha_{\beta}$) for both samples $\beta \in \{ \rmg, \gamma\}$. 
    We also show the rms of shear of galaxies defined in Eq.~(\ref{eq:responsivity}) for shape samples. 
    }
    \label{tab:sample_property}
\end{table*}

\subsection{Shape Sample} 
\label{subsec:shape_sample}
To measure the three-dimensional intrinsic alignment (IA) power spectrum, we need information on \textit{shape} for each galaxy in addition to the spectroscopic redshift. 
In this work, we utilize the shape catalog of SDSS galaxies created and validated in Refs.~\cite{Reyes+2012:shape_catalog,Mandelbaum+2013:clusteringXgglens_dr7,Nakajima+2012:shape_catalog}. 
By cross-matching the shape catalog with the \texttt{CMASSLOWZTOT} catalog, we define our \textit{shape sample} that is a subsample of the \textit{density sample}, where each galaxy has spectroscopic redshift and the precisely-measured ellipticities. 
After this selection, 67.3 (62.4) percent of galaxies survive for NGC low-z (high-z) shape sample, whereas 31.9 (30.6) percent for SGC low-z (high-z) shape sample. 
Note that the significant degradation of the number of available galaxies in the SGC shape sample is mainly due to the $r$-band magnitude cut due to the galactic extinction \citep{Reyes+2012:shape_catalog}. 
To assign the weighted mean number density for each galaxy in the shape sample, we first compute the averaged redshift distribution of the two samples, $p_{\rmg, \gamma}(z)$, and then use the ratio to assign $\bar{n}'_{\gamma, i} \equiv \bar{n}'_{\rmg, i} \times p_\gamma(z_i) / p_{\rmg}(z_i)$ to the $i$-th galaxy in the shape sample. 
Notice that we use the same incompleteness weight $w_\mathrm{c}$ even for the shape sample assuming that the shape measurement failure is independent of the other systematics included in Eq.~(\ref{eq:w_c}).

\begin{figure*}
    \centering
    \includegraphics[width=2.0\columnwidth]{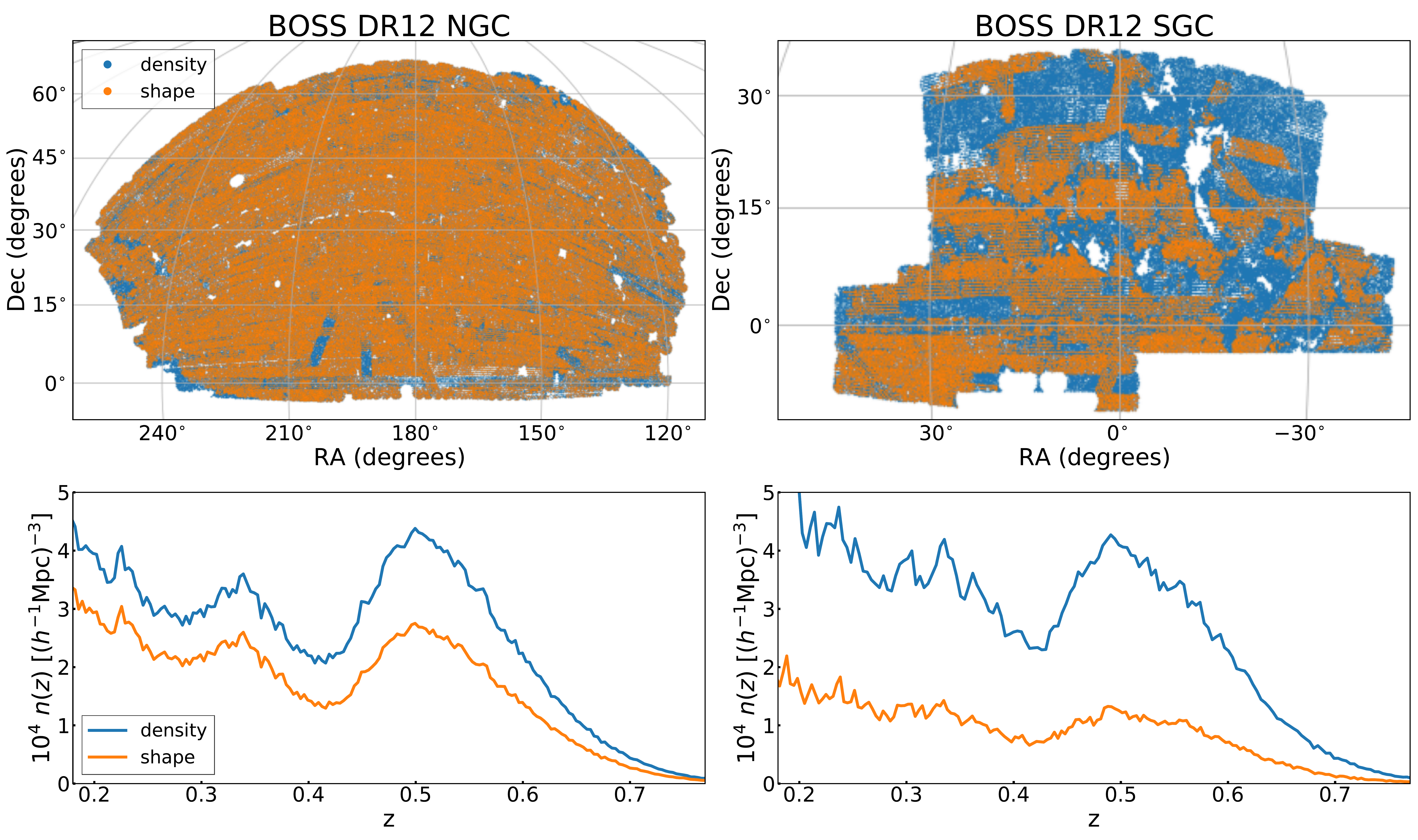}
    \caption{
    The angular distribution (top) and redshift distribution (bottom) of SDSS-III BOSS DR12 galaxy sample in the NGC (left column) and SGC (right).
    The density sample (blue dots/lines) is constructed from the large-scale structure CMASS+LOWZ-combined catalog, and we overplot the shape sample (orange) which is a subsample of the density sample that is obtained by making a cross-matching with the shape catalog constructed from the imaging data. 
    To compute the weighted number density in the bottom panels, we assume flat $\Lambda$CDM cosmology with $\Omega_\rmm=0.31$.
    }
    \label{fig:data}
\end{figure*}

Fig.~\ref{fig:data} shows the angular and redshift distributions of the density and shape samples used in our analysis. 
Since there are non-uniform differences between the distributions of two samples, we need to generate a random catalog that properly mimics the three-dimensional distribution of the shape sample, and will then use the random catalog to compute window convolutions and covariance estimates of the IA power spectrum as we will describe later in detail. 
With the definition of our shape sample in mind, we define the \textit{shape randoms} from a subsample of the \textit{density randoms} using the acceptance-rejection method as follows. 
To address the angular distribution, we first assign each galaxy in both the density and shape samples to the equal-area pixels on the sky using \texttt{HEALPix} code\footnote{\url{http://healpix.sourceforge.net/}} \cite{G'orski+2005:HEALPix} with $N_\mathrm{side}=64$, and compute the ratio between the number counts of the two samples for each pixel ``$p$'': $r_p \equiv \sum_{i\in \mathrm{pix}~p}^{N_\gamma} w_{\mathrm{c},i}/ \sum_{i\in \mathrm{pix}~p}^{N_\rmg} w_{\mathrm{c},i}$. 
After that we perform the same assignment for the density randoms, denoting the number counts of each pixel as $N_{\rmr,p} \equiv \sum_{i\in \mathrm{pix}~p}^{N_{\rmr,\rmg}}$, and then randomly sample $r_p N_{\rmr,p}$ particles from the whole $N_{\rmr,p}$ particles. 
The resultant random sample reproduces the angular distribution of the shape sample, but still obeys $p_\rmg(z)$, not $p_\gamma(z)$, in the redshift direction. 
Thus next to obtain the sample drawn from the desired distribution $p_\gamma(z)$, we further perform the rejection sampling so that the particles in the resultant sample reproduce 
the redshift distribution of the shape sample, $p_\rmg(z)$. 
In this way we obtain the random sample, i.e. shape randoms, for the galaxy shape sample.

\section{Estimators} 
\label{sec:estimator}
\subsection{Density/Shape Fields} 
\label{subsec:field_estimator}

To perform the Fourier space analysis using Fast Fourier Transform (FFT) algorithm, we define the grid-based galaxy density field and galaxy shape field. 
We use the three-dimensional comoving box centered at the observer with $L_\mathrm{box} = 3750~h^{-1}\Mpc$ on a side, which entirely covers the survey volume for each of the four samples, i.e. the low-z/high-z samples in the NGC/SGC field. 
We determine the number of grids such that the Nyquist frequency satisfies $k_\mathrm{Ny}\simeq 1~\hMpci$, 
i.e. $N^3_\mathrm{grid} = 1193^3$.
By assigning galaxies and randoms in the density sample to grid points using the cloud-in-cell (CIC) interpolation scheme \cite{Hockney&Eastwood1981:Simulation}, we make the weighted galaxy density field as 
\begin{equation}
    \hFg(\bx) 
    \equiv w_{\mathrm{FKP},\rmg}(\bx) [n'_\rmg(\bx) - \alpha_\rmg n_{\rmr,\rmg}(\bx)], 
    \label{eq:density_field}
\end{equation}
where $n'_\rmg$ and $n_{\rmr,\rmg}$ are the weighted number density field of the \textit{density} sample and randoms, respectively, and $\alpha_\rmg \equiv N'_\rmg/N_{\rmr, \rmg}$ is the ratio of the weighted number of galaxies and randoms. 
We adopt flat $\Lambda$CDM cosmology with $\Omega_\rmm=0.31$, as a reference cosmology, to convert the angular position and the redshift of galaxy to the comoving coordinates throughout this paper. 

Similarly, we construct the galaxy shape field as
\begin{equation}
    \hFgam(\bx) 
    \equiv w_{\mathrm{FKP},\gamma}(\bx) n'_\gamma(\bx) \gamma(\bx), 
    \label{eq:gamma_field}
\end{equation}
where we adopt a complex representation for the shear of galaxy shapes, 
$\gamma \equiv \gamma_1 + i\gamma_2$. 
The shear is estimated from the measured ellipticities $(e_1, e_2)$ via the shear responsivity $\mathcal{R} \equiv 1-e^2_\mathrm{rms}$ \cite{Bernstein&Jarvis2002:Responsivity}:
\begin{align}
    (\gamma_1, \gamma_2) \equiv \frac{1}{2\mathcal{R}} (e_1, e_2), 
    \label{eq:responsivity}
\end{align}
where $e_\mathrm{rms}$ is the rms ellipticity of intrinsic shapes of galaxies in the shape sample. 
The indices $1$ and $2$ correspond to the ellipticity that has major axes along the directions of coordinate axes (RA, DEC) and the directions rotated by $45^{\circ}$ from coordinate axes, respectively. 
According to Ref.~\cite{Singh+2015:IA_measurement}, we set $\mathcal{R} = 0.87$. 
$n'_\gamma \equiv w_\mathrm{c} n_\gamma$ is the weighted number density of the {shape} sample. 
We also use a weight for each galaxy in the shape sample:
\begin{align}
    w_{\mathrm{FKP},\gamma}(z) \equiv \frac{1}{\sigma^2_\gamma + \bar{n}'_\gamma(z) P_0^\mathrm{IA}},
    \label{eq:fkp_weight_IA}
\end{align}
where $\sigma_\gamma$ is the rms intrinsic ellipticity of galaxy shapes in terms of $\gamma$ that is computed from $e_{\rm rms}$ taking into account the shear responsivity above. 
This weight can be derived for the IA power spectrum estimation by employing the same assumption as that in Ref.~\cite{Feldman+1994:FKP} for the density power spectrum estimation, $w_{\mathrm{FKP},\rmg}$, which was designed to minimize the statistical errors in the power spectrum measurement balancing sample variance and shot noise contributions. 
Hence we call $w_{{\rm FKP},\gamma}$ as ``FKP'' weight for IA power spectrum here\footnote{Exactly speaking, Eq.~(\ref{eq:fkp_weight_IA}) becomes an optimal weight when we use only IA-auto power spectrum in the analysis, i.e. in absence of galaxy clustering signal, since it only balances the shape noise and the diagonal component of IA-auto covariance. 
If we analyze galaxy clustering and IA simultaneously, there is a non-negligible cross covariance term and then optimal weights should be different even for the density sample in general. 
Nevertheless since the FKP weight for the density sample (Eq.~\ref{eq:fkp_weight}) has already been well established in the literature of galaxy clustering analyses, in this work we keep Eq.~(\ref{eq:fkp_weight}) unchanged for consistency with previous works and use Eq.~(\ref{eq:fkp_weight_IA}) for the shape sample although it becomes a sub-optimal choice for our joint analysis.}. 
We set $P_0^\mathrm{IA} = 1 ~(h^{-1}\Mpc)^3$ taking into account a typical amplitude of the monopole IA-auto spectrum. 

Note that we do not use further weights such as the inverse-variance weight considering both $e_\mathrm{rms}$ and the shape measurement error for each galaxy, $\sigma_e$; $w_{\mathrm{iv},i} \equiv (e^2_\mathrm{rms} + \sigma^2_{e,i})^{-1}$, which is often used in the weak lensing analysis. 
Galaxies in the shape sample, after matching with the spectroscopic density sample, tend to be brighter than typical galaxies in the original catalog based on the imaging data, so the shape measurement error is small and the shape weight is effectively uniform over the entire shape sample \cite{Singh+2015:IA_measurement,Singh&Mandelbaum2016:IA_measurement}. 

Also notice that we do not perform any subtraction with the shape randoms ($n_{\rmr, \gamma}$) when making the shape field unlike the density field because the isotropic condition, $\bar{\gamma} = \sum_{i=1}^{N_\gamma} \gamma^i / N_\gamma = 0$, holds well for the average of all galaxy shapes in each shape sample (low-z or high-z in the NGC or SGC field). 
Nevertheless, we will use it to compute the normalization factor (see below).

\subsection{Power Spectrum Estimators} 
\label{subsec:power_spectrum_estimator}

For the auto-power spectrum of the galaxy density field, we employ the local plane-parallel (LPP) estimator, so-called Yamamoto estimator \cite{Yamamoto+2006:estimator}, with the endpoint approximation \cite{Bianchi+2015:estimator,Scoccimarro2015:estimator,Hand+2017:estimator}:
\begin{align}
    \hat{P}^{(\ell)}_{\rmg\rmg}(k_b)
    \equiv \frac{2\ell+1}{\rmI_{\rmg\rmg}}
    \int_{\hbk_b} \hFg^{(\ell)}(\bk) \hFg(-\bk) - S, 
    \label{eq:lpp_estimator_GG}
\end{align}
where
\begin{align}
    \hFg^{(\ell)}(\bk) \equiv \int_{\bx} \hFg(\bx) e^{-i\bk \cdot \bx} \mathcal{L}_\ell (\hbk \cdot \hbx), 
    \label{eq:Fg_ell}
\end{align} 
and we have introduced an abbreviated notation for the binned average over the $b$-th spherical shell:
\begin{align}
    \int_{\hbk_b} \equiv \frac{1}{N_b} \sum_{\bk \in \mathrm{bin}~b},~ 
    k_b \equiv \int_{\hbk_b} |\bk|, 
\end{align} 
with $N_b$ is the number of Fourier modes within the $b$-th bin. 
In this work we employ the linearly-equally spacing from 0 to 0.25$~h\Mpc^{-1}$ with 50 bins, i.e. $\Delta k = 0.005~h\Mpc^{-1}$.
We have also introduced notation for the normalization constant and window function:
\begin{align*}
    \rmI_{\alpha\beta} \equiv \int_\bx W^\alpha_{11}(\bx) W^\beta_{11}(\bx),~ 
    \label{eq:def_of_I}
\end{align*}
with
\begin{align*}
    W^\alpha_{ij}(\bx) \equiv \bar{n}^i_\alpha(\bx) w^j_{\mathrm{tot},\alpha}(\bx), 
\end{align*} 
where $\alpha, \beta \in \{\rmg,\gamma \}$ is the label of the galaxy density or shape field, $\bar{n}$ is the mean number density,and $w_\mathrm{tot} \equiv w_\mathrm{c} w_\mathrm{FKP}$ is the total weight for each galaxy.
In practice, we compute it by taking the $r\rightarrow0$ limit of the window-auto correlation function monopole $Q_0(r)$ estimated by the random catalog (see Section~\ref{subsubsec:window_convolution} for details about the window function) as suggested in Ref.~\cite{Beutler&McDonald2021:UnifiedPgg}, not by replacing the integral, $\int_\bx \bar{n}'(\bx)\cdots$, with the summation over the random particles, $\alpha \sum_{i=1}^{N_\rmr}\cdots$. 
We employ the FFT-based method proposed in Refs.~\cite{Bianchi+2015:estimator,Scoccimarro2015:estimator} to efficiently compute Eq.~(\ref{eq:Fg_ell}) by decomposing the Legendre polynomials into the sum of the products of $\hbx$ and $\hbk$. 
We calculate the Poisson noise $S$ in the monopole moment that arises from the discrete nature of galaxies and randoms: 
\begin{align}
    S \equiv \sum_{i=1}^{N_\rmg} w^2_{\mathrm{c},i} w^2_{\mathrm{FKP},\rmg,i} 
    + \alpha^2_\rmg \sum_{i=1}^{N_{\rmr,\rmg}} w^2_{\mathrm{FKP},\rmg,i}. 
\end{align}

For the IA-galaxy cross power spectrum, we measure the multipole moments in terms of the associated Legendre polynomials, $\mathcal{L}^{m=2}_L~(L\geq2)$, by using the LPP power spectrum estimator for IA recently developed in Ref.~\cite{Kurita&Takada2022:AnalysisIAPS}. 
This choice is convenient when both measuring the power spectrum multipoles and evaluating the window convolution on the theoretical model with FFT-based implementations. 
The estimator is given by 
\begin{align}
    \hat{P}^{(L)}_{\gamma \rmg}(k_b) 
    = \frac{2L+1}{\rmI_{\gamma \rmg}} \frac{(L-2)!}{(L+2)!} 
    \int_{\hbk_b} \hFgam^{(L)}(\bk) \hFg(-\bk), 
    \label{eq:lpp_estimator_IG}
\end{align}
where 
\begin{align}
    \hFgam^{(L)}(\bk) 
    &\equiv 
    \int_{\bx} \hFgam(\bx) e^{-2i\phi_{\hbk,\hbx}} e^{-i\bk \cdot \bx} \mathcal{L}^{m=2}_L(\hbk \cdot \hbx) \nonumber\\
    &\equiv 
    \left[\int_{\bx} \hFgam(\bx) 2e^*_{ij}(\hbx) e^{-i\bk \cdot \bx} \tilde{\mathcal{L}}^{m=2}_L(\hbk \cdot \hbx) \right] \hk_i\hk_j. 
    \label{eq:Fgam_ell}
\end{align}
$e^{-2i\phi_{\hbk,\hbx}}$ is the phase factor that is needed to rotate the shape field on the plane perpendicular to the LOS direction $\hbx$ in Fourier space and to obtain the coordinate-independent quantities, i.e. $E$ and $B$ modes. 
In the second line we have used the definition: $e^{-2i\phi_{\hbk,\hbx}} \equiv 2e^*_{ij}(\hbx) \hk_i\hk_j / (1-(\hbk \cdot \hbx)^2)$ with the complex conjugate of the polarization tensor $e^*_{ij}$. 
Here we have also defined the \textit{scaled} associated Legendre polynomials with the projection factor $1-\mu^2$: 
$\tilde{\mathcal{L}}^{m=2}_L(\mu) \equiv \mathcal{L}^{m=2}_L (\mu) / (1-\mu^2)$. 
Since Eq.~(\ref{eq:Fgam_ell}) also takes the form of products of $\hbx$ and $\hbk$, we can compute it by using FFTs as in the density case (Eq.~\ref{eq:Fg_ell}). 
Note that although there is no shot noise or shape noise terms in the estimated IA-galaxy cross spectrum due to the isotropy $\avrg{\gamma}=0$, we will see that its statistical errors are dominated by these noise terms in Section~\ref{subsec:covariance}.

\section{Analysis Method} 
\label{sec:analysis_method}

In this section we describe theoretical templates to model the multipole moments of 
density auto-power spectrum and density-IA cross power spectrum that we use for the cosmological analysis, and 
describe details of the cosmology inference method. 

\subsection{Model}
\label{subsec:model}
\subsubsection{Linear theory with local PNGs}
\label{subsubsec:linear_theory}

For the theoretical templates, we employ the linear theory based model due to the following reasons. 
(i)~The linear theory of structure formation gives an accurate model that can be safely applied to any clustering observable, at least in $k$ bins in the linear regime. 
(ii)~There is no well-validated model of the IA power spectrum including the effect of redshift space distortion (RSD) effect \cite{Kaiser1987:RSD} on scales beyond the linear regime. Note that the cosmological analysis using the IA power spectrum in this paper is the first of its kind to be performed, and the previous works focused on the projected correlation function of the IA effect, often denoted as $w_{\rmg +}(r_\mathrm{p})$ where $r_\mathrm{p}$ is the projected comoving separation.
(iii)~Main focus of this paper is to constrain the local PNGs from the measured density and IA power spectra, which induces scale-dependent modifications in the power spectra at very small $k$, such as $k^{-2}$, where the linear theory is valid. 
Nevertheless, we still want to use the power spectrum information up to relatively high $k$, just before the quasi nonlinear regime, in order for us to have a sufficient constraining power of the linear density and shape bias parameters that are needed to constrain the PNG parameters  (see later for details). Hence we will below make a careful choice of the $k$ range used for the parameter inference. 

The local PNG we focus on is characterized by its bispectrum: 
\begin{align}
    &B_\Phi(\bk_1,\bk_2,\bk_3) \nonumber\\
    &=2 \sum_{\ell=0,1,2,\cdots} f_\mathrm{NL}^{s=\ell} 
    \left[ \mathcal{L}_\ell(\hbk_1\cdot\hbk_2) P_\phi(k_1)P_\phi(k_2)+2~\mathrm{perms.} \right], 
    \label{eq:bispectrum_def}
\end{align} 
where $\Phi$ is the primordial non-Gaussian potential field, $\phi$ is the Gaussian field and $f_\mathrm{NL}^{s=\ell}$ is an amplitude parameter for each order $\ell$\footnote{Our amplitude parameters of PNG are related to the \textit{Planck} convention \cite[e.g.][]{Shiraishi+2013:fNL_aniso,Planck2018:PNG} as $2f_\mathrm{NL}^{s=\ell} = c_{L=\ell}$ for any $\ell$. 
Note that the ``NL'' parameter in Ref.~\cite{Planck2018:PNG} is thus different from ours. 
For example, $f_\mathrm{NL}^{s=2}~\mathrm{(this~work)} = -8 f_\mathrm{NL}^{L=2}~\mathrm{(the~\textit{Planck}~paper)}$ for $\ell=2$.}. 
In this work we consider the lowest two components, $s=0$ and $s=2$, which have large amplitudes in the squeezed limit. 

The \textit{isotropic} component, $s=0$, has been well studied in the literature \citep[e.g.][]{Bartolo+2004:PNG_review,Dalal+2008:PNG_halo}. 
This type of bispectrum can be realized by the nonlinear transformation in configuration space: 
\begin{align}
    \Phi(\bx) = \phi(\bx) + f_\mathrm{NL}^{s=0} \left( \phi^2(\bx) - \avrg{\phi^2}  \right).
    \label{eq:nonlinear_trs_iso}
\end{align}
In the presence of this local PNG, a modulation in the local matter power spectrum due to the mode-coupling between the long-mode primordial potential field and the small-scale density fluctuation leads to a change of the local number density of galaxies. 
As a result, the linear galaxy bias has an additional scale-dependent term given by:
\begin{align}
    b_1(k;f_\mathrm{NL}^{s=0}) = b_1 + b_\phi f_\mathrm{NL}^{s=0} \mathcal{M}^{-1}(k,z), 
    \label{eq:scale_dependent_bias_b1}
\end{align}
where we denote the response of galaxy number density to the PNG as $b_\phi$ (PNG bias parameter), and $\mathcal{M}(k,z)$ is the transfer function which relates the matter density to the primordial potential in the linear regime as $\delta_\rmm(\bk) = \mathcal{M}(k,z) \Phi(\bk)$, where $\mathcal{M}(k,z) \equiv (2/3) k^2T(k) D(z) / (\Omega_\rmm H_0^2)$, with $T(k)$ and $D(z)$ denoting the transfer function and the linear growth factor, respectively. 
Since $T(k)\rightarrow k^0$ at $k\ll k_{\rm eq}$, where $k_{\rm eq}$ is the wavenumber corresponding to the horizon scale of matter-radiation equality, this PNG induces a scale-dependent modification given by $k^{-2}$ at very small $k$ scales. 

Based on the above background, we adopt the linear model of galaxy power spectrum with the local PNG: 
\begin{align}
    P_{\rmg\rmg}(k,\mu) = \left[ b_1(k;f_\mathrm{NL}^{s=0}) + f\mu^2 \right]^2 P(k) + \frac{c_\mathrm{np}}{\bar{n}}, 
    \label{eq:Pgg_underlying_model}
\end{align}
where $f \equiv \rmd \mathrm{ln}\,D / \rmd \mathrm{ln}\,a$ is the linear growth rate, $P(k)$ is the linear matter power spectrum, and $c_\mathrm{np}$ is a parameter to model the residual shot noise.

The \textit{anisotropic} component, $s=2$, also can be realized by the nonlinear transformation \cite{Akitsu+2021:IA_PNG} in a similar way to Eq.~(\ref{eq:nonlinear_trs_iso}): 
\begin{align}
    \Phi(\bx) = \phi(\bx) + \frac{2}{3} f_\mathrm{NL}^{s=2}
    \sum_{i,j} \left[ (\psi_{ij})^2(\bx)-\avrg{(\psi_{ij})^2} \right],
    \label{eq:nonlinear_trs_aniso}
\end{align}
with the trace-less auxiliary function $\psi_{ij} \equiv 3/2 (\partial_i \partial_j / \partial^2 - \delta_{ij}^\mathrm{K}/3) \phi$.
This PNG also modulates the local matter power spectrum, but in an anisotropic (quadrupolar) way, producing a modulation in the quadrupolar shape of objects. 
Thus the linear shape bias has a scale-dependent term in this case \cite{Schmidt+2015:IA_PNG,Akitsu+2021:IA_PNG}: 
\begin{align}
    b_K(k;f_\mathrm{NL}^{s=2}) = b_K + b_\psi f_\mathrm{NL}^{s=2} \mathcal{M}^{-1}(k),
    \label{eq:scale_dependent_bias_bK}
\end{align}
where we denote the response of shapes to the anisotropic PNG as $b_\psi$\footnote{In this work we adopt a different notation of the PNG-induced shape bias $b_\psi$ from that of Ref.~\cite{Akitsu+2021:IA_PNG} by a factor of 12.}. 
Hence our linear model of the IA-galaxy cross power spectrum is given by
\begin{align}
    P_{\gamma\rmg}(k,\mu) = \frac{1-\mu^2}{2} b_K(k;f_\mathrm{NL}^{s=2}) \left[ b_1(k;f_\mathrm{NL}^{s=0}) + f\mu^2 \right] P(k). 
\end{align}
The geometrical factor, $(1-\mu^2)$, arises from the fact that we can measure only the {\it projected} shapes to the plane perpendicular to the line-of-sight direction, leading the power spectrum to arise from Fourier components that are perpendicular to the line-of-sight direction. 
In simpler words, for Fourier modes with $\mu\pm 1$ that correspond to the modes parallel to the line-of-sight direction, the above power spectrum is vanishing, while Fourier modes with $\mu=0$, the modes perpendicular to the line-of-sight direction, maximize the power spectrum amplitude in a given $k(=|\bk|)$ bin. 

Note that the shape field estimated by Eq.~(\ref{eq:gamma_field}) becomes a density-weighted field as $n_\gamma(\bx) \gamma(\bx) = \bar{n}_\gamma(\bx) (1+\delta_\gamma(\bx)) \gamma(\bx)$. 
Since the corrections due to the density weighting should be higher order effects, $\sim\mathcal{O}(\delta\gamma)$ and we only focus on the signals in the linear regime in our analysis, we ignore this effect hereafter.

\subsubsection{Window convolution}
\label{subsubsec:window_convolution}

\begin{figure*}
    \centering
    \includegraphics[width=2.0\columnwidth]{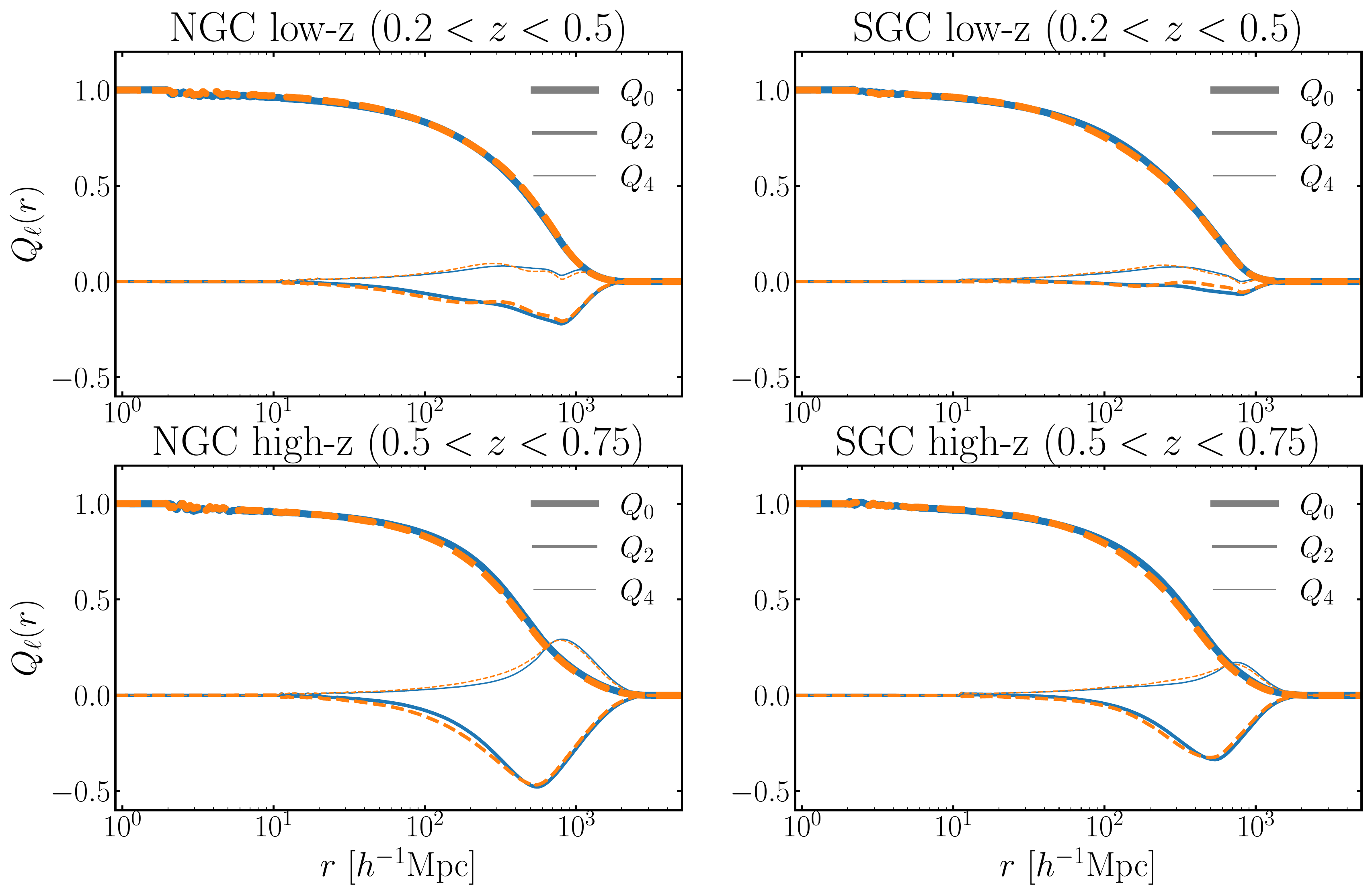}
    \caption{
    Window functions measured from the random catalogs for the different regions. 
    The blue (solid) line is for the window function of the galaxy power spectrum $Q^{\rmg\rmg}_\ell$ and the orange (dot-dashed) line is for that of the IA power spectrum $Q^{\gamma\rmg}_\ell$. 
    The thick, medium, and thin lines correspond to the multipole moments $\ell=0,2,4$, respectively. 
    }
    \label{fig:window}
\end{figure*}

The measured power spectra with the estimators in Eqs.~(\ref{eq:lpp_estimator_GG}) and (\ref{eq:lpp_estimator_IG}) are affected by the window effects due to finite survey volume and spatially-varying weights. 
To implement the window convolutions on the theoretical power spectrum models, we employ the rapid and precise method based on pair-counting approach developed in Refs.~\cite{Wilson+2017,Beutler+2017} for galaxy clustering, and extended to IA in Ref.~\cite{Kurita&Takada2022:AnalysisIAPS}. 
The main steps of the strategy are as follows: 
\begin{itemize}
    \item[(i)] precompute the multipole moments of the window correlation functions for the samples $\alpha, \beta \in \{\rmg,\gamma \}$:
    \begin{align*}
        Q^{\alpha\beta}_{\ell''}(r) 
        \equiv (2\ell''+1) \int \frac{\rmd \Omega_{\hbr}}{4\pi} \int_{\bx} W^\alpha_{11}(\bx) W^\beta_{11}(\bx+\br) \mathcal{L}_{\ell''}(\hbr\cdot\hbx),
    \end{align*}
    by counting random particle pairs in the catalog(s). 
    We show the measured window functions for each data chunks in Fig.~\ref{fig:window}. 
    Notice that the $r\rightarrow 0$ limit of the monopole corresponds to the normalization factor in Eq.~(\ref{eq:def_of_I}). 
    We use this limit value as $\rmI_{\alpha\beta}$ to normalize the measured power spectrum to keep a consistency with the theory side.
    \item[(ii)] compute the correlation function multipoles $\xi^{(\ell')}_{\alpha\beta}$ by the inverse Hankel transforms of model power spectrum multipoles $P^{(\ell')}_{\alpha\beta}$ with the spherical Bessel functions $j_{\ell'}$: 
    \begin{align*}
        \xi^{(\ell')}_{\alpha\beta}(r) 
        = i^{\ell'} \int_0^\infty \frac{k'^2 \mathrm{d}k'}{2\pi^2} j_{\ell'}(k'r)~ P^{(\ell')}_{\alpha\beta}(k'). 
    \end{align*}
    \item[(iii)] multiply $Q^{\alpha\beta}_{\ell''}$ and $\xi^{(\ell')}_{\alpha\beta}$ together and sum up them with the coefficients $c^{\alpha\beta}_{\ell \ell' \ell''}$ for the target multipole $\ell$: 
    \begin{align*}
        \tilde{\xi}^{(\ell)}_{\alpha\beta}(r) \equiv \sum_{\ell',\ell''} c^{\alpha\beta}_{\ell \ell' \ell''} Q^{\alpha\beta}_{\ell''}(r) \xi^{(\ell')}_{\alpha\beta}(r),
    \end{align*}
    where
    \begin{align*}
        c^{\alpha\beta}_{\ell \ell' \ell''} &\equiv
        (2\ell+1) \sqrt{\frac{(\ell-m_{\alpha\beta})!}{(\ell+m_{\alpha\beta})!} \frac{(\ell'+m_{\alpha\beta})!}{(\ell'-m_{\alpha\beta})!}} \\
        &\quad\quad\quad\quad\times
        \begin{pmatrix}
            \ell'' & \ell & \ell'\\
            0 & 0 & 0
        \end{pmatrix}
        \begin{pmatrix}
            \ell'' & \ell & \ell'\\
            0 & m_{\alpha\beta} & -m_{\alpha\beta}
        \end{pmatrix},
    \end{align*}
    with $m_{\rmg\rmg}=0$ and $m_{\gamma\rmg}=2$, respectively.
    The $2\times3$ matrix form represents the Wigner $3j$ symbol. 
    Note that $c^{\rmg\rmg}_{\ell \ell' \ell''}$ is the same coefficient used in the clustering analyses \cite{d'Amico+2020:EFTofLSS_PS_W,Beutler&McDonald2021:UnifiedPgg}.
    \item[(iv)] perform the Hankel transform of order $\ell$ to obtain the window-convolved power spectrum multipole as
    \begin{align*}
        \tilde{P}^{(\ell)}_{\alpha\beta}(k) 
        = 4\pi(-i)^\ell \int_0^\infty r^2 \mathrm{d}r j_{\ell}(kr)~ \tilde{\xi}^{(\ell)}_{\alpha\beta}(r). 
    \end{align*}
\end{itemize}
We use the public code \texttt{CAMB} \cite{Lewis+2000:CAMB} to compute the linear matter power spectrum (more exactly, the transfer function) for a given cosmological model. 
To evaluate the (inverse) Hankel transforms, we use the publicly available \texttt{FFTLog} code \cite{Fang+2020:fftlog}. 

\subsubsection{Integral constraint}
\label{subsubsec:integral_constraint}

Since we always define the density fluctuation using the total number of galaxies observed within a finite survey region, the \textit{integral constraint} (IC) is imposed on the measured galaxy correlation function and power spectrum: 
\begin{align}
    \int_{V_\mathrm{s}} \hat{\xi}_\mathrm{gg}^\mathrm{obs}(\br) \rmd \br = 0 ~\Leftrightarrow~\hat{P}_\mathrm{gg}^\mathrm{obs}(\bk\to\mathbf{0})=0.
\end{align}
Therefore, we must impose this condition on the theoretical model used in the analysis.
We correct for IC in the model prediction of power spectrum by subtracting the DC mode from the naive power spectrum estimator \citep[also see][]{Peacock&Nicholson1991:IC,Wilson+2017}: 
\begin{align}
    \tilde{P}^{(\ell)}_{\mathrm{gg,IC}\mathchar`-\mathrm{corrected}}(k) 
    = \tilde{P}_\mathrm{gg}^{(\ell)}(k) - \frac{Q_\ell(k)}{Q_0(k\to 0)} \tilde{P}_\mathrm{gg}^{(0)}(k\to 0), 
    \label{eq:Pgg_IC_corrected}
\end{align}
where $Q_\ell(k)$ is the $\ell$-th order Hankel transform of $Q_\ell(r)$. 

Note that this correction is usually small enough and also cause no numerical problem in the case of the usual $\Lambda$CDM cosmological analyses with the Gaussian initial condition because the underlying power spectrum ($P_\mathrm{gg} \propto k^{n_\mathrm{s}}$ at small $k$) already satisfies IC and then the DC limit of the window-convolved monopole power spectrum $\tilde{P}_\mathrm{gg}^{(0)}(k\to0)$ is finite and sufficiently small for a large-volume survey. 
However, the situation is different in the presence of the local PNG. 
The local PNG is imprinted on the galaxy number fluctuation as the scale-dependent bias, which causes the additional terms proportional to $f^{s=0}_\mathrm{NL}k^{n_\mathrm{s}-2}$ and $(f^{s=0}_\mathrm{NL})^2k^{n_\mathrm{s}-4}$ at small $k$ for the galaxy power spectrum as in Eq.~(\ref{eq:Pgg_underlying_model}). 
In particular, for the latter, its inverse Fourier transform leads to an IR divergence:
$\propto \int j_0(kr) \times k^{n_\mathrm{s}-2}\rmd k$. 
Although the observational IC imposed as the subtraction in Eq.~(\ref{eq:Pgg_IC_corrected}) ensures the exact cancellation of this divergence \cite{Wands&Slosar2009:PNGwithGR}, the numerical implementation should be carefully done because the results might easily depend on the choice of the minimum wavenumber of the $k$-integral, $k^\mathrm{th}_\mathrm{min}$. 
We checked that our model predictions after the IC correction are consistent at sub-percent level in the $k$-range of interest ($k>0.01~h\Mpc^{-1}$) even if we change $k^\mathrm{th}_\mathrm{min}$ by an order of magnitude, compared to our fiducial choice of $k^{\rm th}_{\rm min}=3\times 10^{-5}~\hMpci$.

\subsubsection{Weak lensing effects} 
\label{subsubsec:weak_lensing} 

The observed spatial fluctuation of the galaxy number density is affected by weak lensing effect due to the foreground large-scale structure along the same line-of-sight direction to the BOSS galaxies, the so-called magnification bias. 
The observed ellipticity of galaxy image is also distorted from its original shape by the weak lensing distortion due to the same foreground structure. 
The leading order contributions of these weak lensing effects, 
which we hereafter label as ``WL'', can be written, to a good approximation in the weak lensing regime, as 
\begin{align}
    \delta_\rmg^\mathrm{obs}(\bx) &= \delg(\bx) + \delta_\rmg^\mathrm{WL}(\bx), \label{eq:delta_WL}\\
    \gamma^\mathrm{obs}(\bx) &= \gamma^\mathrm{IA}(\bx) + \gamma^\mathrm{WL}(\bx), \label{eq:gamma_WL}
\end{align}
with $\delta_\rmg^\mathrm{WL} \equiv 2(\alpha_\mathrm{mag}-1) \kappa^\mathrm{WL}$. 
$\kappa^\mathrm{WL}$ and $\gamma^\mathrm{WL}$ are the weak lensing convergence and shear fields 
given by 
\begin{align}
    \left(\kappa^\mathrm{WL}(\bx), \gamma^\mathrm{WL}(\bx)\right) 
    \equiv \frac{1}{2} \left(\hat{\nabla}^2, \eth^2 \right) \phi^\mathrm{WL}(\bx), 
\end{align}
where $\hat{\nabla}^2$ is the Laplacian on the sphere, $\eth^2 \equiv 2e_{ij}(\hbx) \hat{\nabla}_i\hat{\nabla}_j$ with $e_{ij}$ being the polarization tensor, and $\phi^\mathrm{WL}$ is the lensing potential. 
$\alpha_\mathrm{mag}$ is defined with the slope of the cumulative galaxy number counts for galaxies brighter than magnitude $m$: 
\begin{align*}
    \alpha_\mathrm{mag} \equiv \frac{5}{2} \frac{\rmd {\ln N(<m)}}{\rmd m}, 
\end{align*}
and $\alpha_{\rm mag}$ depends on the selection function of galaxy sample (we will discuss this issue later). 

The two-point correlations of the observed fields Eqs.~(\ref{eq:delta_WL}) and (\ref{eq:gamma_WL}) then have the following three WL-related correlations in general in addition to the \textit{intrinsic} galaxy-galaxy and galaxy-IA correlations: 
\begin{align}
    \avrg{\delta_\rmg^\mathrm{obs} \delta_\rmg^\mathrm{obs}} &= \avrg{\delg \delta_\rmg} \nonumber\\
    &+ \avrg{\delg \delta_\rmg^\mathrm{WL}} + \avrg{\delta_\rmg^\mathrm{WL} \delta_\rmg} + \avrg{\delta_\rmg^\mathrm{WL} \delta_\rmg^\mathrm{WL}}, \label{eq:delg_obs-delg_obs}\\
    \avrg{\gamma^\mathrm{obs} \delg^\mathrm{obs}} &= \avrg{\gamma^\mathrm{IA} \delg} \nonumber\\
    &+ \avrg{\gamma^\mathrm{WL} \delg} + \avrg{\gamma^\mathrm{IA} \delg^\mathrm{WL}} + \avrg{\gamma^\mathrm{WL} \delg^\mathrm{WL}}. \label{eq:delg_obs-gamma_obs}
\end{align}
The second and third terms arise due to the finite radial width of our galaxy samples, i.e. the breakdown of the thin redshift shell approximation, and the last term is for the pure weak lensing ``auto''-correlation arising from the foreground structures at different redshifts from those of BOSS galaxies. 
The weak lensing effects on the galaxy density power spectrum (Eq.~\ref{eq:delg_obs-delg_obs}) have been derived and discussed by e.g. Refs.~\cite{Hui+2007:MagReal,Hui+2008:MagFourier}. 
In this work, we derive the weak lensing terms on the density-shape power spectrum (Eq.~\ref{eq:delg_obs-gamma_obs}) in a similar way including the actual survey window effects by developing the rapid convolution method. 
We show the full derivation and window convolution method for the WL-related power spectrum in Appendix~\ref{sec:weak_lensing}. 
We find that the WL contributions are not negligible for the IA power spectrum and thus we add th em (Eqs.~\ref{eq:PSWL:delg_WL-gamma_WL}, \ref{eq:PSWL:delg-gamma_WL} and \ref{eq:PSWL:delg_WL-gamma_IA}) to the model template of the IA power spectrum as
\begin{align}
    &\tilde{P}^{(L)}_{\gamma^\mathrm{obs} \rmg^\mathrm{obs}}(k) = \tilde{P}^{(L)}_{\gamma^\mathrm{IA} \rmg}(k) \nonumber\\
    &+ \tilde{P}^{(L)}_{\gamma^\mathrm{WL} \rmg}(k) 
    + \tilde{P}^{(L)}_{\gamma^\mathrm{IA} \rmg^\mathrm{WL}}(k; \alpha_\mathrm{mag}) 
    + \tilde{P}^{(L)}_{\gamma^\mathrm{WL} \rmg^\mathrm{WL}}(k; \alpha_\mathrm{mag}). 
\end{align}

The estimation of $\alpha_\mathrm{mag}$ for our galaxy samples is not straightforward because the 
magnitude and multi-color dependent cuts, used for targeting BOSS galaxies \citep{Alam+2015:SDSS-III_final_data}, make it difficult to estimate the true slope of the number counts as a function of the absolute magnitude. 
Ref.~\cite{Kramsta+2021:MagBias} carefully estimated the magnification bias for the exactly same galaxy sample (\texttt{CMASSLOWZTOT}) and redshift binning definition ($0.2<z<0.5$,~$0.5<z<0.75$) as ours by using realistic mock data built from the MICE2 simulation \cite{Fosalba+2015:MICE-I,Crocce+2015:MICE-II,Fosalba+2015:MICE-III}. 
They obtained 
$\alpha^\mathrm{low\mathchar`-z}_\mathrm{mag} = 1.93 \pm 0.05$ and 
$\alpha^\mathrm{high\mathchar`-z}_\mathrm{mag} = 2.62 \pm 0.28$, 
which were used in the Kilo-Degree Survey (KiDS-1000) cosmological inference of the joint weak lensing and galaxy clustering analysis \cite{Joachimi+2021:KiDS1000_Methodology,Heymans+2021:KiDS_3x2pt}. 
In this work, we adopt these estimations as the prior information of $\alpha_\mathrm{mag}$. 
Since the final result almost does not depend on whether we use the normal distribution or fix it at the best-fit values because of small error bars, we report the results with fixed $\alpha_\mathrm{mag}$ throughout this paper.

\subsection{Covariance} 
\label{subsec:covariance}
As far as we know, there currently does not exist a suite of realistic and well physically motivated mock catalogs for galaxy IA unlike the galaxy clustering such as the MultiDark-Patchy mock catalogs (hereafter Patchy mocks) \cite{Kitaura+2016:PatchyMock}. 
Therefore in this work, we derive an analytic covariance for IA power spectrum, following Ref.~\cite{Wadekar&Scoccimarro2020:CovPT} who derived and validated the analytic covariance for galaxy power spectrum. 
Since we only use the measured power spectrum in the linear regime, we only consider the Gaussian and shot/shape noise terms with the survey window effects, and ignore other higher order non-Gaussian terms such as the beat-coupling and local-average effects.
In Appendix~\ref{sec:derivation_covariance}, we show the detail derivation, numerical implementation, and validation tests for our covariance. 
We here summarize the formulae and show the results.
The full covariance matrix of our analysis is
\begin{align*} 
    \bC = 
    \begin{pmatrix}
        \mathrm{Cov}\left[ P_{\gamma\rmg}, P_{\gamma\rmg} \right] & \mathrm{Cov}\left[ P_{\gamma\rmg}, P_\mathrm{gg} \right]\\
        \mathrm{Cov}\left[ P_\mathrm{gg}, P_{\gamma\rmg} \right]& \mathrm{Cov}\left[ P_\mathrm{gg}, P_\mathrm{gg} \right]
    \end{pmatrix} \equiv 
    \begin{pmatrix}
        \bC^\mathrm{II} & \bC^\mathrm{IG}\\
        {}^t\bC^\mathrm{IG} & \bC^\mathrm{GG}
    \end{pmatrix},
\end{align*}
where each component has the continuous part and the shot/shape noise-related part:
\begin{align*}
    \bC^\mathrm{XY} \equiv \bC^\mathrm{XY(cont.)} + \bC^\mathrm{XY(SN)},
\end{align*}
with $\mathrm{X,Y} \in \{ \mathrm{G,I} \}$. 
The results of galaxy-auto covariance $\bC^\mathrm{GG}$ was derived in Ref.~\cite{Wadekar&Scoccimarro2020:CovPT} as
\begin{align}
    &\bC^\mathrm{GG(cont.)}_{\ell_1\ell_2}(k_1,k_2) \nonumber\\
    &= 
    \sum_{\ell'_1,\ell'_2} P^{(\ell'_1)}_\mathrm{gg}(k_1) P^{(\ell'_2)}_\mathrm{gg}(k_2) 
    \mathcal{W}^{\mathrm{GG}(1)}_{\ell_1,\ell_2,\ell'_1,\ell'_2}(k_1,k_2), \\
    &\bC^\mathrm{GG(SN)}_{\ell_1\ell_2}(k_1,k_2) \nonumber\\
    &=
    \sum_{\ell'} \left[P^{(\ell')}_\mathrm{gg}(k_1) \mathcal{W}^{\mathrm{GG}(2)}_{\ell_1,\ell_2,\ell'}(k_1,k_2) + (k_1 \leftrightarrow k_2) \right] \nonumber\\
    &\quad+ \mathcal{W}^{\mathrm{GG}(3)}_{\ell_1,\ell_2}(k_1,k_2),
\end{align}
where the window functions $\mathcal{W}^{\mathrm{GG}(i)}_{\ell_1,\ell_2,\cdots}~(i=1,2,3)$ are defined in Eqs.~(\ref{eq:cov_PggPgg_cont}), (\ref{eq:cov_W_GG_2}) and (\ref{eq:cov_W_GG_3}). 
The indices $i=1,2,3$ represent the continuous ($P\times P$), continuous-shot noise ($P\times 1/\bar{n}$) and shot noise-shot noise ($1/\bar{n} \times 1/\bar{n}$) terms, respectively. 
Similarly, we derive the IA-auto covariance $\bC^\mathrm{II}$: 
\begin{align}
    &\bC^\mathrm{II(cont.)}_{L_1L_2}(k_1,k_2) \nonumber\\
    &= \sum_{\ell'_1,\ell'_2} P^{(\ell'_1)}_{\mathrm{Eg}}(k_1) P^{(\ell'_2)}_{\mathrm{Eg}}(k_2) 
    \mathcal{W}^{\mathrm{II}(1,\mathrm{A})}_{L_1,L_2,\ell'_1,\ell'_2}(k_1,k_2) \nonumber\\
    &+ \sum_{\ell'_1,\ell'_2} \left[ 
    P^{(\ell'_1)}_{\mathrm{gg}}(k_1) P^{(\ell'_2)}_\mathrm{EE}(k_2) 
    \mathcal{W}^{\mathrm{II}(1,\mathrm{B})}_{L_1,L_2,\ell'_1,\ell'_2}(k_1,k_2) +  (k_1 \leftrightarrow k_2) \right], \\
    &\bC^\mathrm{II(SN)}_{L_1L_2}(k_1,k_2) \nonumber\\
    &= \sum_{\ell'} \bigg[ \left\{P^{(\ell')}_\mathrm{gg}(k_1) \mathcal{W}^{\mathrm{II}(2,\mathrm{shape})}_{L_1,L_2,\ell'}(k_1,k_2) \right . \nonumber\\
    &\left . \quad\quad\quad\quad+ P^{(\ell')}_\mathrm{EE}(k_1) \mathcal{W}^{\mathrm{II}(2,\mathrm{shot})}_{L_1,L_2,\ell'}(k_1,k_2) \right\} 
    + (k_1 \leftrightarrow k_2) \bigg] \nonumber\\
    &\quad\quad\quad+ \mathcal{W}^{\mathrm{II}(3)}_{L_1,L_2}(k_1,k_2),
\end{align}
where the window functions $\mathcal{W}^{\mathrm{II}(i)}_{L_1,L_2,\cdots}~(i=1,2,3)$ are defined in Eqs.~(\ref{eq:cov_ia_PartA}), (\ref{eq:WII_1_B}), and (\ref{eq:W_II_2_shape})--(\ref{eq:W_II_3}). 
`(2,shape)' and `(2,shot)' are the continuous-shape noise ($P_{\rmg\rmg} \times \sigma^2_\gamma/\bar{n}$) and the continuous-shot noise ($P_\mathrm{EE} \times 1/\bar{n}$) terms, respectively. 

We also derive the IA-galaxy cross covariance $\bC^\mathrm{IG}$: 
\begin{align}
    &\bC^\mathrm{IG(cont.)}_{L_1\ell_2}(k_1,k_2) \nonumber\\
    &= \sum_{\ell'_1,\ell'_2} \left[ P^{(\ell'_1)}_{\mathrm{gg}}(k_1) P^{(\ell'_2)}_{\mathrm{Eg}}(k_2) 
    \mathcal{W}^{\mathrm{IG}(1)}_{L_1,\ell_2,\ell'_1,\ell'_2}(k_1,k_2) 
    + (k_1 \leftrightarrow k_2) \right], \\
    &\bC^\mathrm{IG(SN)}_{L_1\ell_2}(k_1,k_2) \nonumber\\
    &= \sum_{\ell'} \left[ 
    P^{(\ell'_2)}_{\mathrm{Eg}}(k_2) 
    \mathcal{W}^{\mathrm{IG}(2)}_{L_1,\ell_2,\ell'}(k_1,k_2) 
    + (k_1 \leftrightarrow k_2) \right],
\end{align}
where the window functions $\mathcal{W}^{\mathrm{IG}(i)}_{L_1,\ell_2,\cdots}~(i=1,2)$ are defined in Eqs.~(\ref{eq:W_IG_1}) and (\ref{eq:W_IG_2}). 

Since each window function $\mathcal{W}^{\mathrm{XY}}$, which is the quartic function of $W_{11}$, has a multi-dimensional integration $\int_{\hbk_1,\hbk_2,\bx_1,\bx_2} \cdots$, the direct evaluation by the sum of the random particles would be computationally expensive. 
In this work, we employ the grid-based implementation to utilize FFTs as proposed by Ref.~\cite{Wadekar&Scoccimarro2020:CovPT} (see Appendix~\ref{sec:derivation_covariance} for numerical implementation and validation test). 

\begin{figure*}
    \centering
    \includegraphics[width=2.0\columnwidth]{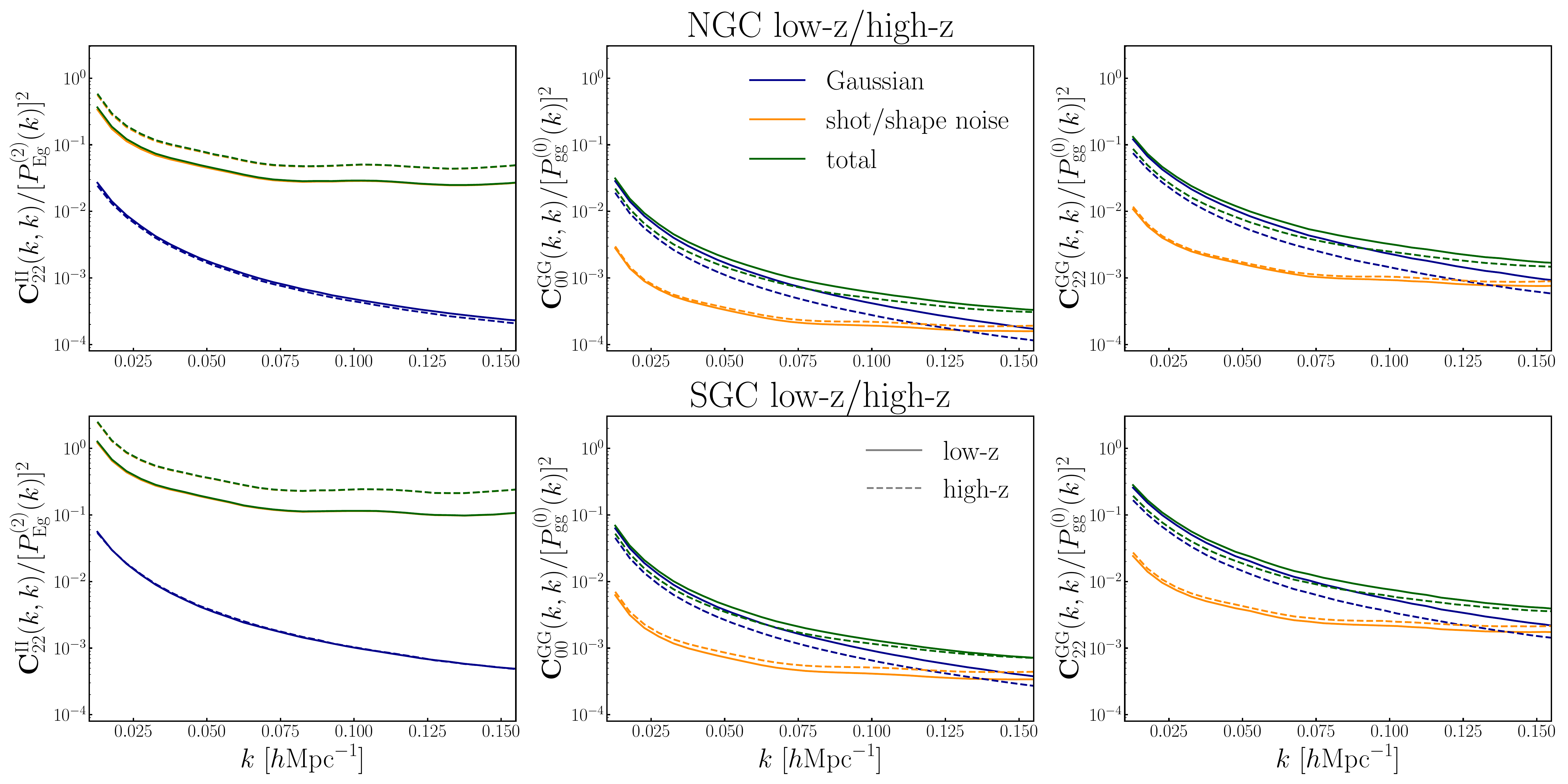}
    \caption{Relative error of the IA power spectrum (left panel) and of the monopole (middle) or quadrupole (right)
    moment of the galaxy density power spectrum, respectively. 
    For the galaxy power spectra, we use the mean power spectra of the Patchy mocks instead of the linear model. 
    We use the monopole power spectrum as the denominator in the case of the quadrupole to avoid the zero-crossing. 
    The blue, orange and green curves correspond to the Gaussian term, SN-related term and total covariance as indicated by legend. 
    The upper (lower) panel is for the NGC (SGC) and the solid (dashed) line is for the low-z (high-z), respectively.
    }
    \label{fig:covariance_diagonal}
\end{figure*}
Fig.~\ref{fig:covariance_diagonal} shows the fractional error of each power spectrum, which is defined by the diagonal elements of the covariance matrix divided by the square of the power spectrum. 
The contribution from the Gaussian component is similar among the three power spectra because its fractional amplitudes are almost determined by the number of independent Fourier modes with the cancellation of the absolute amplitude (of linear bias) of the power spectrum. 
On the other hand, the error of IA power spectrum is dominated by the shape noise component at all scales unlike the galaxy power spectrum due to the lower number density of the shape sample (see Section~\ref{sec:data}) and the smaller amplitude of the IA correlation: $b_K^2P_\rmm(k) \ll \sigma_\gamma^2/\bar{n}$. 

\begin{figure*}
    \centering
    \includegraphics[width=2.0\columnwidth]{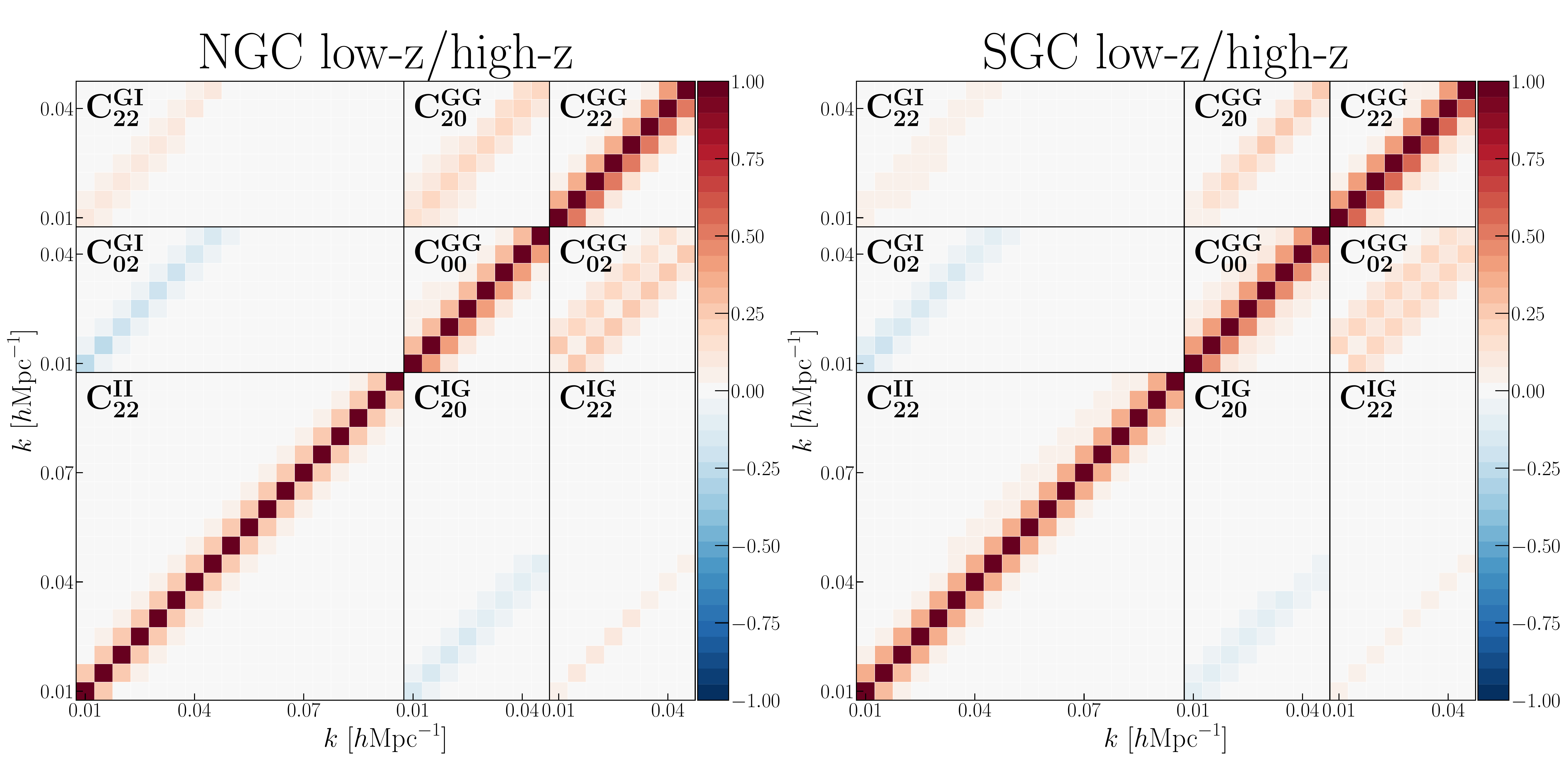}
    \caption{Correlation coefficients of the covariance matrix, defined as 
    $r_{ij} \equiv \bC_{ij}/\sqrt{\bC_{ii} \bC_{jj}}$. 
    Here we show only the elements for the IA power spectrum $P^{(2)}_\mathrm{Eg}~(0.01 \leq k < 0.1~\hMpci)$ and the galaxy power spectra $P^{(0)}_\mathrm{gg}$ and $P^{(2)}_\mathrm{gg}~(0.01 \leq k < 0.05~\hMpci)$ for the NGC (left plot) and SGC (right plot). 
    In each plot the upper-left elements are for the low-z sample and the lower-right elements are for the high-z sample, respectively. 
    The red (blue) color indicates positive (negative) correlation. 
    }
    \label{fig:correlation_matrix}
\end{figure*}
Fig.~\ref{fig:correlation_matrix} shows the corresponding correlation matrices of our analytic covariance defined by $r_{ij} \equiv \bC_{ij}/\sqrt{\bC_{ii} \bC_{jj}}$ for each galaxy sample. 
Since we adopt the analytic approach with the Gaussian and the shot/shape noise terms, the non-zero off-diagonal elements of each sub-matrix arise from the pure window smearing of the BOSS survey footprints, $\delta k \lesssim 0.03~\hMpci \sim 1/R_\mathrm{survey}$. 
Hence all the other elements beyond $\delta k$ are zero. 
This approximation would be valid for our linear-scale analysis. 
Actually, we checked our analytic covariance of the galaxy power spectrum is in good agreement with the covariance computed from the Patchy mocks up to $k_\mathrm{max} = 0.05~\hMpci$. 
Since the covariance of the IA power spectrum is dominated by the shape noise and the non-Gaussian corrections should be subdominant even at quasi-nonlinear scales, we adopt our ``linear'' covariance up to $k_\mathrm{max} = 0.1~h\Mpc^{-1}$ for the IA power spectrum in our analysis. 
Note that this $k_\mathrm{max} = 0.1~h\Mpc^{-1}$ corresponds to the acceptable maximum wavenumber for our linear \textit{model} in the analysis to obtain an unbiased constraint on the $f^{s=2}_\mathrm{NL}$ parameter based on the results of the validation test described in Appendix~\ref{sec:validation_test_for_analysis}.

\subsection{Parametrization and Priors} 
\label{subsec:params_and_priors}
In this work, we consider four types of analyses for different purposes as follows. 
We summarize the parameters and priors for each case in Table~\ref{tab:params_priors}. 
\begin{itemize}
    \item \textit{Gaussian analysis} (Section~\ref{subsec:gaussian_analysis}): 
    We set $f^{s}_\mathrm{NL}=0~(s=0,2)$ in this analysis.
    The main motivation is to determine the linear shape bias $b_K$ (or $A_\mathrm{IA}$) of our galaxy samples and check consistency with the previous work \cite{Singh+2015:IA_measurement}.
    \item \textit{PNG analysis without bias relations} (Section~\ref{subsec:png_analysis_sod}): 
    We explore the presence of PNGs, however, we do not impose any assumption on the PNG-induced linear bias parameters; we constrain a direct observable of the PNG effect, i.e. the parameter combinations, $\left(b_\phi f^{s=0}_\mathrm{NL}\right)$ and $\left(b_\psi f^{s=2}_\mathrm{NL}\right)$. 
    Although we cannot constrain the amplitude of the PNG parameter alone ($f_\mathrm{NL}^s$) in this case, this analysis can tell us a significance of the positive- or null-detection. 
    \item \textit{PNG analysis with $b_\psi (b_K)$} (Section~\ref{subsec:png_analysis_bkbpsi}): 
    We constrain $f^{s=2}_\mathrm{NL}$ assuming the relation between the linear shape bias $b_K$ and the PNG shape bias $b_\psi$. 
    We adopt $b_\psi (b_K) = 2.04\,b_K$ obtained in Ref.~\cite{Akitsu+2021:IA_PNG} for dark matter halos. 
    For the isotropic PNG $f^{s=0}_\mathrm{NL}$, on the other hand, we keep the combination $\left(b_\phi f^{s=0}_\mathrm{NL}\right)$ unchanged to constrain $f^{s=2}_\mathrm{NL}$ based on the minimum assumption. 
    This analysis corresponds to our baseline analysis. 
    \item \textit{PNG analysis with $b_\phi (b_1)$ and $b_\psi (b_K)$} (Section~\ref{subsec:png_analysis_b1bphi_bkbpsi}): 
    We constrain $f^{s=0}_\mathrm{NL}$ and $f^{s=2}_\mathrm{NL}$ simultaneously by assuming the relation between the linear galaxy bias $b_1$ and the PNG bias $b_\phi$ as well as the relation $b_\psi (b_K)$ assumed in the previous analysis setup. 
    We adopt $b_\phi (b_1) = 2\delta_\mathrm{c}(b_1-p)$ with $\delta_\mathrm{c} = 1.686$ and $p=0.55$ suggested by Ref.~\cite{Barreira+2020:PNG_bias_TNG}. 
\end{itemize}
\begin{table}
    \centering
    \begin{tabular}{cccccc}
    \toprule\midrule
    Parameter & Prior & \multicolumn{4}{c}{Analysis (Section)}\\ 
    && \ref{subsec:gaussian_analysis} & \ref{subsec:png_analysis_sod} & \ref{subsec:png_analysis_bkbpsi} & \ref{subsec:png_analysis_b1bphi_bkbpsi} \\
    \midrule
    $b_1$ & $\mathcal{U}(1,4)$ & $\checkmark$ & $\checkmark$ & $\checkmark$ & $\checkmark$\\
    $b_K$ & $\mathcal{U}(-0.2,0.2)$ & $\checkmark$ & $\checkmark$ & $\checkmark$ & $\checkmark$\\
    $c_\mathrm{np}$ & $\mathcal{N}(0.0,0.1)$ & $\checkmark$ & $\checkmark$ & $\checkmark$ & $\checkmark$\\
    \midrule
    $b_\phi f^{s=0}_\mathrm{NL}$ & $\mathcal{U}(-2500,2500)$ & - & $\checkmark$ & $\checkmark$ & -\\
    $b_\psi f^{s=2}_\mathrm{NL}$ & $\mathcal{U}(-500,500)$ & - & $\checkmark$ & - & -\\
    \midrule
    $f^{s=0}_\mathrm{NL}$ & $\mathcal{U}(-500,500)$ & - & - & - & $\checkmark$\\
    $f^{s=2}_\mathrm{NL}$ & $\mathcal{U}(-1000,1000)$ & - & - & $\checkmark$ & $\checkmark$\\
    \midrule\bottomrule
    \end{tabular}
    \caption{Model parameters and priors used in each of our analysis setups (whose results are given in  
    Sections~\ref{subsec:gaussian_analysis}--\ref{subsec:png_analysis_b1bphi_bkbpsi}, respectively, as indicated in the table header). 
    The mark ``$\checkmark$'' means that the parameter is included in the parameter inference of the corresponding setup. 
    $\mathcal{U}(a,b)$ denotes a flat prior with range $[a,b]$, while $\mathcal{N}(\mu,\sigma)$ denotes a Gaussian prior with mean $\mu$ and width $\sigma$. 
    }
    \label{tab:params_priors}
\end{table}

Except the $f^{s=0,2}_\mathrm{NL}$ parameters, we fix other cosmological parameters to the values that are consistent with the \textit{Planck} CMB data \citep{Planck2018_cosmo}: 
$\Omega_\rmm = 0.3153$, $\omega_\rmb = 0.02237$, $\omega_\rmc = 0.1200$, $n_\rms = 0.9649$, and $\ln(10^{10}A_\rms) = 3.044$. 

\subsection{Likelihood Analysis}

We estimate the parameters based on the Bayesian inference: 
\begin{align}
    p(\bp|\bd) \propto \mathcal{L}(\bd|\bp) \pi(\bp), 
\end{align} 
where $p$ is the posterior distribution of the model parameters $\bp$, $\pi$ is the prior distribution described in Section~\ref{subsec:params_and_priors}, and $\mathcal{L}(\bd|\bp)$ is the likelihood of the data vector $\bd$ given the model that is specified by a set of parameters (${\bf p}$). 
We assume the Gaussian likelihood: 
\begin{align}
    -2 \mathrm{ln}~\mathcal{L}(\bd|\bp) = {}^t[\bd - \mathbf{m}(\bp)] \bC^{-1} [\bd - \mathbf{m}(\bp)], 
\end{align} 
where $\mathbf{m}(\bp)$ is the model prediction that is given by ${\bf p}$ and $\bC$ is the covariance matrix defined in Section~\ref{subsec:covariance}. 
We omit the normalization factor. 
Note that we ignore the cross-covariance among four different galaxy samples. 

We include the lowest order multipole of the IA power spectrum, $P^{(2)}_\mathrm{Eg}$, and the monopole and quadrupole of the galaxy power spectrum, $P^{(0)}_\mathrm{gg}$ and $P^{(2)}_\mathrm{gg}$, over $k_\mathrm{min} < k < k_\mathrm{max}$ with $k_\mathrm{min} = 0.01~\hMpci$ for all the power spectra, $k_\mathrm{max} = 0.1~\hMpci$ for the IA power spectrum, and $k_\mathrm{max} = 0.05~\hMpci$ for the galaxy power spectrum, respectively. 
As we employ $\Delta k = 0.005~\hMpci$ for the $k$-bin width, the dimension of data vector is $n_\mathrm{data} = (18+2\times 8) = 34$ for each sample. 
Hence we have 136 data points of the power spectra in total for the four samples. 
In Appendix~\ref{sec:validation_test_for_analysis}, we give a validation test to determine a conservative choice of $k_\mathrm{max}$ for our linear IA model to obtain an unbiased constraint on the $f^{s=2}_\mathrm{NL}$ parameter by using the mock data of IA halo power spectrum that is computed for the halo sample in $N$-body simulation. 

We adopt the nested sampling algorithm \texttt{MultiNest} \cite{Feroz&Hobson2008:MultiNest,Feroz+2009:MultiNest,Feroz+2019:MultiNest} to obtain the posterior distributions. 
We make the plots and calculate the statistics of the marginalized posterior distributions by using the public python package \texttt{GetDist} \cite{Lewis2019:GetDist}.

\section{Results} 
\label{sec:results}
We show the results of the measurements and the systematic tests (Section~\ref{subsec:measurements}) and the various likelihood analyses in Sections~\ref{subsec:gaussian_analysis}--\ref{subsec:png_analysis_b1bphi_bkbpsi} (while 
we summarize the setups in Section~\ref{subsec:params_and_priors}).

\subsection{Measurements}
\label{subsec:measurements}

\begin{figure*}
    \centering
    \includegraphics[width=2.0\columnwidth]{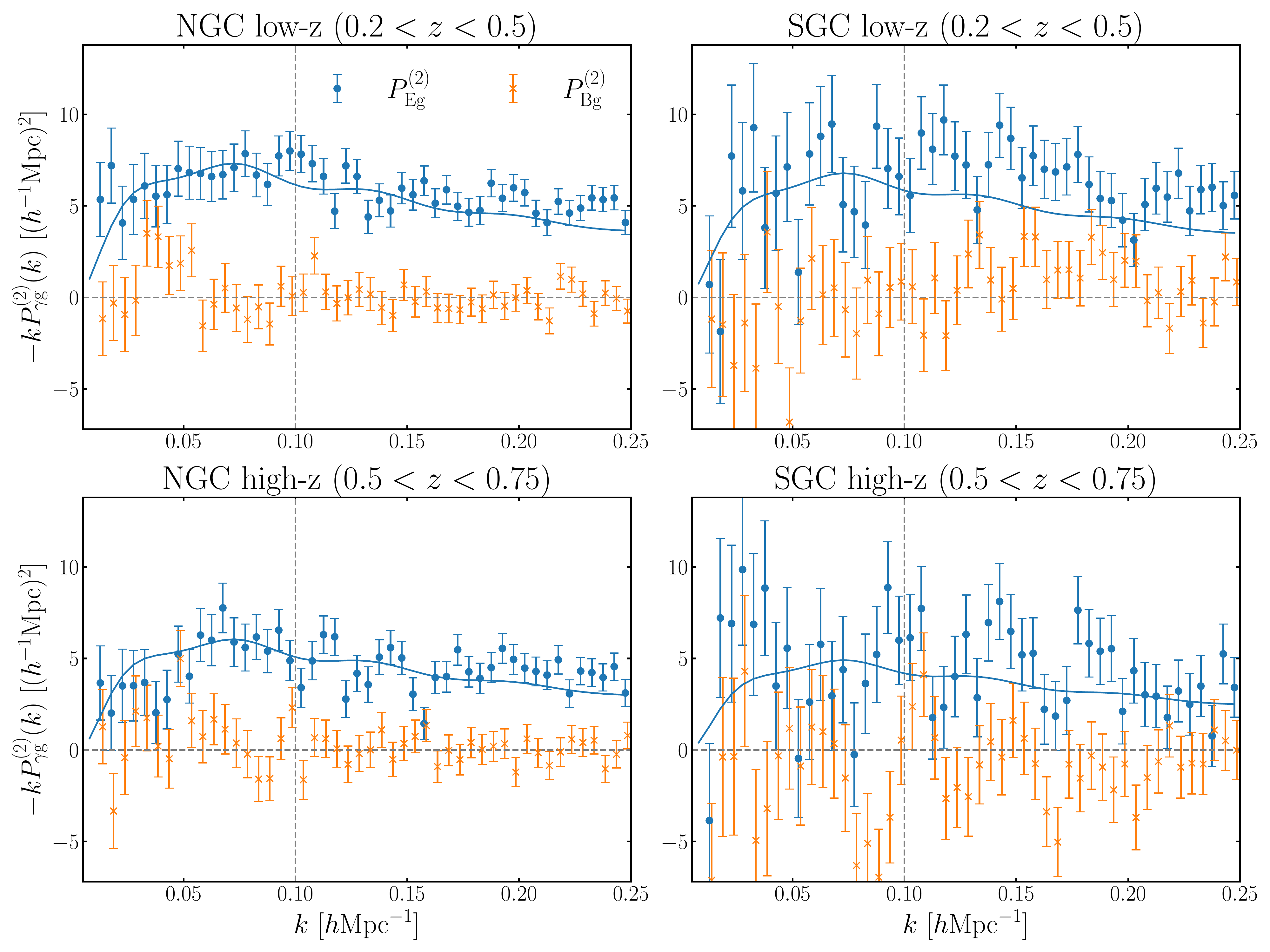}
    \caption{
    Measurements of the $E$-mode (blue) and $B$-mode (orange) IA power spectra for each sample. 
    The blue line denotes the best-fit model prediction at the \textit{maximum a posteriori} (MAP) of the Bayesian parameter inference for our ``Gaussian'' analysis setup (Section~\ref{subsec:gaussian_analysis}), 
    where we assume the linear shape bias parameter and the Gaussian initial condition for the flat $\Lambda$CDM model
    and include the measured power spectrum up to $k_\mathrm{max}=0.1~\hMpci$ (vertical dashed-line) in the parameter inference. 
    }
    \label{fig:measurement}
\end{figure*}
Fig.~\ref{fig:measurement} shows the measured IA power spectra ($E$-mode and $B$-mode cross-power spectra) for the four samples. 
We multiply all the measurements by minus one for illustrative purpose. 
The negative sign of the $E$-mode power spectrum means that the major axis of galaxy shape tends to align with the minor axis of the surrounding LSS which corresponds to the stretching axis or the direction of the filament structure. 

\begin{figure*}
    \centering
    \includegraphics[width=2.0\columnwidth]{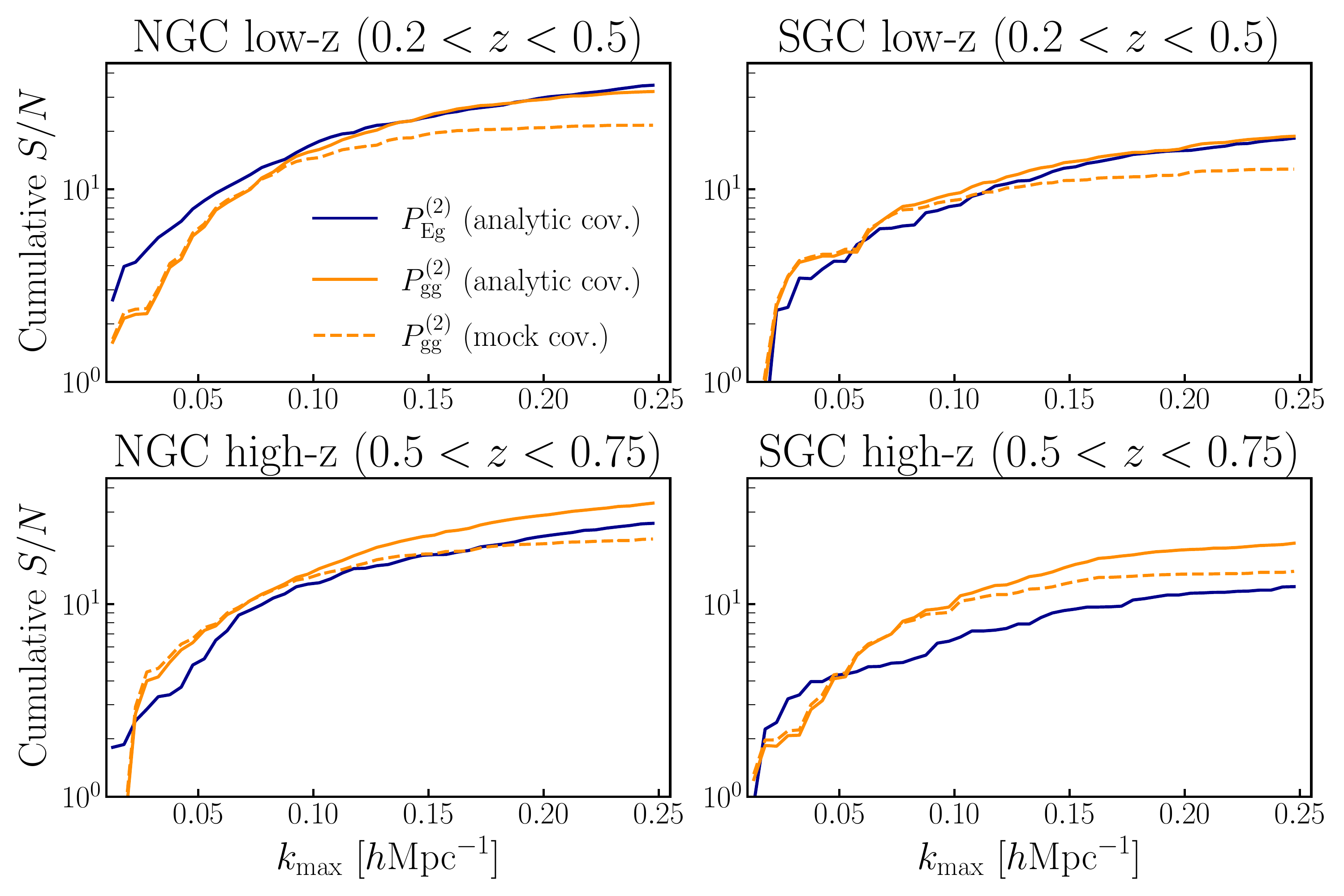}
    \caption{Cumulative signal-to-noise ratio ($S/N$) of the IA power spectrum (blue) for each sample as a function of the maximum wavenumber ($k_{\rm max}$) up to which we include the power spectrum information in the $S/N$ calculation. 
    We also show the two estimations of $S/N$ for the quadrupole moment of the galaxy power spectrum using the analytic covariance (orange-solid) or covariance estimated from the Patchy mocks (orange-dashed) for comparison purpose. 
    }
    \label{fig:signal-to-noise}
\end{figure*}
Fig.~\ref{fig:signal-to-noise} shows the cumulative signal-to-noise ratio ($S/N$) of the spectra that is defined by the square root of the chi-square statistics assuming a null-signal hypothesis (therefore this gives a significance to exclude a null signal): 
$S/N \equiv \sqrt{\chi_0^2}$ with
\begin{align}
    \chi_0^2
    = \sum_{b, b'}^{k_\mathrm{max}} 
    \hat{P}^{(\ell)}_{\alpha\beta}(k_b) \mathrm{Cov}^{-1}\left[ P^{(\ell)}_{\alpha\beta}(k_b), P^{(\ell)}_{\alpha\beta}(k_{b'}) \right] \hat{P}^{(\ell)}_{\alpha\beta}(k_{b'}). 
\end{align} 
We clearly detect the $E$-mode spectra for all the samples; the total $S/N=48.7$, for $k_{\rm max}=0.25~\hMpci$. 
We find that the $S/N$ of the IA power spectrum is comparable with that of the quadrupole moment of the galaxy power spectrum for each sample except for the SGC high-z sample partly because of larger rms ellipticity $\sigma_\gamma$ (see Table~\ref{tab:sample_property}) and relatively small amplitude ($b_K$) of IA (see the next section for our fit). 

For the systematic tests, we use the $B$-mode spectrum which is expected to be zero at all scales from the symmetry. 
As shown in Fig.~\ref{fig:measurement}, we have a null detection of the $B$-mode signal with $41.5<\chi_0^2<51.6$, with 48 bins for each sample, giving $p$-values within $0.33<p<0.74$. 
In Table~\ref{tab:sn_and_bmode}, we summarize the $S/N$ and the null tests for our measurements. 
\begin{table}
    \centering
    \begin{tabular}{ccccc}
    \toprule\midrule
    Sample & \multicolumn{2}{c}{$E$-mode $S/N~(k_\mathrm{max})$} & \multicolumn{2}{c}{$B$ mode}\\ 
    & \chead{$0.1~h\Mpc^{-1}$} & \chead{$0.25~h\Mpc^{-1}$} & \chead{$\chi_\mathrm{red}^2$} & \chead{$p$-value} \\
    \midrule
    NGC low-z & 16.6 & 34.6 & 0.865 & 0.733\\
    SGC low-z & 8.1 & 18.4 & 0.907 & 0.656\\
    \midrule
    NGC high-z & 12.7 & 26.2 & 1.074 & 0.336\\
    SGC high-z & 6.4 & 12.3 & 0.944 & 0.583\\
    \midrule
    total & 23.3 & 48.7 & 0.948 & 0.687\\
    \midrule\bottomrule
    \end{tabular}
    \caption{Summary of the signal-to-noise ratio ($S/N$) estimation for the $E$-mode power spectrum and a significance of null-signal hypothesis for the $B$-mode power spectrum (see text for details). 
    We set the minimum wave vector as $k_\mathrm{min}=0.01~\hMpci$ for both and $k_\mathrm{max}=0.25~\hMpci$ for the null test. 
    The reduced chi-square is defined as $\chi_\mathrm{red}^2 = \chi_0^2 / N_\mathrm{bin}$ with $N_\mathrm{bin}=48$ for each sample. 
    }
    \label{tab:sn_and_bmode}
\end{table}

\subsection{Gaussian Analysis}
\label{subsec:gaussian_analysis} 

In this analysis, we assume the linear model including the linear galaxy bias $b_1$, the linear shape bias $b_K$ and the residual shot noise $c_\mathrm{np}$ for each galaxy sample. 
The reduced chi-square of our model for the $E$-mode signal at the \textit{maximum a posteriori} (MAP) is 
$\chi^2/N_\mathrm{dof} = 125.57/(136-12) = 1.013$ corresponding to $p$-value with $p=0.444$, which implies that our model is acceptable to the data. 
This model is shown in Fig.~\ref{fig:measurement}, together with the measurements. 
We summarize the results in Table~\ref{tab:result_gaussian_analysis}. 
Note that previous studies on IA and WL have often used the parameter $A_\mathrm{IA}$ for convention to characterize the linear IA amplitude which is linearly related to our $b_K$ as
\begin{align}
    A_\mathrm{IA} = - \frac{\bar{D}(z)}{2 C_1\rho_\mathrm{cri} \Omega_\mathrm{m}} b_K, 
    \label{eq:A_IA_vs_b_K}
\end{align}
where $\bar{D}$ is the linear growth factor normalized to unity at $z=0$ and we set $C_1 \rho_\mathrm{cri} = 0.0134$ following Ref.~\cite{Joachimi+2011:AIA_C1}. 
In Table~\ref{tab:result_gaussian_analysis}, we also show our estimations of $A_\mathrm{IA}$. 
\begin{table}
    \centering
    \begin{tabular}{crrr}
    \toprule\midrule
    Sample & \multicolumn{1}{c}{$b_K \times 10^2$} & \multicolumn{1}{c}{$A_\mathrm{IA}$}& \multicolumn{1}{c}{$b_1$}\\ 
    & \chead{68\%CI} & \chead{68\%CI} & \chead{68\%CI} \\
    \midrule
    NGC low-z & $-5.14_{-0.31}^{+0.31}$ & $ 4.97_{-0.30}^{+0.30}$ & $2.03_{-0.03}^{+0.03}$\\
    SGC low-z &  $-4.90_{-0.70}^{+0.74}$ & $ 4.74_{-0.67}^{+0.72}$ & $2.08_{-0.05}^{+0.04}$\\
    \midrule
    NGC high-z & $-4.67_{-0.42}^{+0.36}$ & $ 4.02_{-0.36}^{+0.31}$ & $2.17_{-0.04}^{+0.04}$\\
    SGC high-z & $-4.26_{-0.96}^{+1.06}$ & $ 3.66_{-0.83}^{+0.92}$ & $2.17_{-0.05}^{+0.05}$\\
    \midrule\bottomrule
    \end{tabular}
    \caption{Results of the ``Gaussian'' analysis (Section~\ref{subsec:gaussian_analysis}). 
    We show the central value (mode) and 68\% credible interval of the 1D posterior distribution for 
    each bias parameter ($b_K$ and $b_1$) including marginalization over other parameters. 
    The result for $A_{\rm IA}$, an alternative convention to characterize the IA amplitude, is computed from the result of $b_K$ using Eq.~(\ref{eq:A_IA_vs_b_K}). 
    }
    \label{tab:result_gaussian_analysis}
\end{table}

Here we check consistency of our measurement and analysis with the previous studies. 
In Ref.~\cite{Singh+2015:IA_measurement}, they measured the projected correlation function of IA, often denoted as $w_\mathrm{g+}$, from BOSS LOWZ galaxy sample ($0.16 < z < 0.36$) and reported the estimation of $A_{\rm IA}$ and the linear density parameter $b_1$ as $A_\mathrm{IA}=4.6 \pm 0.5$ and $b_1=1.77\pm 0.04$ (see Table~2 in their paper). 
To compare with this result, we make an additional subsample ``LOWZ'' with redshift-cut of $0.16 < z < 0.36$ for our galaxy sample (\texttt{CMASSLOWZTOT}) and do the same analysis in this section, i.e. measuring the IA power spectrum and fitting our linear model to the signal. 
Although our low-z sample is constructed from the combined sample of BOSS CMASS and LOWZ, galaxies within $0.16 < z < 0.36$ almost belong to the original LOWZ sample. 
Note that our LOWZ sample used in this section is very similar, but not exactly the same as that in Ref.~\cite{Singh+2015:IA_measurement}, because we used the BOSS DR12 sample compared to the DR11 sample in Ref.~\cite{Singh+2015:IA_measurement}. 

\begin{figure}
    \centering
    \includegraphics[width=1.0\columnwidth]{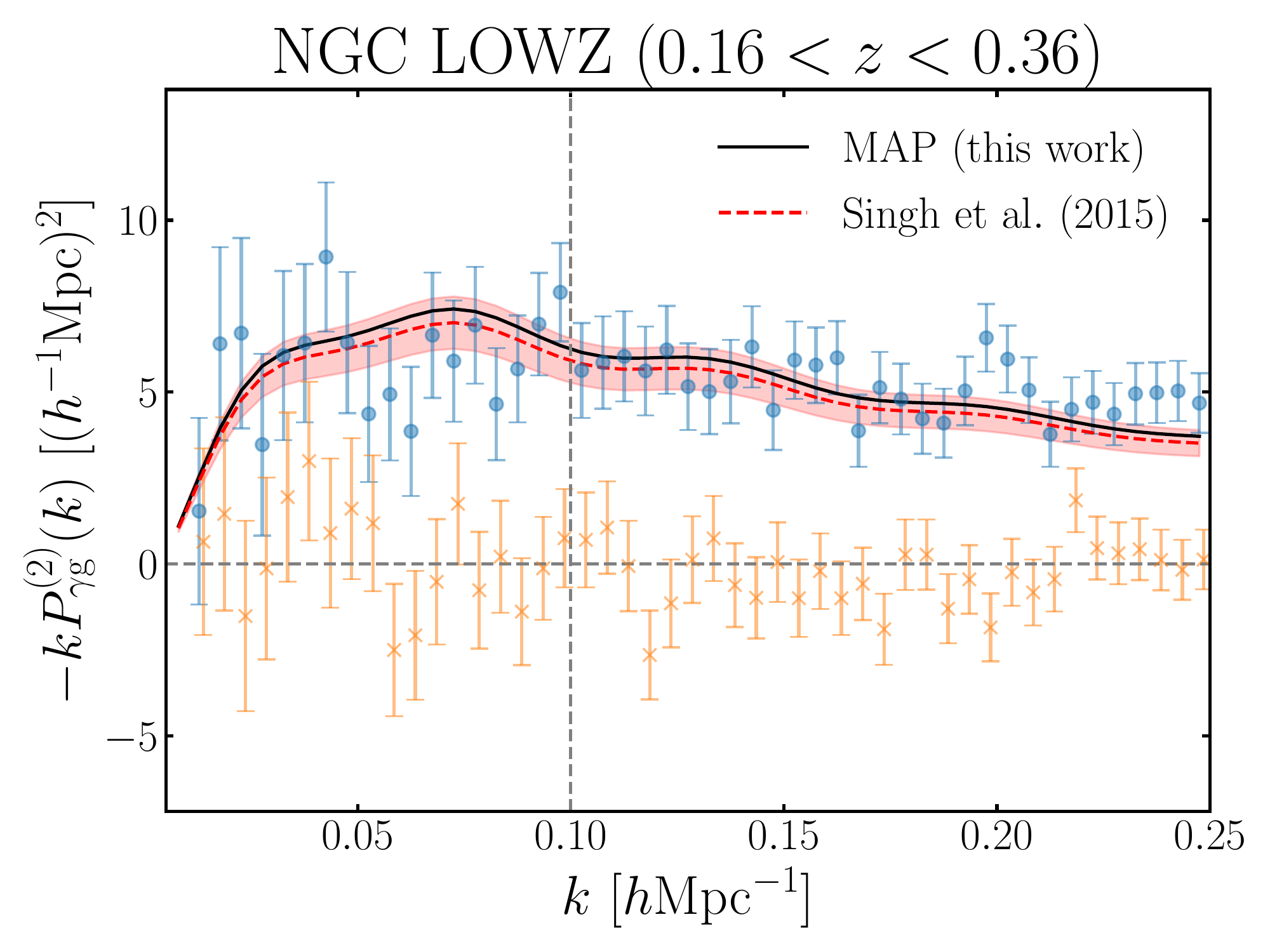}
    \caption{To compare our power spectrum analysis with the previous work, we also measure the IA power spectrum for the different galaxy sample, from our fiducial sample, where the different one is constructed from BOSS galaxies at $0.16 < z < 0.36$. 
    The black line shows the best-fit model prediction computed at MAP for the Gaussian analysis. 
    For comparison, the red-dashed line shows the best-fit model of Ref.~\cite{Singh+2015:IA_measurement}, which used the projected correlation function $w_{\rmg +}$ for the very similar galaxy sample, where the red-color shaded region denotes the model predictions allowed by the $1\sigma$ statistical errors of $A_{\rm IA}$ in their fitting results. 
    }
    \label{fig:measurement_LOWZ}
\end{figure}

Fig.~\ref{fig:measurement_LOWZ} shows the measured power spectrum from our ``LOWZ'' sample and the comparison of our MAP model and the linear model with the bias parameters estimated in Ref.~\cite{Singh+2015:IA_measurement}. 
Our measured power spectrum is in good agreement with the result of Ref.~\cite{Singh+2015:IA_measurement}
to within a $1\sigma$ level. 
The 1D posterior of each parameter gives 
$b_K=-0.0426_{-0.0040}^{+0.0041}$ 
(or equivalently $A_\mathrm{IA}=4.34_{-0.40}^{+0.42}$) 
and $b_1=1.85_{-0.05}^{+0.04}$, respectively. 
Note that the slightly different values of $b_1$ and $b_K$ between our results and Ref.~\cite{Singh+2015:IA_measurement} are probably due to the different analysis setups:
(i) the BOSS DR12 and DR11 galaxy samples, 
(ii) the FKP weighting scheme employed in this work, and
(iii) analyses in Fourier- and configuration-space.
Hence we conclude that there is no systematic uncertainty in our Fourier-space analysis.
Although we use the linear-scale signal only up to $k<0.1~\hMpci$, the fractional error of $b_K$ is improved: $b_K/\sigma(b_K) = 10.5$ compared to $= 9.2$ in Ref.~\cite{Singh+2015:IA_measurement}, even if the previous work includes 
the information of the projected correlation function ($w_{{\rm g}+}$) down to $R=6~\hiMpc$, which is in the quasi nonlinear regime. 
This implies that the three-dimensional IA correlation function, the power spectrum in our case, indeed contains more information on the IA amplitude parameter than in the projected correlation function. 
If we include $P_\mathrm{Eg}$ information down to $k=0.2~\hMpci$ (still keeping the same information of $P_{\rm gg}$ up to $k=0.05~\hMpci$), we find further improvements in the parameters as given by $b_K=-0.0459_{-0.0026}^{+0.0026}$ (and $b_1=1.85_{-0.04}^{+0.04}$),
yielding $b_K/\sigma(b_K)\simeq 17.7$. 
This corresponds to an about twofold improvement in $b_K/\sigma(b_K)$ compared to the projected correlation function. 
The reduced chi-square value for the MAP model
$\chi^2/N_\mathrm{dof} = 47.27/(54-3) = 0.927$, 
meaning that the MAP model is still acceptable (the $p$-value is 0.623). 
In summary our method using the three-dimensional IA power spectrum gives a promising route to constraining the IA amplitude for a given galaxy sample.

\subsection{PNG Analysis without Bias Relations}
\label{subsec:png_analysis_sod}

\begin{figure}
    \centering
    \includegraphics[width=1.0\columnwidth]{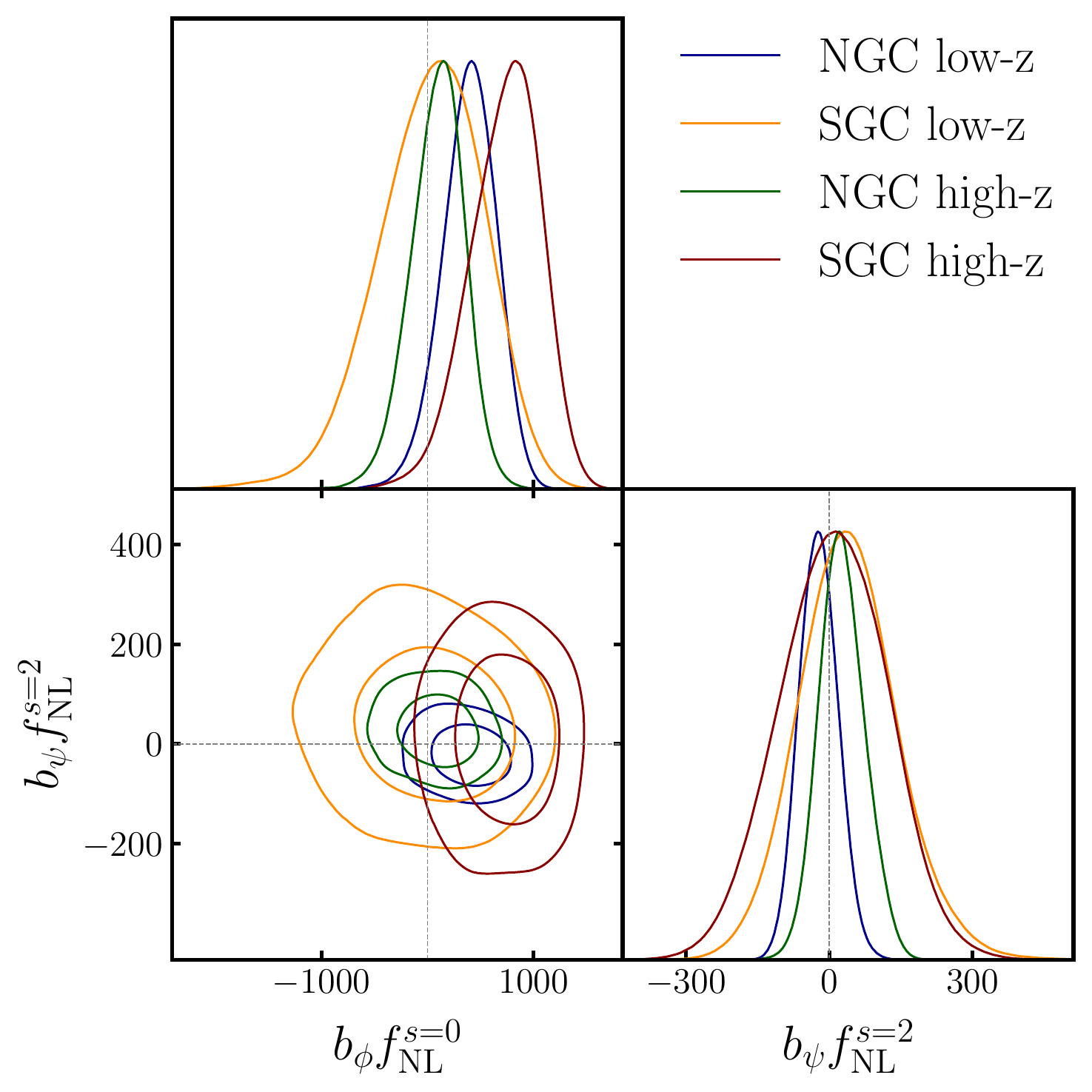}
    \caption{Posterior distributions in $(b_\phi f^{s=0}_\mathrm{NL})$-$(b_\psi f^{s=2}_\mathrm{NL})$ plane for the different galaxy samples.
    }
    \label{fig:bphifnl0_bpsifnl2_post2d}
\end{figure}

From this section, we explore whether the BOSS spectra exhibit the isotropic and anisotropic local PNGs characterized by $f^{s=0}_\mathrm{NL}$ and $f^{s=2}_\mathrm{NL}$. 
Here we do not assume any additional relation between the linear bias ($b_1,b_K$) and the PNG bias ($b_\phi,b_\psi$), that is, we regard the combinations $(b_\phi f^{s=0}_\mathrm{NL})$ or $(b_\psi f^{s=2}_\mathrm{NL})$ as one parameter. 
Hence the analysis here gives a significance of the PNG signal, if exists. 
We show the 2D posterior distributions of $(b_\phi f^{s=0}_\mathrm{NL})$ and $(b_\psi f^{s=2}_\mathrm{NL})$ for each sample in Fig.~\ref{fig:bphifnl0_bpsifnl2_post2d} and summarize the results in Table~\ref{tab:result_png_analysis_sod}. 
Note that $b_\psi$ should be different among the sample like $b_K$, we cannot obtain a unified constraint on the combination $(b_\psi f^{s=2}_\mathrm{NL})$ by combining the constraints for the different galaxy samples. 
We find no significant evidence for both types of PNG. 
\begin{table}
    \centering
    \begin{tabular}{crrrr}
    \toprule\midrule
    Sample & \multicolumn{2}{c}{$b_\psi f^{s=2}_\mathrm{NL}$} & \multicolumn{2}{c}{$b_\phi f^{s=0}_\mathrm{NL}$}\\ 
    & \chead{MAP} & \chead{68\%CI} & \chead{MAP} & \chead{68\%CI} \\
    \midrule
    NGC low-z & $ -25.0 $ & $ -21.8_{-40.6}^{+37.2}$ & $ 246.3 $ & $ 436.1_{-251.8}^{+252.1}$\\
    SGC low-z & $ 12.7 $ & $ 32.8_{-93.4}^{+95.5}$ & $ -190.9 $ & $ 118.6_{-501.1}^{+436.1}$\\
    \midrule
    NGC high-z & $ -47.3 $ & $ 25.2_{-45.5}^{+42.6}$ & $ 8.2 $ & $ 135.0_{-259.9}^{+238.3}$\\
    SGC high-z & $ -33.6 $ & $ 15.2_{-101.0}^{+90.0}$ & $ 618.2 $ & $ 720.4_{-313.6}^{+328.7}$\\
    \midrule\bottomrule
    \end{tabular}
    \caption{Results of the PNG analysis where we do not adopt the assumption on the bias relations (see Section~\ref{subsec:png_analysis_sod} for details). 
    A non-zero value of ($b_\phi f_\mathrm{NL}$ or $b_\psi f^{s=2}_\mathrm{NL}$) means a detection of 
    the PNG signals in the IA or galaxy power spectrum. 
    }
    \label{tab:result_png_analysis_sod}
\end{table}

\subsection{PNG Analysis with $b_\psi (b_K)$}
\label{subsec:png_analysis_bkbpsi}
\begin{figure*}
    \begin{minipage}{.45\textwidth}
        \centering
        \includegraphics[width=1.0\columnwidth]{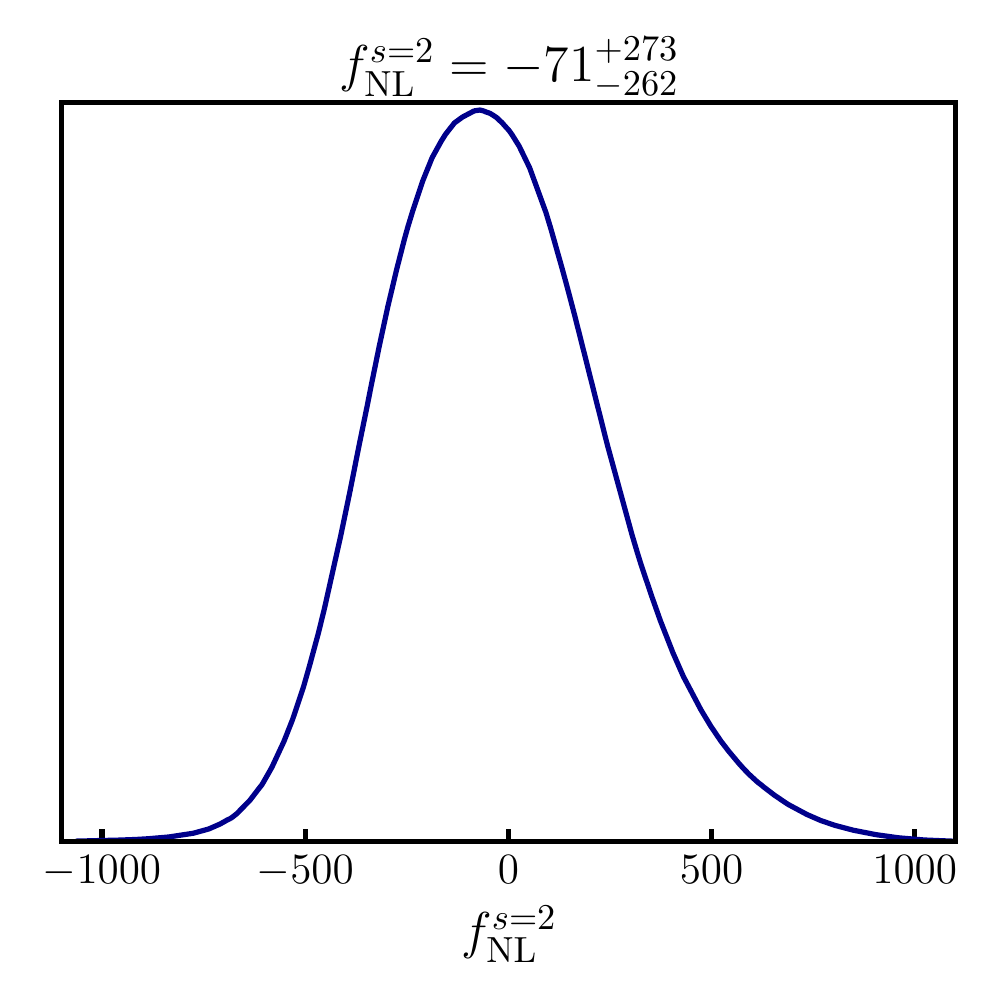}
    \end{minipage}
    \begin{minipage}{.45\textwidth}
        \centering
        \includegraphics[width=1.0\columnwidth]{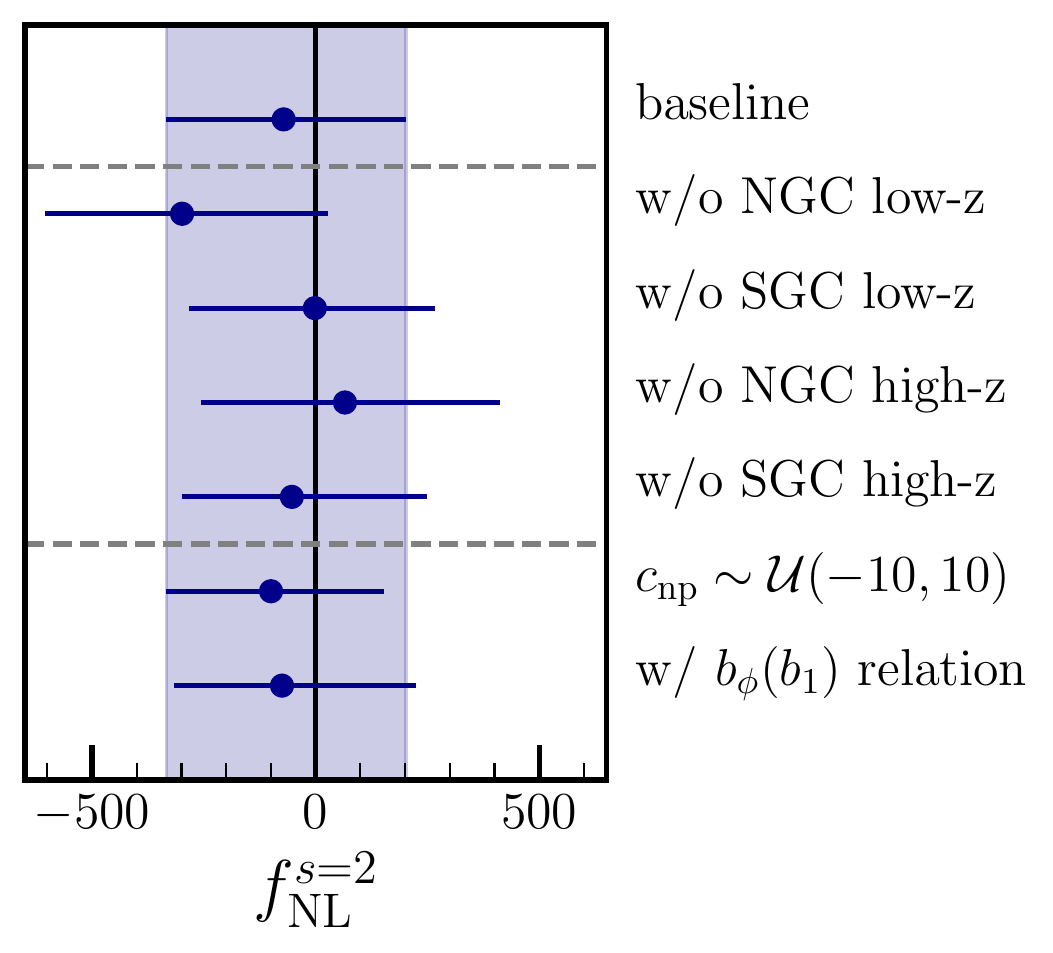}
    \end{minipage}
    \centering
    \caption{
    \textit{Left panel}: Posterior distribution of $f^{s=2}_\mathrm{NL}$ in our ``baseline'' analysis
    (see Section~\ref{subsec:png_analysis_bkbpsi} for details), where we assume that the PNG shape bias parameter $b_\psi$ is specified by the Gaussian linear shape bias parameter $b_K$ and here we adopt the relation $b_\psi=2.04\,b_K$, the empirical relation found from halo catalogs in $N$-body simulations. 
    \textit{Right panel}: Constraints on $f^{s=2}_\mathrm{NL}$ for the different setups.
    In addition to the baseline analysis, we performed the internal consistency tests; 
    we re-did the analysis removing one galaxy sample out of the four samples or 
    assuming a uniform prior $\mathcal{U}(-10,10)$ for the residual shot noise parameter $c_{\rm np}$ instead of our fiducial Gaussian prior $\mathcal{N}(0.0,0.1)$ as described in Table~\ref{tab:params_priors}.
    We also show the result of Section~\ref{subsec:png_analysis_b1bphi_bkbpsi} in the row labeled by ``w/ $b_\phi(b_1)$ relation'', where we further adopt the assumption that the PNG density bias $b_\phi$ is specified by the Gaussian 
    linear bias $b_1$. 
    }
    \label{fig:fnl2_post1d_baseline}
\end{figure*}
Here we assume the bias relation between $b_K$ and $b_\psi$ to obtain a constraint on $f^{s=2}_\mathrm{NL}$. 
We adopt $b_\psi (b_K) = 2.04\,b_K$ obtained in Ref.~\cite{Akitsu+2021:IA_PNG} for dark matter halos. 
Although Ref.~\cite{Akitsu+2021:IA_PNG} found that this relation does not vary with redshift and mass of halo samples and simulation resolution, it has to be carefully stuided whether this assumption holds for galaxies. 
Nevertheless we adopt it since it is the only known relation for the anisotropic PNG-induced bias. 
On the other hand, we keep the combination of $(b_\phi f^{s=0}_\mathrm{NL})$ as a parameter to obtain the constraint on $f^{s=2}_\mathrm{NL}$ based on the minimum assumption. 
Hence we call the analysis in this section as the baseline analysis for a direct constraint on $f^{s=2}_\mathrm{NL}$. 
We show the 1D posterior distribution in Fig.~\ref{fig:fnl2_post1d_baseline} and obtain the constraint 
\begin{align}
    f^{s=2}_\mathrm{NL} = -71_{-262}^{+273},
\end{align}
with the mode and 68\% credible interval. 

In addition to this baseline analysis, we perform alternative analyses with the different setups to check the internal consistency of our constraint described in Fig.~\ref{fig:fnl2_post1d_baseline}. 
In summary, we do not find any significant detection of $f^{s=2}_\mathrm{NL}$, due to the anisotropic local PNG. 

Our constraint from the BOSS galaxies is about 13 times larger than that from the \textit{Planck} CMB data $\sigma (f^{s=2}_\mathrm{NL}) \sim 19.2$ \cite{Planck2018:PNG}. 
As we will discuss in the next section, there should be room to improve the constraint even with the same dataset by employing a more optimal sample selection.

\subsection{PNG Analysis with $b_\phi (b_1)$ and $b_\psi (b_K)$}
\label{subsec:png_analysis_b1bphi_bkbpsi}
\begin{figure}
    \centering
    \includegraphics[width=1.0\columnwidth]{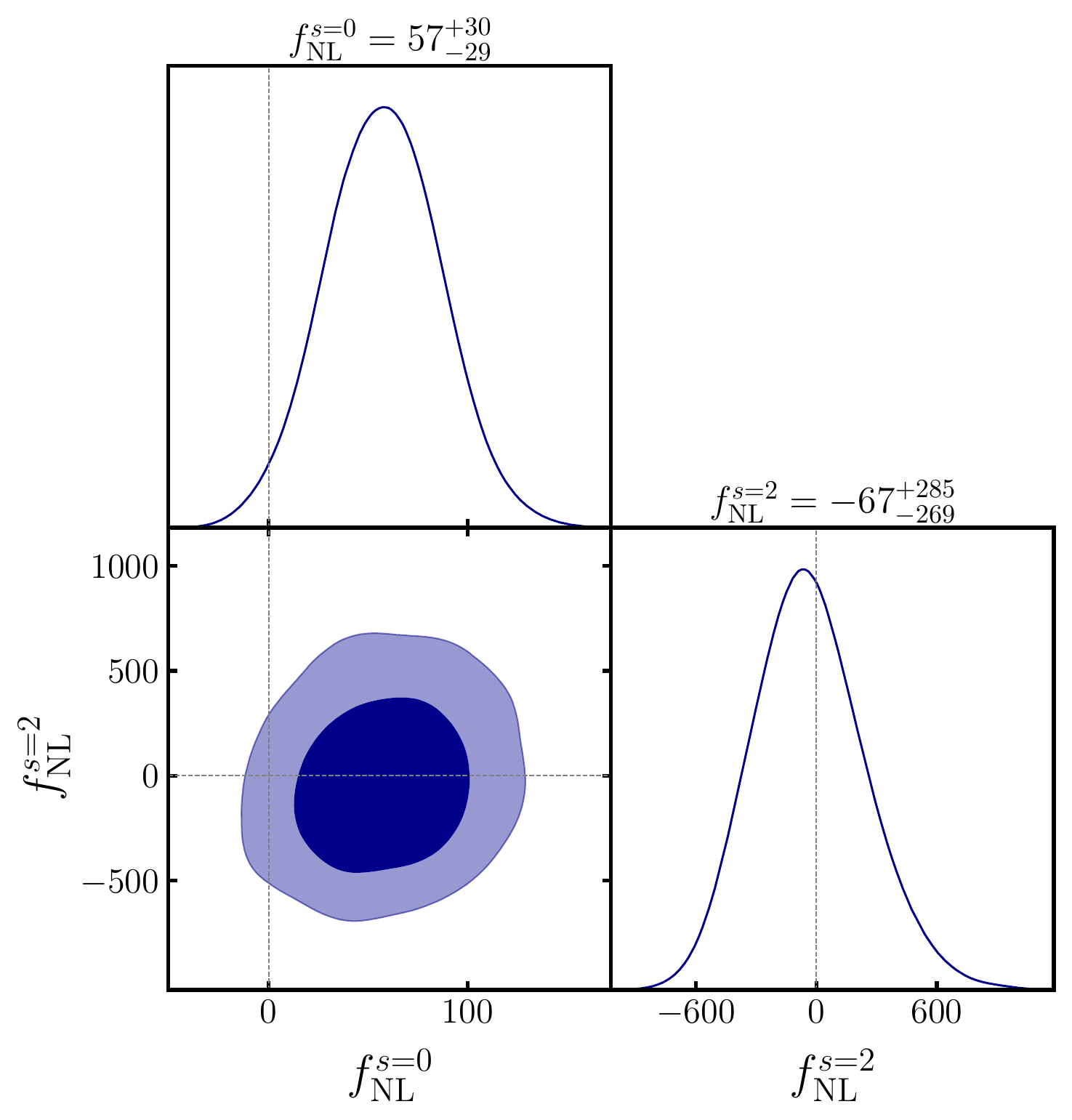}
    \caption{Posterior distributions in $f^{s=0}_\mathrm{NL}$-$f^{s=2}_\mathrm{NL}$ plane for the PNG analysis with $b_\psi(b_K)$ and $b_\phi(b_1)$ (see Section~\ref{subsec:png_analysis_b1bphi_bkbpsi} for details). 
    }
    \label{fig:fnl0_fnl2_post2d}
\end{figure}
In addition to the relation of the linear shape bias $b_\psi (b_K)$, we assume that of the linear galaxy bias $b_\phi (b_1)$. 
In this work, we employ the relation estimated by Ref.~\cite{Barreira+2020:PNG_bias_TNG}: $b_\phi (b_1) = 2\delta_\mathrm{c}(b_1-p)$ with $\delta_\mathrm{c} = 1.686$ and $p=0.55$. 
We show the resulting constraint on the two PNG parameters in Fig.~\ref{fig:fnl0_fnl2_post2d}. 
For the $f^{s=0}_\mathrm{NL}$ parameter, our constraint $f^{s=0}_\mathrm{NL} = 57_{-29}^{+30}$ is ($1\sigma$) consistent with, but slightly tighter than that of Ref.~\cite{Barreira2022:PNGbias} which performed similar linear-scale analyses to ours for the same BOSS galaxy sample using the galaxy density power spectrum and obtained $f^{s=0}_\mathrm{NL} = 33_{-34}^{+32}$ (see Table 1 in their paper). 
We expect that this difference arises mainly from the difference of the data vector, i.e. the measurement method of the power spectrum. 
In this work we employ the conventional ``FKP'' estimator that is affected by the window effect.
On the other hand, Ref.~\cite{Barreira2022:PNGbias} used the power spectrum measured by the ``window-free'' estimator \citep{Philcox2021:Quadratic_Estimator}. 
The data points of the binned spectrum of the former estimator are considered to include contributions from the underlying Fourier modes outside the $k$ range due to the window convolution. 
Hence our method includes some contributions from larger-scale modes $k<k_\mathrm{min}=0.01~h{\rm Mpc}^{-1}$ and yields a slightly tighter constraint on $f_\mathrm{NL}^{s=0}$ than the constraint from the window-free power spectrum used in Ref.~\cite{Barreira2022:PNGbias}.

\section{Conclusion} 
\label{sec:conclusion}

In this paper we have measured the three-dimensional IA power spectrum from the spectroscopic BOSS galaxy sample. 
We then used the measurements to obtain constraints on the local PNG parameters $f^{s=0,2}_\mathrm{NL}$ from a joint analysis of the galaxy density and IA power spectra.
To do these, we used the publicly available large-scale structure catalog \cite{Reid+2016:LSScatalog} as the density sample and defined the shape sample by cross-matching the shape catalog generated in Ref.~\cite{Reyes+2012:shape_catalog} with the density catalog. 
We also generated a random catalog for the shape sample by using the acceptance-rejection method. 
We measured the IA power spectrum using an efficient estimator developed in Ref.~\cite{Kurita&Takada2022:AnalysisIAPS}. 
For likelihood analysis, we newly derived an analytic covariance for the estimated power spectrum.
Our measurement and cosmological analysis using the IA power spectra are the first of its kind to be performed for the actual galaxy survey data. 

We clearly detected the $E$-mode IA spectra and found that the $S/N$ of the IA power spectrum is comparable with that of the quadrupole moment of the galaxy density power spectrum for each of the four galaxy samples that are constructed from the NGC and SGC regions in the two redshift ranges of $0.2<z<0.5$ and $0.5<z<0.75$, respectively. 
The measured $B$-mode spectra are consistent with a null signal for all the galaxy samples. 
In the Gaussian analysis, we confirmed that our measured IA power spectrum is in good agreement with the previous results. 
Nevertheless, we would like to emphasize that the three-dimensional power spectrum gives an improved precision of the $b_K$ estimation even if our analysis uses only the information in the linear regime, while the previous works using the two-dimensional (projected) correlation functions include the information down to the nonlinear scales. 
Hence our method would give a promising means of constraining the IA effect for a given galaxy sample. 
To constrain the PNG parameters, we performed joint analyses using several different analysis setups, e.g. those with or without the assumption on the relations between the linear bias and the PNG-induced bias.
We did not find any significant evidence for both types of PNG for all the analyses.

The detection of the IA power spectrum with high $S/N$ implies that we can potentially extract further cosmological information from the IA signals. 
In fact there is still room for improvements in terms of both theory and observation. 
For example, cosmological information in the linear-scale power spectrum is basically limited to that in the broadband spectrum, and thus there is little information about baryonic acoustic oscillations (BAO) in the present linear-scale analysis. 
Therefore, to obtain tighter constraints on the standard cosmological parameters such as $\Omega_m$, $\sigma_8$ and $H_0$, it is necessary to prepare sufficiently accurate theoretical templates of the IA power spectrum and its covariance which are valid down to quasi-nonlinear scales beyond the linear alignment model such as the ``Tidal Alignment + Tidal Torque'' (TATT) model and the effective field theory of intrinsic alignments \citep[e.g.][]{Blazek+2019:IA_TATT,Vlah+2020:IA_EFT}. 
The formulation of these models in the nonlinear regime has been done only on real space, not on redshift space that is relevant to an actual galaxy survey \citep[but see][for recent developments of a general formalism with the integrated perturbation theory]{Matsubara2022:IA_iPT_I,Matsubara2022:IA_iPT_II}.
Hence it would be necessary to extend them to include realistic observational effects such as the nonlinear redshift distortion effect beyond the Kaiser factor \citep{Kaiser1987:RSD} and carefully examine a valid wavenumber- ($k$-) range of them to obtain unbiased constraints on cosmological parameters as in the case of usual galaxy clustering analysis \citep[e.g.][for such a study of the galaxy density power spectrum]{Nishimichi+2020:blinded_challenge,Kobayashi+2022:FullShape}.

On the observational side, we have used all the galaxies in the entire sample for the IA power spectrum measurement in order to reduce the shot and shape noise contribution as much as possible. 
However, several studies reported that the amplitude of the IA signal, therefore the $S/N$ of IA, depends on properties and environments of galaxies, and also the shape measurement methods used to estimate the individual galaxy shapes. 
For example, Ref.~\cite{Singh+2015:IA_measurement} reported that the amplitude of IA is a monotonically increasing function of their luminosity. 
These results suggest that if we use a sample of only bright galaxies, we could obtain a higher amplitude IA signal. 
On the other hand, the lower number density due to the sample selection obviously leads to the higher Poisson shot/shape noise in the statistical errors. 
Therefore an optimal sample or weighting scheme for the IA measurements is not clear yet and worth exploring. 

Finally, to do a joint analysis described in this work, we need both spectroscopic and imaging data for the same cosmological survey volume, where the imaging data is needed to characterize shapes of individual galaxies and the spectroscopic data is needed to measure distances to galaxies for the three-dimensional power spectrum analysis. 
This is indeed the case for ongoing/future galaxy surveys; 
the Dark Energy Spectrograph Instrument \citep[DESI,][]{DESI_2016:DESI}, Subaru Prime Focus Spectrograph \citep[PFS,][]{Takada+2014:PFS}, Euclid \citep{Laureijs+2011:Euclid}, NASA Nancy Grace Roman Telescope \citep{Spergel+2015:Roman}, and the Spectro-Photometer for the History of the Universe and Ices Explorer \citep[SPHEREx,][]{Dore2014:SPHEREx} for spectroscopic surveys; 
the Subaru HSC survey \citep{Aihara+2018:HSC}, the Kilo-Degree survey \citep[KiDS,][]{Kuijken+2015:KiDS_WL}, the Dark Energy Survey \citep[DES,][]{Abbott+2018:DESY1_cosmo,Becker+2016:DESY1_Verification}, the Vera Rubin Observatory's Legacy Survey of Space and Time \citep[LSST,][]{LSST2009:LSST}, Euclid, and Roman Space Telescope for imaging surveys.
We believe that the method established in this work helps to extract as much cosmological information as possible from these current and upcoming datasets.


\acknowledgments
We would like to thank Rachel~Mandelbaum for allowing us to use the shape catalog in this work. 
We are very thankful to Yosuke~Kobayashi for useful discussions. 
We also thank the Yukawa Institute for Theoretical Physics at Kyoto University for discussions during the YITP workshop YITP-W-22-16 on ``New Frontiers in Cosmology with the Intrinsic Alignments of Galaxies'' during which this work was partly done. 
This work was supported in part by World Premier International Research Center Initiative (WPI Initiative), MEXT, Japan, and JSPS KAKENHI Grants No. JP20J22055, No. JP20H05850, No. JP20H05855, No. JP19H00677, and by Basic Research Grant (Super AI) of Institute for AI and Beyond of the University of Tokyo. 
T.K. is supported by JSPS Research Fellowship for Young Scientists. 
The simulations for the validation tests in this work were carried out on Cray XC50 at Center for Computational Astrophysics, National Astronomical Observatory of Japan.

\onecolumngrid
\appendix

\section{Derivation and Validation of Covariance Matrices} 
\label{sec:derivation_covariance}

In this section, we describe the derivation of our analytic covariance matrices in Sections~\ref{subsec:cov_PggPgg}--\ref{subsec:cov_PsgPgg}, the numerical implementation in Section~\ref{subsec:cov_implementation}, and the validation tests in Section~\ref{subsec:cov_validation}, respectively.
We introduce an abbreviated notation for the window (weight) functions following Ref.~\cite{Wadekar&Scoccimarro2020:CovPT} as
\begin{align*}
    W^\alpha_{ij}(\bx) \equiv \bar{n}^i_\alpha(\bx) w^j_\alpha(\bx),
\end{align*}
where $\bar{n}$ is the mean number density, $w$ is the weight for each galaxy and $\alpha \in \{\rmg,\gamma \}$ is the label of the galaxy density or shape field for later convenience.

\subsection{Galaxy Clustering Auto Covariance: $\mathrm{Cov}\left[ P_\mathrm{gg}, P_\mathrm{gg} \right]$}
\label{subsec:cov_PggPgg}
We first reproduce the results for the galaxy power spectrum covariance, $\mathrm{Cov}\left[ P_\mathrm{gg}, P_\mathrm{gg} \right]$, derived in Ref.~\cite{Wadekar&Scoccimarro2020:CovPT}. 
We here omit the derivation (see Ref.~\cite{Wadekar&Scoccimarro2020:CovPT} for detail derivation), however we explicitly show the derivation of the IA-related covariances, such as $\mathrm{Cov}\left[ P_{\gamma {\rmg}}, P_{\gamma {\rmg}} \right]$ and $\mathrm{Cov}\left[ P_{\gamma {\rmg}}, P_\mathrm{gg} \right]$, in detail in the next subsection.
Since we need the field labels, the shape field ($\gamma$) and the density field ($\rmg$), we will use some slightly different notations from Ref.~\cite{Wadekar&Scoccimarro2020:CovPT} as defined below.
Also since we restrict ourselves to the linear regime, we will ignore the non-Gaussian, beat-coupling and the local-average effects in the covariance.
We thus take into account only the Gaussian term and shot/shape noise terms including the survey window effects.

We use the following notations for the observed galaxy density field:
\begin{align*}
    \hdelg(\bx) \equiv 
    \frac{\bar{n}_{\rmg}(\bx)w_{\rmg}(\bx)}{\sqrt{\mathrm{I}_\mathrm{gg}}} \frac{n_{\rmg}(\bx) - \alpha n_{\rmr}(\bx)}{\bar{n}_{\rmg}(\bx)}
    = \frac{W^{\rmg}_{11}(\bx)}{\sqrt{\mathrm{I}_\mathrm{gg}}} \delg(\bx),
\end{align*}
where the normalization factor for the density field:
\begin{align}
    \mathrm{I}_\mathrm{gg} \equiv \int_\bx \bar{n}^2_{\rmg}(\bx)w^2_{\rmg}(\bx) 
    = \int_\bx [W^{\rmg}_{11}(\bx)]^2 = \int_\bx W^{\rmg}_{22}(\bx) 
    ~(\equiv \mathrm{I}_{22} \mbox{\rm ~in~Ref.~\cite{Wadekar&Scoccimarro2020:CovPT}}).
\end{align}

We decompose the full Gaussian covariance into the (pure) ``continuous'' component and the shot noise-related components for convenience:
\begin{align}
    \mathrm{Cov}\left[ P^{(\ell_1)}_\mathrm{gg}, P^{(\ell_2)}_\mathrm{gg} \right] \equiv \bC^\mathrm{GG(cont.)}_{\ell_1\ell_2} + \bC^\mathrm{GG(SN)}_{\ell_1\ell_2}. 
    \label{eq:cov_GG_tot}
\end{align}

The continuous component (Eq.~57 in Ref.~\cite{Wadekar&Scoccimarro2020:CovPT}) can be written as 
\begin{align}
    \bC^\mathrm{GG(cont.)}_{\ell_1\ell_2}(k_1,k_2) 
    &= N^\mathrm{G}_{\ell_1}N^\mathrm{G}_{\ell_2}
    \int_{\hbk_1,\hbk_2,\bx_1,\bx_2} P^\mathrm{local}_\mathrm{gg}(\bk_2;\bx_1) P^\mathrm{local}_\mathrm{gg}(\bk_1;\bx_2) \nonumber\\
    &\quad\quad\quad\quad\quad\quad\quad\quad\times
    W_{22}(\bx_1)W_{22}(\bx_2) e^{-i(\bk_1-\bk_2)\cdot(\bx_1-\bx_2)}
    \mathcal{L}_{\ell_1}(\hbk_1\cdot\hbx_1)
    \left[ \mathcal{L}_{\ell_2}(\hbk_2\cdot\hbx_2) + \mathcal{L}_{\ell_2}(\hbk_2\cdot\hbx_1) \right] \nonumber\\
    &\simeq \sum_{\ell'_1,\ell'_2} P^{(\ell'_1)}_\mathrm{gg}(k_1) P^{(\ell'_2)}_\mathrm{gg}(k_2)
    \bigg\{ N^\mathrm{G}_{\ell_1}N^\mathrm{G}_{\ell_2}
    \int_{\hbk_1,\hbk_2,\bx_1,\bx_2} W_{22}(\bx_1)W_{22}(\bx_2) e^{-i(\bk_1-\bk_2)\cdot(\bx_1-\bx_2)} \nonumber\\
    &\quad\quad\quad\quad\quad\quad\quad\quad\times 
    \mathcal{L}_{\ell_1}(\hbk_1\cdot\hbx_1)
    \left[ \mathcal{L}_{\ell_2}(\hbk_2\cdot\hbx_2) + \mathcal{L}_{\ell_2}(\hbk_2\cdot\hbx_1) \right]
    \mathcal{L}_{\ell'_1}(\hbk_1\cdot\hbx_2) \mathcal{L}_{\ell'_2}(\hbk_2\cdot\hbx_1)
    \bigg\} \nonumber\\
    &\equiv \sum_{\ell'_1,\ell'_2} P^{(\ell'_1)}_\mathrm{gg}(k_1) P^{(\ell'_2)}_\mathrm{gg}(k_2) 
    \mathcal{W}^{\mathrm{GG}(1)}_{\ell_1,\ell_2,\ell'_1,\ell'_2}(k_1,k_2). 
    \label{eq:cov_PggPgg_cont}
\end{align}
Here we have used the normalization factor $N^\mathrm{G}_{\ell} \equiv (2\ell+1)/\mathrm{I}_\mathrm{gg}$, and defined the local galaxy power spectrum in the direction $\bx$ by
\begin{align}
    P^\mathrm{local}_\mathrm{gg}(\bk;\bx) \equiv \int_{\bs} \xi_\mathrm{gg}(\bs;\bx) e^{-i\bk\cdot\bs} \equiv \int_{\bs} \avrg{\delg(\bx) \delg(\bx-\bs)} e^{-i\bk\cdot\bs}.
\end{align}

The shot noise terms (Eq.~B12 in Ref.~\cite{Wadekar&Scoccimarro2020:CovPT}) are
\begin{align}
    \bC^\mathrm{GG(SN)}_{\ell_1\ell_2}(k_1,k_2) 
    \equiv \sum_{\ell'} \left[P^{(\ell')}_\mathrm{gg}(k_1) \mathcal{W}^{\mathrm{GG}(2)}_{\ell_1,\ell_2,\ell'}(k_1,k_2) + (k_1 \leftrightarrow k_2) \right]+ \mathcal{W}^{\mathrm{GG}(3)}_{\ell_1,\ell_2}(k_1,k_2), 
    \label{eq:cov_PggPgg_SN}
\end{align}
where
\begin{align}
    &\mathcal{W}^{\mathrm{GG}(2)}_{\ell_1,\ell_2,\ell'}(k_1,k_2) \equiv
    \frac{1+\alpha}{2} N^\mathrm{G}_{\ell_1}N^\mathrm{G}_{\ell_2} 
    \int_{\hbk_1,\hbk_2,\bx_1,\bx_2} W_{22}(\bx_1)W_{12}(\bx_2) e^{-i(\bk_1-\bk_2)\cdot(\bx_1-\bx_2)} 
    \mathcal{L}_{\ell'}(\hbk_1\cdot\hbx_1) \nonumber\\
    &\quad\times 
    \left[ \mathcal{L}_{\ell_1}(\hbk_1\cdot\hbx_1) \mathcal{L}_{\ell_2}(\hbk_2\cdot\hbx_1)
    + \mathcal{L}_{\ell_1}(\hbk_1\cdot\hbx_2) \mathcal{L}_{\ell_2}(\hbk_2\cdot\hbx_2) 
    + \mathcal{L}_{\ell_1}(\hbk_1\cdot\hbx_1) \mathcal{L}_{\ell_2}(\hbk_2\cdot\hbx_2) 
    + \mathcal{L}_{\ell_1}(\hbk_1\cdot\hbx_2) \mathcal{L}_{\ell_2}(\hbk_2\cdot\hbx_1) \right], \label{eq:cov_W_GG_2}\\
    &\mathcal{W}^{\mathrm{GG}(3)}_{\ell_1,\ell_2}(k_1,k_2) \equiv
    (1+\alpha)^2 N^\mathrm{G}_{\ell_1}N^\mathrm{G}_{\ell_2} 
    \int_{\hbk_1,\hbk_2,\bx_1,\bx_2} W_{12}(\bx_1)W_{12}(\bx_2) e^{-i(\bk_1-\bk_2)\cdot(\bx_1-\bx_2)} \nonumber\\ &\quad\quad\quad\quad\quad\quad\quad\quad\quad\quad\quad\quad\quad\quad\quad\quad\quad\quad\quad\quad\quad\quad\quad\quad\quad\quad\quad\quad\times 
    \mathcal{L}_{\ell_1}(\hbk_1\cdot\hbx_1) 
    \left[ \mathcal{L}_{\ell_2}(\hbk_2\cdot\hbx_1) + \mathcal{L}_{\ell_2}(\hbk_2\cdot\hbx_2) \right]. 
    \label{eq:cov_W_GG_3}
\end{align}
Note that once we obtain the quartic functions of the window function, $\mathcal{W}^{\mathrm{GG}(i)}_{\ell,\ell',\cdots}~(i=1,2,3)$ from the random catalog, we can immediately construct the full Gaussian covariance with multiplications by the theoretical power spectrum multipoles as in Eqs.~(\ref{eq:cov_PggPgg_cont}) and (\ref{eq:cov_PggPgg_SN}).
The indices $i=1,2,3$ represent the functions for the continuous-continuous, continuous-SN, SN-SN components, respectively.

\subsection{Intrinsic Alignments Auto Covariance: $\mathrm{Cov}\left[ P_{\gamma{\rmg}}, P_{\gamma{\rmg}} \right]$} 
\label{subsec:cov_PsgPsg}
We next derive the IA-IA auto covariance.
We use the following notations:
\begin{align*}
    \hat{\gamma}(\bx) \equiv 
    \frac{\bar{n}_{\gamma}(\bx)w_{\gamma}(\bx)}{\sqrt{\mathrm{I}_{\gamma\gamma}}} \frac{n_{\gamma}(\bx) \gamma(\bx)}{\bar{n}_{\gamma}(\bx)} 
    = \frac{W^{\gamma}_{11}(\bx)}{\sqrt{\mathrm{I}_{\gamma\gamma}}} [1+\delta_\gamma(\bx)] \gamma(\bx) 
    \equiv \frac{W^{\gamma}_{11}(\bx)}{\sqrt{\mathrm{I}_{\gamma\gamma}}} \tilde{\gamma}(\bx),
\end{align*}
where the normalization factor for the shape field:
\begin{align*}
    \mathrm{I}_{\gamma\gamma} \equiv \int_\bx \bar{n}^2_{\gamma}(\bx) w^2_{\gamma}(\bx)
    = \int_\bx W^{\gamma}_{22}(\bx).
\end{align*}
In the following, we assume the density-weighted shape field $\tilde{\gamma}(\bx) \equiv [1+\delta_\gamma(\bx)] \gamma(\bx)$ is a Gaussian field and simply denote it as $\gamma(\bx)$. 
Also we introduce another normalization factor:
\begin{align*}
    \mathrm{I}_{\gamma{\rmg}} \equiv \int_\bx w_{\gamma}(\bx)\bar{n}_{\gamma}(\bx) w_{\rmg}(\bx)\bar{n}_{\rmg}(\bx) 
    = \int_\bx W^{\gamma}_{11}(\bx) W^{\rmg}_{11}(\bx),
\end{align*}
to define the unbiased estimator of the IA-galaxy cross power spectrum as we will see in the next.

\subsubsection{Local-Plane Parallel Estimator}
Our definition of the local-plane parallel estimator for IA power spectrum with the endpoint approximation \cite{Kurita&Takada2022:AnalysisIAPS} is
\begin{align}
    \hat{P}^{(L)}_{\gamma {\rmg}} (k) &\equiv (2L+1)\frac{(L-2)!}{(L+2)!} \frac{\sqrt{\mathrm{I}_{\gamma\gamma}}\sqrt{\mathrm{I}_\mathrm{gg}}}{\mathrm{I}_{\gamma {\rmg}}} \int_{\hbk,\bx,\bx'} 
    \hat{\gamma}(\bx) \hdelg(\bx') 
    e^{-2i\phi_{\hbk,\hbx}} e^{-i\bk\cdot(\bx-\bx')} \mathcal{L}^{m=2}_{L}(\hbk\cdot\hbx) \nonumber\\
    &= (2L+1)\frac{(L-2)!}{(L+2)!} \frac{1}{\mathrm{I}_{\gamma {\rmg}}} \int_{\hbk} 
    \left[ \int_\bx W^\gamma_{11}(\bx) \gamma(\bx) e^{-2i\phi_{\hbk,\hbx}} e^{-i\bk\cdot\bx} \mathcal{L}^{m=2}_{L}(\hbk\cdot\hbx) \right]
    \left[ \int_{\bx'} W^{\rmg}_{11}(\bx') \delg(\bx') e^{i\bk\cdot\bx'} \right] \nonumber\\
    &\equiv (2L+1)\frac{(L-2)!}{(L+2)!} \frac{1}{\mathrm{I}_{\gamma {\rmg}}} \int_{\hbk} 
    \hFgamij^{(L)}(\bk) \hk_i\hk_j \hFg^{(0)}(-\bk),
    \label{eq:LPP_Est_Psg}
\end{align}
where we have used $e^{-2i\phi_{\hbk,\hbx}} = 2e^*_{ij}(\hbx) \hk_i\hk_j / (1-(\hbk \cdot \hbx)^2)$ with $e^*_{ij}$ is the complex conjugate of the polarization tensor. 
The ensemble average of this estimator becomes
\begin{align}
    \avrg{\hat{P}^{(L)}_{\gamma {\rmg}}(k)} 
    &= (2L+1)\frac{(L-2)!}{(L+2)!}\frac{1}{\mathrm{I}_{\gamma {\rmg}}} \int_{\hbk,\bx,\bx'} 
    W^{\gamma}_{11}(\bx)W^{\rmg}_{11}(\bx') \avrg{\gamma(\bx)\delg(\bx')}
    e^{-2i\phi_{\hbk,\hbx}} e^{-i\bk\cdot(\bx-\bx')} \mathcal{L}^{m=2}_{L}(\hbk\cdot\hbx) \nonumber\\
    &= (2L+1)\frac{(L-2)!}{(L+2)!}\frac{1}{\mathrm{I}_{\gamma {\rmg}}} \int_{\hbk,\bx,\bs} 
    W^{\gamma}_{11}(\bx)W^{\rmg}_{11}(\bx-\bs) \xi_{\gamma {\rmg}}(\bs,\bx) 
    e^{2i\phi_{\hbs;\hbx}-2i\phi_{\hbk,\hbx}}  e^{-i\bk\cdot\bs} \mathcal{L}^{m=2}_{L}(\hbk\cdot\hbx) \nonumber\\
    &\simeq (2L+1)\frac{(L-2)!}{(L+2)!}\frac{1}{\mathrm{I}_{\gamma {\rmg}}} \int_{\hbk,\bx} 
    W^{\gamma}_{11}(\bx) W^{\rmg}_{11}(\bx)
    P^\mathrm{local}_{\gamma {\rmg}} (\bk;\bx) \mathcal{L}^{m=2}_{L}(\hbk\cdot\hbx) \nonumber\\
    &\simeq (2L+1)\frac{(L-2)!}{(L+2)!}\frac{1}{\mathrm{I}_{\gamma {\rmg}}} \int_{\hbk,\bx} 
    W^{\gamma}_{11}(\bx) W^{\rmg}_{11}(\bx)
    \left[ \sum_{L'\geq2} P^{(L')}_{\gamma {\rmg}}(k) \mathcal{L}^{m=2}_{L'}(\hbk\cdot\hbx)\right]
    \mathcal{L}^{m=2}_{L}(\hbk\cdot\hbx) \nonumber\\
    &= P^{(L)}_{\gamma {\rmg}}(k).
\end{align}
In the first approximation, we have defined the local IA power spectrum as
\begin{align}
    P^\mathrm{local}_{\gamma {\rmg}} (\bk;\bx) 
    \equiv \int_{\bs} \xi_{\gamma {\rmg}}(\bs;\bx) e^{2i\phi_{\hbs;\hbx}-2i\phi_{\hbk,\hbx}} e^{-i\bk\cdot\bs} 
    \equiv \int_{\bs} \avrg{\gamma(\bx) \delg(\bx-\bs)} e^{2i\phi_{\hbs;\hbx}-2i\phi_{\hbk,\hbx}} e^{-i\bk\cdot\bs}, 
\end{align}
and assumed that the IA power spectrum is a smooth function within each $k$-bin and the survey window is much larger than the wavevector we are interested in, i.e. 
\begin{align}
    \int_\bs W(\bx-\bs) \xi_{\gamma {\rmg}}(\bs;\bx) e^{2i\phi_{\hbs;\hbx}-2i\phi_{\hbk,\hbx}} e^{-i\bk\cdot\bs}
    = \int_\bq W^{\rmg}_{11}(\bq) e^{i\bq\cdot\bx} 
    P^\mathrm{local}_{\gamma {\rmg}} (\bk+\bq;\bx) e^{2i\phi_{\widehat{\bk+\bq},\hbx}} e^{-2i\phi_{\hbk,\hbx}} 
    \simeq W(\bx) P^\mathrm{local}_{\gamma {\rmg}} (\bk;\bx). 
    \label{eq:smooth_Pk_approx}
\end{align}
In the second approximation, we have ignored higher order wide-angle corrections to the local power spectrum:
\begin{align}
    P^\mathrm{local}_{\gamma {\rmg}} (\bk;\bx) 
    = P_{\gamma {\rmg}} (k,\hbk\cdot\hbx;kx) 
    \simeq P_{\gamma {\rmg}} (k,\hbk\cdot\hbx) 
    = \sum_{L\geq2} P^{(L)}_{\gamma {\rmg}}(k) \mathcal{L}^{m=2}_{L}(\hbk\cdot\hbx). 
    \label{eq:wide_angle_approx}
\end{align}
Thus our estimator is an unbiased estimator. 

\subsubsection{Gaussian covariance: continuous component}
We derive the continuous component of the Gaussian covariance for the estimated IA cross power spectrum, $\hat{P}^{(L)}_\mathrm{Eg}(k)$.
First, we define the $E$-mode estimator as the real part of Eq.~(\ref{eq:LPP_Est_Psg}):
\begin{align}
    \hat{P}^{(L)}_\mathrm{Eg}(k) 
    &\equiv \mathrm{Re}\left[ \hat{P}^{(L)}_{\gamma {\rmg}}(k) \right] 
    = \frac{2L+1}{2} \frac{(L-2)!}{(L+2)!} \frac{\sqrt{\mathrm{I}_{\gamma\gamma}}\sqrt{\mathrm{I}_\mathrm{gg}}}{\mathrm{I}_{\gamma {\rmg}}} 
    \int_{\hbk,\bx,\bx'} 
    \left[ \hat{\gamma}(\bx)e^{-2i\phi_{\hbk,\hbx}}  + \hat{\gamma}^*(\bx)e^{2i\phi_{\hbk,\hbx}} \right] \hdelg(\bx') 
    e^{-i\bk\cdot(\bx-\bx')} \mathcal{L}^{m=2}_{L}(\hbk\cdot\hbx) \nonumber\\
    &\equiv N^\mathrm{I}_L 
    \int_{\hbk} 
    \left[ \hFgamij^{(L)}(\bk)\hk_i\hk_j + \hFgamij^{(L)*}(\bk)\hk_i\hk_j \right]  \hFg^{(0)}(-\bk).
    \label{eq:LPP_est_PEg}
\end{align}
We have replaced $\hat{\gamma}(\bx)e^{-2i\phi_{\hbk,\hbx}}$ in Eq.~(\ref{eq:LPP_Est_Psg}) with $\frac{1}{2} \left[ \hat{\gamma}(\bx)e^{-2i\phi_{\hbk,\hbx}} + \hat{\gamma}^*(\bx)e^{2i\phi_{\hbk,\hbx}} \right]$ and defined $N^\mathrm{I}_L \equiv \frac{(2L+1)}{2}\frac{(L-2)!}{(L+2)!} \frac{1}{\mathrm{I}_{\gamma {\rmg}}}$ for convenience of discussion. 
The continuous component of the auto covariance becomes
\begin{align*}
    &\bC^\mathrm{II(cont.)}_{L_1L_2} \equiv \avrg{\hat{P}^{(L_1)}_\mathrm{Eg}(k_1) \hat{P}^{(L_2)}_\mathrm{Eg}(k_2)} 
    - \avrg{\hat{P}^{(L_1)}_\mathrm{Eg}(k_1)} \avrg{\hat{P}^{(L_2)}_\mathrm{Eg}(k_2)} \\
    &= N^\mathrm{I}_{L_1}N^\mathrm{I}_{L_2}
    \int_{\hbk_1, \hbk_2} 
    \avrg{\left[ \hFgamij^{(L_1)}(\bk_1)\hk^i_1\hk^j_1 + \hFgamij^{(L_1)*}(\bk_1)\hk^i_1\hk^j_1 \right]  \hFg^{(0)}(-\bk_1)
    \left[ \hFgamkl^{(L_2)}(\bk_2)\hk^k_2\hk^l_2 + \hFgamkl^{(L_2)*}(\bk_2)\hk^k_2\hk^l_2 \right]  \hFg^{(0)}(-\bk_2)} \\
    &\overset{\mathrm{Gaussian}}{\simeq}
    N^\mathrm{I}_{L_1}N^\mathrm{I}_{L_2}
    \int_{\hbk_1, \hbk_2} 
    \bigg\{\avrg{\left[ \hFgamij^{(L_1)}(\bk_1)\hk^i_1\hk^j_1 + \hFgamij^{(L_1)*}(\bk_1)\hk^i_1\hk^j_1 \right]  \hFg^{(0)}(-\bk_2)}
    \avrg{\hFg^{(0)}(-\bk_1) \left[ \hFgamkl^{(L_2)}(\bk_2)\hk^k_2\hk^l_2 + \hFgamkl^{(L_2)*}(\bk_2)\hk^k_2\hk^l_2 \right]} \\
    &\quad\quad\quad\quad\quad\quad\quad\quad+
    \avrg{\left[ \hFgamij^{(L_1)}(\bk_1)\hk^i_1\hk^j_1 + \hFgamij^{(L_1)*}(\bk_1)\hk^i_1\hk^j_1 \right] 
    \left[ \hFgamkl^{(L_2)}(\bk_2)\hk^k_2\hk^l_2 + \hFgamkl^{(L_2)*}(\bk_2)\hk^k_2\hk^l_2 \right]}
    \avrg{\hFg^{(0)}(-\bk_1)  \hFg^{(0)}(-\bk_2)} \bigg\} \\
    &\equiv 
    \bigg\{\mathrm{A:} \avrg{\gamma \delta}\avrg{\gamma \delta} \bigg\} + \bigg\{ \mathrm{B:} \avrg{\gamma \gamma}\avrg{\delta \delta} \bigg\},
\end{align*}
with
\begin{align}
    \bigg\{\mathrm{A:} \avrg{\gamma \delta}\avrg{\gamma \delta} \bigg\} 
    &\equiv N^\mathrm{I}_{L_1}N^\mathrm{I}_{L_2} 
    \int_{\hbk_1,\hbk_2,\bx_1,\bx'_1,\bx_2,\bx'_2} 
    e^{-i\bk_1\cdot(\bx_2-\bx'_1)} e^{-i\bk_2\cdot(\bx_1-\bx'_2)} e^{-i(\bk_1-\bk_2)\cdot(\bx_1-\bx_2)} \nonumber\\
    &\quad\quad\times 
    W^{\gamma}_{11}(\bx_1)W^{\rmg}_{11}(\bx'_1)W^{\gamma}_{11}(\bx_2)W^{\rmg}_{11}(\bx'_2) 
    \mathcal{L}^{m=2}_{L_1}(\hbk_1 \cdot \hbx_1) \mathcal{L}^{m=2}_{L_2}(\hbk_2 \cdot \hbx_2) \nonumber\\
    &\quad\quad\times 
    \avrg{\left[ \gamma(\bx_1)e^{-2i\phi_{\hbk_1,\hbx_1}}  + \gamma^*(\bx_1)e^{2i\phi_{\hbk_1,\hbx_1}} \right] \delg(\bx'_2)}
    \avrg{\left[ \gamma(\bx_2)e^{-2i\phi_{\hbk_2,\hbx_2}}  + \gamma^*(\bx_2)e^{2i\phi_{\hbk_2,\hbx_2}} \right] \delg(\bx'_1)} \nonumber\\
    &\overset{\bs_1\equiv \bx_1-\bx'_2,~\bs_2\equiv \bx_2-\bx'_1}{=} N^\mathrm{I}_{L_1}N^\mathrm{I}_{L_2} 
    \int_{\hbk_1,\hbk_2,\bx_1,\bs_2,\bx_2,\bs_1} 
    e^{-i\bk_1\cdot \bs_2} e^{-i\bk_2\cdot \bs_1} e^{-i(\bk_1-\bk_2)\cdot(\bx_1-\bx_2)} \nonumber\\
    &\quad\quad\times 
    W^{\gamma}_{11}(\bx_1)W^{\rmg}_{11}(\bx_2-\bs_2)W^{\gamma}_{11}(\bx_2)W^{\rmg}_{11}(\bx_1-\bs_1) 
    \mathcal{L}^{m=2}_{L_1}(\hbk_1 \cdot \hbx_1) \mathcal{L}^{m=2}_{L_2}(\hbk_2 \cdot \hbx_2) \nonumber\\
    &\quad\quad\times 
    \bigg[ \xi_{\gamma {\rmg}}(\bs_1;\bx_1)e^{2i\phi_{\hbs_1,\hbx_1}-2i\phi_{\hbk_1,\hbx_1}} + \xi^*_{\gamma {\rmg}}(\bs_1;\bx_1)e^{-2i\phi_{\hbs_1,\hbx_1}+2i\phi_{\hbk_1,\hbx_1}} \bigg] \nonumber\\
    &\quad\quad\times
    \bigg[ \xi_{\gamma {\rmg}}(\bs_2;\bx_2)e^{2i\phi_{\hbs_2,\hbx_2}-2i\phi_{\hbk_2,\hbx_2}} + \xi^*_{\gamma {\rmg}}(\bs_2;\bx_2)e^{-2i\phi_{\hbs_2,\hbx_2}+2i\phi_{\hbk_2,\hbx_2}} \bigg] \nonumber\\
    &\overset{\mathrm{Eq.~(\ref{eq:smooth_Pk_approx})}}{\simeq} N^\mathrm{I}_{L_1}N^\mathrm{I}_{L_2} 
    \int_{\hbk_1,\hbk_2,\bx_1,\bx_2} 
    e^{-i(\bk_1-\bk_2)\cdot(\bx_1-\bx_2)} \nonumber\\
    &\quad\quad\times 
    W^{\gamma}_{11}(\bx_1)W^{\rmg}_{11}(\bx_2)W^{\gamma}_{11}(\bx_2)W^{\rmg}_{11}(\bx_1) 
    \mathcal{L}^{m=2}_{L_1}(\hbk_1 \cdot \hbx_1) \mathcal{L}^{m=2}_{L_2}(\hbk_2 \cdot \hbx_2) \nonumber\\
    &\quad\quad\times 
    \bigg[ P^\mathrm{local}_{\gamma {\rmg}} (\bk_2;\bx_1) + P^\mathrm{local*}_{\gamma {\rmg}} (\bk_2;\bx_1) \bigg] 
    \bigg[ P^\mathrm{local}_{\gamma {\rmg}} (\bk_1;\bx_2) + P^\mathrm{local*}_{\gamma {\rmg}} (\bk_1;\bx_2) \bigg] \nonumber\\
    &= 4N^\mathrm{I}_{L_1}N^\mathrm{I}_{L_2} 
    \int_{\hbk_1,\hbk_2,\bx_1,\bx_2} 
    e^{-i(\bk_1-\bk_2)\cdot(\bx_1-\bx_2)} \nonumber\\
    &\quad\quad\times 
    W^{\gamma}_{11}(\bx_1)W^{\rmg}_{11}(\bx_2)W^{\gamma}_{11}(\bx_2)W^{\rmg}_{11}(\bx_1) 
    \mathcal{L}^{m=2}_{L_1}(\hbk_1 \cdot \hbx_1) \mathcal{L}^{m=2}_{L_2}(\hbk_2 \cdot \hbx_2) 
    P^\mathrm{local}_{\mathrm{Eg}} (\bk_2;\bx_1) P^\mathrm{local}_{\mathrm{Eg}} (\bk_1;\bx_2) \nonumber\\
    &\overset{\mathrm{Eq.~(\ref{eq:wide_angle_approx})}}{\simeq} \sum_{\ell'_1,\ell'_2} P^{(\ell'_1)}_{\mathrm{Eg}}(k_1) P^{(\ell'_2)}_{\mathrm{Eg}}(k_2)
    ~4N^\mathrm{I}_{L_1}N^\mathrm{I}_{L_2}  \int_{\hbk_1,\hbk_2,\bx_1,\bx_2} 
    e^{-i(\bk_1-\bk_2)\cdot(\bx_1-\bx_2)} \nonumber\\
    &\quad\quad\times 
    W^{\gamma}_{11}(\bx_1)W^{\rmg}_{11}(\bx_2)W^{\gamma}_{11}(\bx_2)W^{\rmg}_{11}(\bx_1) 
    \mathcal{L}^{m=2}_{L_1}(\hbk_1 \cdot \hbx_1) \mathcal{L}^{m=2}_{L_2}(\hbk_2 \cdot \hbx_2) 
    \mathcal{L}_{\ell'_1}(\hbk_2 \cdot \hbx_1) \mathcal{L}_{\ell'_2}(\hbk_1 \cdot \hbx_2) \nonumber\\
    &\equiv \sum_{\ell'_1,\ell'_2} P^{(\ell'_1)}_{\mathrm{Eg}}(k_1) P^{(\ell'_2)}_{\mathrm{Eg}}(k_2) 
    \mathcal{W}^{\mathrm{II}(1,\mathrm{A})}_{L_1,L_2,\ell'_1,\ell'_2}(k_1,k_2). 
    \label{eq:cov_ia_PartA}
\end{align}
and 
\begin{align}
    \bigg\{\mathrm{B:} \avrg{\gamma \gamma}\avrg{\delta \delta} \bigg\} &\equiv 
    N^\mathrm{I}_{L_1}N^\mathrm{I}_{L_2} 
    \int_{\hbk_1,\hbk_2,\bx_1,\bx'_1,\bx_2,\bx'_2} 
    e^{-i\bk_1\cdot(\bx_1-\bx'_1)} e^{-i\bk_2\cdot(\bx_2-\bx'_2)} \nonumber\\
    &\quad\quad\times 
    W^{\gamma}_{11}(\bx_1)W^{\rmg}_{11}(\bx'_1)W^{\gamma}_{11}(\bx_2)W^{\rmg}_{11}(\bx'_2) 
    \mathcal{L}^{m=2}_{L_1}(\hbk_1 \cdot \hbx_1) \mathcal{L}^{m=2}_{L_2}(\hbk_2 \cdot \hbx_2) \nonumber\\
    &\quad\quad\times 
    \avrg{\left[ \gamma(\bx_1)e^{-2i\phi_{\hbk_1,\hbx_1}}  + \gamma^*(\bx_1)e^{2i\phi_{\hbk_1,\hbx_1}} \right] \left[ \gamma(\bx_2)e^{-2i\phi_{\hbk_2,\hbx_2}}  + \gamma^*(\bx_2)e^{2i\phi_{\hbk_2,\hbx_2}} \right]}
    \avrg{ \delg(\bx'_1) \delg(\bx'_2) } \nonumber\\
    &\overset{\bx_2\leftrightarrow\bx'_2,~\hbk_2\rightarrow-\hbk_2,~L_2:~\mathrm{even}}{=} N^\mathrm{I}_{L_1}N^\mathrm{I}_{L_2} 
    \int_{\hbk_1,\hbk_2,\bx_1,\bx'_1,\bx_2,\bx'_2} 
    e^{-i\bk_1\cdot(\bx_2-\bx'_1)} e^{-i\bk_2\cdot(\bx_1-\bx'_2)} e^{-i(\bk_1-\bk_2)\cdot(\bx_1-\bx_2)} \nonumber\\
    &\quad\quad\times 
    W^{\gamma}_{11}(\bx_1)W^{\rmg}_{11}(\bx'_1)W^{\gamma}_{11}(\bx'_2)W^{\rmg}_{11}(\bx_2) 
    \mathcal{L}^{m=2}_{L_1}(\hbk_1 \cdot \hbx_1) \mathcal{L}^{m=2}_{L_2}(\hbk_2 \cdot \hbx'_2) \nonumber\\
    &\quad\quad\times 
    \avrg{\left[ \gamma(\bx_1)e^{-2i\phi_{\hbk_1,\hbx_1}}  + \gamma^*(\bx_1)e^{2i\phi_{\hbk_1,\hbx_1}} \right] \left[ \gamma(\bx'_2)e^{-2i\phi_{\hbk_2,\hbx'_2}} + \gamma^*(\bx'_2)e^{2i\phi_{\hbk_2,\hbx'_2}} \right]}
    \avrg{ \delg(\bx'_1) \delg(\bx_2) } \nonumber\\
    &\overset{\bs_1\equiv \bx_1-\bx'_2,~\bs_2\equiv \bx_2-\bx'_1}{=} N^\mathrm{I}_{L_1}N^\mathrm{I}_{L_2} 
    \int_{\hbk_1,\hbk_2,\bx_1,\bs_2,\bx_2,\bs_1} 
    e^{-i\bk_1\cdot \bs_2} e^{-i\bk_2\cdot \bs_1} e^{-i(\bk_1-\bk_2)\cdot(\bx_1-\bx_2)} \nonumber\\
    &\quad\quad\times 
    W^{\gamma}_{11}(\bx_1)W^{\rmg}_{11}(\bx_2-\bs_2)W^{\gamma}_{11}(\bx_1-\bs_1)W^{\rmg}_{11}(\bx_2) 
    \mathcal{L}^{m=2}_{L_1}(\hbk_1 \cdot \hbx_1) \mathcal{L}^{m=2}_{L_2}(\hbk_2 \cdot \widehat{\bx_1-\bs_1}) \nonumber\\
    &\quad\quad\times 
    \bigg[ \xi_{-}(\bs_1;\bx_1) e^{4i\phi_{\hbs_1,\hbx_1}} e^{-2i\phi_{\hbk_1,\hbx_1}}e^{-2i\phi_{\hbk_2,\widehat{\bx_1-\bs_1}}} + \xi^*_{-}(\bs_1;\bx_1) e^{-4i\phi_{\hbs_1,\hbx_1}} e^{2i\phi_{\hbk_1,\hbx_1}}e^{2i\phi_{\hbk_2,\widehat{\bx_1-\bs_1}}} \nonumber\\
    &\quad\quad\quad\quad
    + \xi_{+}(\bs_1;\bx_1) e^{-2i\phi_{\hbk_1,\hbx_1}}e^{2i\phi_{\hbk_2,\widehat{\bx_1-\bs_1}}} + \xi^*_{+}(\bs_1;\bx_1) e^{2i\phi_{\hbk_1,\hbx_1}}e^{-2i\phi_{\hbk_2,\widehat{\bx_1-\bs_1}}} \bigg] 
    \xi_\mathrm{gg}(\bs_2;\bx_2) \nonumber\\
    &\overset{\widehat{\bx_1-\bs_1} \rightarrow \hbx_1,~\mathrm{Eq.~(\ref{eq:smooth_Pk_approx})}}{\simeq} N^\mathrm{I}_{L_1}N^\mathrm{I}_{L_2} 
    \int_{\hbk_1,\hbk_2,\bx_1,\bx_2} 
    e^{-i(\bk_1-\bk_2)\cdot(\bx_1-\bx_2)} \nonumber\\
    &\quad\quad\times 
    W^{\gamma}_{22}(\bx_1)W^{\rmg}_{22}(\bx_2)
    \mathcal{L}^{m=2}_{L_1}(\hbk_1 \cdot \hbx_1) \mathcal{L}^{m=2}_{L_2}(\hbk_2 \cdot \hbx_1) \nonumber\\
    &\quad\quad\times 
    \bigg[ P^\mathrm{local}_{-} (\bk_2;\bx_1) e^{-2i\phi_{\hbk_1,\hbx_1}}e^{2i\phi_{\hbk_2,\hbx_1}} + P^\mathrm{local*}_{-} (\bk_2;\bx_1) e^{2i\phi_{\hbk_1,\hbx_1}}e^{-2i\phi_{\hbk_2,\hbx_1}} \nonumber\\
    &\quad\quad\quad\quad
    + P^\mathrm{local}_{+} (\bk_2;\bx_1) e^{-2i\phi_{\hbk_1,\hbx_1}}e^{2i\phi_{\hbk_2,\hbx_1}} + P^\mathrm{local*}_{+} (\bk_2;\bx_1) e^{2i\phi_{\hbk_1,\hbx_1}}e^{-2i\phi_{\hbk_2,\bx_1}} \bigg] 
    P^\mathrm{local}_\mathrm{gg} (\bk_1;\bx_2) \nonumber\\
    &= 2N^\mathrm{I}_{L_1}N^\mathrm{I}_{L_2} 
    \int_{\hbk_1,\hbk_2,\bx_1,\bx_2} 
    e^{-i(\bk_1-\bk_2)\cdot(\bx_1-\bx_2)} \nonumber\\
    &\quad\quad\times 
    W^{\gamma}_{22}(\bx_1)W^{\rmg}_{22}(\bx_2)
    \mathcal{L}^{m=2}_{L_1}(\hbk_1 \cdot \hbx_1) \mathcal{L}^{m=2}_{L_2}(\hbk_2 \cdot \hbx_1) \nonumber\\
    &\quad\quad\times 
    \bigg\{ e^{-2i\phi_{\hbk_1,\hbx_1}}e^{2i\phi_{\hbk_2,\hbx_1}} + e^{2i\phi_{\hbk_1,\hbx_1}}e^{-2i\phi_{\hbk_2,\hbx_1}} \bigg\} 
    P^\mathrm{local}_\mathrm{EE} (\bk_2;\bx_1) P^\mathrm{local}_\mathrm{gg} (\bk_1;\bx_2) \nonumber\\
    &\overset{\mathrm{Eq.~(\ref{eq:wide_angle_approx})}}{\simeq} 
    \sum_{\ell'_1,\ell'_2} P^{(\ell'_1)}_{\mathrm{gg}}(k_1) P^{(\ell'_2)}_\mathrm{EE}(k_2)~2N^\mathrm{I}_{L_1}N^\mathrm{I}_{L_2}
    \int_{\hbk_1,\hbk_2,\bx_1,\bx_2} 
    e^{-i(\bk_1-\bk_2)\cdot(\bx_1-\bx_2)} \nonumber\\
    &\quad\quad\times 
    W^{\gamma}_{22}(\bx_1)W^{\rmg}_{22}(\bx_2)
    \mathcal{L}^{m=2}_{L_1}(\hbk_1 \cdot \hbx_1) \mathcal{L}^{m=2}_{L_2}(\hbk_2 \cdot \hbx_1) \mathcal{L}_{\ell'_1}(\hbk_1 \cdot \hbx_2) \mathcal{L}_{\ell'_2}(\hbk_2 \cdot \hbx_1) \nonumber\\
    &\quad\quad\times
    \bigg\{ e^{-2i\phi_{\hbk_1,\hbx_1}}e^{2i\phi_{\hbk_2,\hbx_1}} + e^{2i\phi_{\hbk_1,\hbx_1}}e^{-2i\phi_{\hbk_2,\hbx_1}} \bigg\}, 
    \label{eq:cov_ia_PartB_nonsym}
\end{align}
where we have defined the local IA-auto power spectra:
\begin{align}
    P^\mathrm{local}_\mathrm{EE} (\bk;\bx) 
    &\equiv \frac{1}{2} \mathrm{Re} \left[P^\mathrm{local}_{+} (\bk;\bx) + P^\mathrm{local}_{-} (\bk;\bx) \right], \\
    P^\mathrm{local}_{+} (\bk;\bx) 
    &\equiv \int_{\bs} \xi_{+}(\bs;\bx) e^{-i\bk\cdot\bs}, \\
    P^\mathrm{local}_{-} (\bk;\bx) 
    &\equiv \int_{\bs} \xi_{-}(\bs;\bx) e^{4i\phi_{\hbs;\hbx}-4i\phi_{\hbk,\hbx}} e^{-i\bk\cdot\bs}.
\end{align}
In the second line, we have changed the dummy variable $\bx_2$ with $\bx'_2$ and also used that $L_2$ is even, i.e. $\mathcal{L}^{m=2}_{L_2}(\mu) = \mathcal{L}^{m=2}_{L_2}(-\mu)$, after replacing $\hbk_2 \rightarrow -\hbk_2$. 
Eq.~(\ref{eq:cov_ia_PartB_nonsym}) apparently breaks the symmetry under $(L_1,k_1) \leftrightarrow (L_2,k_2)$ due to the choice of the LOS direction associated with the endpoint approximation. 
To avoid this inconsistency, we symmetrize the result in Eq.~(\ref{eq:cov_ia_PartB_nonsym}) as 
\begin{align}
    \bigg\{\mathrm{B:} \avrg{\gamma \gamma}\avrg{\delta \delta} \bigg\} 
    \equiv 
    \sum_{\ell'_1,\ell'_2} \left[ 
    P^{(\ell'_1)}_{\mathrm{gg}}(k_1) P^{(\ell'_2)}_\mathrm{EE}(k_2) 
    \mathcal{W}^{\mathrm{II}(1,\mathrm{B})}_{L_1,L_2,\ell'_1,\ell'_2}(k_1,k_2) +  (k_1 \leftrightarrow k_2) \right], 
    \label{eq:cov_ia_PartB}
\end{align}
where
\begin{align}
    \mathcal{W}^{\mathrm{II}(1,\mathrm{B})}_{L_1,L_2,\ell'_1,\ell'_2}(k_1,k_2) &\equiv
    N^\mathrm{I}_{L_1}N^\mathrm{I}_{L_2}
    \int_{\hbk_1,\hbk_2,\bx_1,\bx_2} 
    e^{-i(\bk_1-\bk_2)\cdot(\bx_1-\bx_2)} W^{\gamma}_{22}(\bx_1)W^{\rmg}_{22}(\bx_2) 
    \mathcal{L}_{\ell'_1}(\hbk_1 \cdot \hbx_2) \mathcal{L}_{\ell'_2}(\hbk_2 \cdot \hbx_1) \nonumber\\
    &\quad\quad\times 
    \mathcal{L}^{m=2}_{L_1}(\hbk_1 \cdot \hbx_1) \mathcal{L}^{m=2}_{L_2}(\hbk_2 \cdot \hbx_1) 
    \bigg\{ e^{-2i\phi_{\hbk_1,\hbx_1}}e^{2i\phi_{\hbk_2,\hbx_1}} + e^{2i\phi_{\hbk_1,\hbx_1}}e^{-2i\phi_{\hbk_2,\hbx_1}} \bigg\}. 
    \label{eq:WII_1_B}
\end{align}
We can obtain the latter case explicitly by changing $\bx_1\leftrightarrow\bx'_1, ~\hbk_1\rightarrow-\hbk_1$ and using that $L_2$ is even in the second line of Eq.~(\ref{eq:cov_ia_PartB_nonsym}) instead. 
Note that we have expanded the IA power spectra by using the Legendre polynomials during the calculation of the covariance, not the associated Legendre polynomials as in the definition of the measurements. 

In summary, the continuous part of the auto covariance for the IA-galaxy power spectrum is given by
\begin{align}
    \bC^\mathrm{II(cont.)}_{L_1L_2}(k_1,k_2) = 
    \sum_{\ell'_1,\ell'_2} P^{(\ell'_1)}_{\mathrm{Eg}}(k_1) P^{(\ell'_2)}_{\mathrm{Eg}}(k_2) 
    \mathcal{W}^{\mathrm{II}(1,\mathrm{A})}_{L_1,L_2,\ell'_1,\ell'_2}(k_1,k_2)
    + \sum_{\ell'_1,\ell'_2} \left[ 
    P^{(\ell'_1)}_{\mathrm{gg}}(k_1) P^{(\ell'_2)}_\mathrm{EE}(k_2) 
    \mathcal{W}^{\mathrm{II}(1,\mathrm{B})}_{L_1,L_2,\ell'_1,\ell'_2}(k_1,k_2) +  (k_1 \leftrightarrow k_2) \right], 
    \label{eq:cov_II_cont_tot}
\end{align}
where $\mathcal{W}^{\mathrm{II}(1,\mathrm{A})}$ and $\mathcal{W}^{\mathrm{II}(1,\mathrm{B})}$ are defined by Eq.~(\ref{eq:cov_ia_PartA}) and (\ref{eq:WII_1_B}), respectively.

\subsubsection{Gaussian covariance: shot/shape noise terms}
To derive the shot noise and shape noise contributions, we first define the fields that are estimated from discrete objects by
\begin{align*}
    &\hFg^{(\ell)}(\bk) = \int_{\bx} \bar{n}_{\rmg}(\bx)w_{\rmg}(\bx) e^{-i\bk\cdot \bx} \mathcal{L}_\ell (\hbk\cdot\hbx) 
    &&\rightarrow 
    \left(\sum_i^{N_{\rmg}} - \alpha \sum_i^{N_{\rmr}} \right) w_{\rmg}(\bx_i) e^{-i\bk\cdot \bx_i} \mathcal{L}_\ell (\hbk\cdot\hbx_i) \\
    &\hFgamij^{(L)}(\bk)\hk^i\hk^j = \int_{\bx} \bar{n}_{\gamma}(\bx)w_{\gamma}(\bx) \gamma(\bx) e^{-2i\phi_{\hbk,\hbx}} e^{-i\bk\cdot \bx} \mathcal{L}^{m=2}_L (\hbk\cdot\hbx) 
    &&\rightarrow 
    \sum_i^{N_\gamma} w_{\gamma}(\bx_i) \gamma(\bx_i) e^{-2i\phi_{\hbk,\hbx_i}} e^{-i\bk\cdot \bx_i} \mathcal{L}^{m=2}_L (\hbk\cdot\hbx_i). 
\end{align*}
Hereafter we distinguish the discrete field from the continuous limit by using the label ``${}^d$'', i.e.  ${}^d\hFg^{(\ell)}$ and ${}^d\hFgamij^{(L)}$.
In general, the two-point correlations of these discrete fields have the shot noise and shape noise terms, respectively:
\begin{align*}
    &\avrg{^d\hFg^{(\ell_1)}(\bk_1) {}^d\hFg^{(\ell_2)}(-\bk_2)} \\
    &\quad\quad= 
    \avrg{\bigg[ \bigg( \sum^{N_{\rmg}}_i - \alpha \sum^{N_{\rmr}}_i \bigg) w_{\rmg}(\bx_i) e^{-i\bk_1\cdot \bx_i} \mathcal{L}_{\ell_1} (\hbk_1\cdot\hbx_i) \bigg]
    \bigg[ \bigg( \sum^{N_{\rmg}}_j - \alpha \sum^{N_{\rmr}}_j \bigg) w_{\rmg}(\bx_j) e^{i\bk_2\cdot \bx_j} \mathcal{L}_{\ell_2} (\hbk_2\cdot\hbx_j) \bigg]} \\
    &\quad\quad= 
    \avrg{ \bigg( \sum^{N_{\rmg}}_i - \alpha \sum^{N_{\rmr}}_i \bigg) \bigg( \sum^{N_{\rmg}}_{j\ne i} - \alpha \sum^{N_{\rmr}}_{j\ne i} \bigg) w_{\rmg}(\bx_i)w_{\rmg}(\bx_j) e^{-i\bk_1\cdot \bx_i} e^{i\bk_2\cdot \bx_j} \mathcal{L}_{\ell_1} (\hbk_1\cdot\hbx_i) \mathcal{L}_{\ell_2} (\hbk_2\cdot\hbx_j)} \\
    &\quad\quad\quad+
    \avrg{ \bigg( \sum^{N_{\rmg}}_i + \alpha^2 \sum^{N_{\rmr}}_i \bigg) w^2_{\rmg}(\bx_i) e^{-i\bk_1\cdot \bx_i} e^{i\bk_2\cdot \bx_i} \mathcal{L}_{\ell_1} (\hbk_1\cdot\hbx_i) \mathcal{L}_{\ell_2} (\hbk_2\cdot\hbx_i) } \\
    &\quad\quad\simeq
    \avrg{\hFg^{(\ell_1)}(\bk_1) \hFg^{(\ell_2)}(-\bk_2)} 
    +
    (1 + \alpha) \int_{\bx} \bar{n}_{\rmg}(\bx) w^2_{\rmg}(\bx) e^{-i(\bk_1 - \bk_2)\cdot \bx} \mathcal{L}_{\ell_1} (\hbk_1\cdot\hbx) \mathcal{L}_{\ell_2} (\hbk_2\cdot\hbx) \\
    &\avrg{^d\hFgamij^{(L_1)}(\bk_1)\hk^i_1\hk^j_1 {}^d\hFgamkl^{(L_2)*}(-\bk_2)\hk^k_2\hk^l_2} \\
    &\quad\quad= \avrg{\sum_i^{N_\gamma} w_{\gamma}(\bx_i) \gamma(\bx_i) e^{-2i\phi_{\hbk_1,\hbx_i}} e^{-i\bk_1\cdot \bx_i} \mathcal{L}^{m=2}_{L_1} (\hbk_1\cdot\hbx_i) 
    \sum_j^{N_\gamma} w_{\gamma}(\bx_j) \gamma^*(\bx_j) e^{+2i\phi_{\hbk_2,\hbx_j}} e^{-i\bk_2\cdot \bx_j} \mathcal{L}^{m=2}_{L_2} (\hbk_2\cdot\hbx_j)} \\
    &\quad\quad= \avrg{\sum_i^{N_\gamma} \sum_{j\ne i}^{N_\gamma} w_{\gamma}(\bx_i) \gamma(\bx_i) e^{-2i\phi_{\hbk_1,\hbx_i}} e^{-i\bk_1\cdot \bx_i} \mathcal{L}^{m=2}_{L_1} (\hbk_1\cdot\hbx_i) 
    w_{\gamma}(\bx_j) \gamma^*(\bx_j) e^{+2i\phi_{\hbk_2,\hbx_j}} e^{-i\bk_2\cdot \bx_j} \mathcal{L}^{m=2}_{L_2} (\hbk_2\cdot\hbx_j)} \\
    &\quad\quad\quad\quad+
    \avrg{\sum_i^{N_\gamma} w^2_{\gamma}(\bx_i) |\gamma(\bx_i)|^2 e^{-2i\phi_{\hbk,\hbx_i}+2i\phi_{\hbk,\hbx_i}} e^{-i(\bk_1-\bk_2)\cdot\bx_i} \mathcal{L}^{m=2}_{L_1} (\hbk\cdot\hbx_i) \mathcal{L}^{m=2}_{L_2} (\hbk\cdot\hbx_i)} \\
    &\quad\quad\simeq \avrg{\hFgamij^{(L_1)}(\bk_1)\hk^i_1\hk^j_1 \hFgamkl^{(L_2)*}(-\bk_2)\hk^k_2\hk^l_2} 
    + 
    2\sigma^2_\gamma \int_{\bx} W^\gamma_{12}(\bx) e^{-2i\phi_{\hbk_1,\hbx}+2i\phi_{\hbk_2,\hbx}} e^{-i(\bk_1-\bk_2)\cdot\bx} \mathcal{L}^{m=2}_{L_1} (\hbk_1\cdot\hbx) \mathcal{L}^{m=2}_{L_2} (\hbk_2\cdot\hbx)
\end{align*} 
Note that since $\avrg{\gamma} = 0$ and $\avrg{\gamma^2} = \avrg{\gamma^2_1} - \avrg{\gamma^2_2} = 0$, the shape noise terms appear in the ``plus'' power spectrum component in the B-part (Eq.~\ref{eq:cov_ia_PartB}): 
$\hFgamij\hFgamkl^* \sim \avrg{|\gamma|^2} = \avrg{\gamma^2_1} + \avrg{\gamma^2_2} \equiv 2\sigma^2_\gamma$. 
For convenience of discussion, we introduce the noise-related correlations:
\begin{align*}
    \mathcal{I}^\mathrm{gg}_{\ell_1\ell_2}(\bk_1,-\bk_2) 
    &\equiv (1 + \alpha) \int_{\bx} W^{\rmg}_{12}(\bx) e^{-i(\bk_1 - \bk_2)\cdot \bx} \mathcal{L}_{\ell_1} (\hbk_1\cdot\hbx) \mathcal{L}_{\ell_2} (\hbk_2\cdot\hbx) \\
    2\mathcal{I}^{\gamma\gamma}_{L_1L_2}(\bk_1,-\bk_2) 
    &\equiv 2\sigma^2_\gamma \int_{\bx} W^\gamma_{12}(\bx) e^{-i(\bk_1-\bk_2)\cdot\bx} \mathcal{L}^{m=2}_{L_1} (\hbk_1\cdot\hbx) \mathcal{L}^{m=2}_{L_2} (\hbk_2\cdot\hbx)
    \bigg\{e^{-2i\phi_{\hbk_1,\hbx}}e^{2i\phi_{\hbk_2,\hbx}} + e^{2i\phi_{\hbk_1,\hbx}}e^{-2i\phi_{\hbk_2,\hbx}}\bigg\}.
\end{align*}
Then in the case of the discrete fields, the B-part  can be rewritten as
\begin{align*}
    &{}^d\bigg\{\mathrm{B:} \avrg{\gamma \gamma}\avrg{\delta \delta} \bigg\} \\
    &\quad= N^\mathrm{I}_{L_1}N^\mathrm{I}_{L_2}
    \int_{\hbk_1, \hbk_2} 
    \avrg{\left[ {}^d\hFgamij^{(L_1)}(\bk_1)\hk^i_1\hk^j_1 + {}^d\hFgamij^{(L_1)*}(\bk_1)\hk^i_1\hk^j_1 \right] 
    \left[ {}^d\hFgamkl^{(L_2)}(-\bk_2)\hk^k_2\hk^l_2 + {}^d\hFgamkl^{(L_2)*}(-\bk_2)\hk^k_2\hk^l_2 \right]} \\
    &\quad\quad\quad\quad\quad\quad\quad\quad\times 
    \avrg{{}^d\hFg(-\bk_1)  {}^d\hFg(\bk_2)} \\
    &\quad\simeq N^\mathrm{I}_{L_1}N^\mathrm{I}_{L_2}
    \int_{\hbk_1, \hbk_2} 
    \bigg\{ \avrg{\left[ \hFgamij^{(L_1)}(\bk_1)\hk^i_1\hk^j_1 + \hFgamij^{(L_1)*}(\bk_1)\hk^i_1\hk^j_1 \right] 
    \left[ \hFgamkl^{(L_2)}(-\bk_2)\hk^k_2\hk^l_2 + \hFgamkl^{(L_2)*}(-\bk_2)\hk^k_2\hk^l_2 \right]} + 2\mathcal{I}^{\gamma\gamma}_{L_1L_2}(\bk_1,-\bk_2) \bigg\} \\
    &\quad\quad\quad\quad\quad\quad\quad\quad\times 
    \bigg\{ \avrg{\hFg(-\bk_1) \hFg(\bk_2)} + \mathcal{I}^\mathrm{gg}_{00}(-\bk_1,\bk_2) \bigg\} \\
    &\quad\equiv 
    {}^\mathrm{cont.}\bigg\{\mathrm{B:} \avrg{\gamma \gamma}\avrg{\delta \delta} \bigg\} \\
    &\quad\quad+ 
    \sum_{\ell'} \bigg[ \left\{P^{(\ell')}_\mathrm{gg}(k_1) \mathcal{W}^{\mathrm{II}(2,\mathrm{shape})}_{L_1,L_2,\ell'}(k_1,k_2) 
    + P^{(\ell')}_\mathrm{EE}(k_1) \mathcal{W}^{\mathrm{II}(2,\mathrm{shot})}_{L_1,L_2,\ell'}(k_1,k_2) \right\} 
    + (k_1 \leftrightarrow k_2) \bigg] 
    + \mathcal{W}^{\mathrm{II}(3)}_{L_1,L_2}(k_1,k_2) \\
    &\quad\equiv 
    {}^\mathrm{cont.}\bigg\{\mathrm{B:} \avrg{\gamma \gamma}\avrg{\delta \delta} \bigg\} 
    + \bC^\mathrm{II(SN)}_{L_1L_2}
\end{align*}
where the first term corresponds to the continuous limit (Eq.~\ref{eq:cov_ia_PartB}) and the window kernels for the shot/shape noise contributions are 
\begin{align}
    &\mathcal{W}^{\mathrm{II}(2,\mathrm{shape})}_{L_1,L_2,\ell'}(k_1,k_2) 
    \equiv
    \sigma^2_\gamma N^\mathrm{I}_{L_1}N^\mathrm{I}_{L_2} \int_{\hbk_1, \hbk_2, \bx_1, \bx_2} 
    W^{\gamma}_{12}(\bx_1)W^{\rmg}_{22}(\bx_2)  e^{-i(\bk_1 - \bk_2)\cdot (\bx_1 - \bx_2)} \nonumber\\
    &\quad\quad\quad\quad\quad\quad\quad\quad\quad\quad\times 
    \mathcal{L}^{m=2}_{L_1}(\hbk_1\cdot \hbx_1) \mathcal{L}^{m=2}_{L_2}(\hbk_2\cdot \hbx_1)
    \bigg\{e^{-2i\phi_{\hbk_1,\hbx_1}}e^{2i\phi_{\hbk_2,\hbx_1}} + e^{2i\phi_{\hbk_1,\hbx_1}}e^{-2i\phi_{\hbk_2,\hbx_1}}\bigg\} 
    \mathcal{L}_{\ell'}(\hbk_1\cdot \hbx_2), \label{eq:W_II_2_shape}\\
    &\mathcal{W}^{\mathrm{II}(2,\mathrm{shot})}_{L_1,L_2,\ell'}(k_1,k_2) 
    \equiv 
    (1+\alpha) N^\mathrm{I}_{L_1}N^\mathrm{I}_{L_2} \int_{\hbk_1, \hbk_2, \bx_1, \bx_2} 
    W^{\gamma}_{22}(\bx_1)W^{\rmg}_{12}(\bx_2)  e^{-i(\bk_1 - \bk_2)\cdot (\bx_1 - \bx_2)} \nonumber\\
    &\quad\quad\quad\quad\quad\quad\quad\quad\quad\quad\quad\quad\quad\quad\quad\quad\quad\quad\quad\quad\quad\quad\quad\quad\quad\quad\quad\quad\times 
    \mathcal{L}^{m=2}_{L_1}(\hbk_1\cdot \hbx_1) \mathcal{L}^{m=2}_{L_2}(\hbk_2\cdot \hbx_1) 
    \mathcal{L}_{\ell'}(\hbk_1\cdot \hbx_1), \label{eq:W_II_2_shape}\\
    &\mathcal{W}^{\mathrm{II}(3)}_{L_1,L_2}(k_1,k_2)
    \equiv 
    2\sigma^2_\gamma (1+\alpha) N^\mathrm{I}_{L_1}N^\mathrm{I}_{L_2} \int_{\hbk_1, \hbk_2, \bx_1, \bx_2} 
    W^{\gamma}_{12}(\bx_1)W^{\rmg}_{12}(\bx_2)  e^{-i(\bk_1 - \bk_2)\cdot (\bx_1 - \bx_2)} \nonumber\\
    &\quad\quad\quad\quad\quad\quad\quad\quad\quad\quad\quad\quad\quad\quad\quad\times 
    \mathcal{L}^{m=2}_{L_1}(\hbk_1\cdot \hbx_1) \mathcal{L}^{m=2}_{L_2}(\hbk_2\cdot \hbx_1)
    \bigg\{e^{-2i\phi_{\hbk_1,\hbx_1}}e^{2i\phi_{\hbk_2,\hbx_1}} + e^{2i\phi_{\hbk_1,\hbx_1}}e^{-2i\phi_{\hbk_2,\hbx_1}}\bigg\}. 
    \label{eq:W_II_3}
\end{align}

\subsection{Intrinsic Alignments-Galaxy Clustering Cross Covariance: $\mathrm{Cov}\left[ P_{\gamma{\rmg}}, P_\mathrm{gg} \right]$} 
\label{subsec:cov_PsgPgg}
Similarly, we derive the cross components of the covariance matrix.
\begin{align}
    \bC^\mathrm{IG}_{L_1L_2} 
    &\equiv \avrg{\hat{P}^{(L_1)}_\mathrm{Eg}(k_1) \hat{P}^{(\ell_2)}_\mathrm{gg}(k_2)} 
    - \avrg{\hat{P}^{(L_1)}_\mathrm{Eg}(k_1)} \avrg{\hat{P}^{(\ell_2)}_\mathrm{gg}(k_2)} \nonumber\\
    &\equiv 
    \sum_{\ell'_1,\ell'_2} \left[ 
    P^{(\ell'_1)}_{\mathrm{gg}}(k_1) P^{(\ell'_2)}_{\mathrm{Eg}}(k_2) 
    \mathcal{W}^{\mathrm{IG}(1)}_{L_1,\ell_2,\ell'_1,\ell'_2}(k_1,k_2) 
    + (k_1 \leftrightarrow k_2) \right] 
    + \sum_{\ell'} \left[ 
    P^{(\ell'_2)}_{\mathrm{Eg}}(k_2) 
    \mathcal{W}^{\mathrm{IG}(2)}_{L_1,\ell_2,\ell'}(k_1,k_2) 
    + (k_1 \leftrightarrow k_2) \right] \nonumber\\ 
    &\equiv \bC^\mathrm{IG(cont.)}_{L_1\ell_2} + \bC^\mathrm{IG(SN)}_{L_1\ell_2}, 
\end{align}
where
\begin{align}
    \mathcal{W}^{\mathrm{IG}(1)}_{L_1,\ell_2,\ell'_1,\ell'_2}(k_1,k_2) 
    &\equiv
    N^\mathrm{I}_{L_1}N^\mathrm{G}_{\ell_2}  \int_{\hbk_1,\hbk_2,\bx_1,\bx_2} 
    W^{\gamma}_{11}(\bx_1)W^{\rmg}_{11}(\bx_1)W^{\rmg}_{22}(\bx_2) e^{-i(\bk_1-\bk_2)\cdot(\bx_1-\bx_2)} \nonumber\\
    &\quad\quad\quad\quad\quad\quad\quad\quad\times 
    \mathcal{L}^{m=2}_{L_1}(\hbk_1 \cdot \hbx_1) 
    \left[ \mathcal{L}_{\ell_2}(\hbk_2 \cdot \hbx_2) + \mathcal{L}_{\ell_2}(\hbk_2 \cdot \hbx_1)\right]
    \mathcal{L}_{\ell'_1}(\hbk_1 \cdot \hbx_2) \mathcal{L}_{\ell'_2}(\hbk_2 \cdot \hbx_1), \label{eq:W_IG_1}\\
    \mathcal{W}^{\mathrm{IG}(2)}_{L_1,\ell_2,\ell'}(k_1,k_2) 
    &\equiv
    N^\mathrm{I}_{L_1}N^\mathrm{G}_{\ell_2}  \int_{\hbk_1,\hbk_2,\bx_1,\bx_2} 
    W^{\gamma}_{11}(\bx_1)W^{\rmg}_{11}(\bx_1)W^{\rmg}_{12}(\bx_2) e^{-i(\bk_1-\bk_2)\cdot(\bx_1-\bx_2)} \nonumber\\
    &\quad\quad\quad\quad\quad\quad\quad\quad\times 
    \mathcal{L}^{m=2}_{L_1}(\hbk_1 \cdot \hbx_1) 
    \left[ \mathcal{L}_{\ell_2}(\hbk_2 \cdot \hbx_2) + \mathcal{L}_{\ell_2}(\hbk_2 \cdot \hbx_1)\right]
    \mathcal{L}_{\ell'}(\hbk_2 \cdot \hbx_1), \label{eq:W_IG_2}
\end{align}

\subsection{Numerical Implementation} 
\label{subsec:cov_implementation}
To obtain the elements of analytic covariance matrices, we show how to evaluate the quartic functions in terms of the window functions, 
$\mathcal{W}^{\mathrm{XY}(i)}_{\ell_1,\ell_2,\cdots}~(\mathrm{X,Y}\in \{ \mathrm{G,I}\},~i=1,2,3)$, from given random catalogs.
Since there is a multi-dimensional integration, $\int_{\hbk_1,\hbk_2,\bx_1,\bx_2} \cdots$, in all $\mathcal{W}$s and thus particle-based direct calculations are considered to be not efficient, we employed the grid-based implementation with FFT algorithm following Ref.~\cite{Wadekar&Scoccimarro2020:CovPT}. 
For example, in the case of the IA-IA covariance, the contributions from the monopoles, $(\ell'_1,\ell'_2) = (0,0)$, to the continuous parts of the lowest order multipole, $(L_1,L_2)=(2,2)$, can be rewritten as (Eq.~\ref{eq:cov_ia_PartA} and \ref{eq:cov_ia_PartB})
\begin{align*}
    &\mathcal{W}^{\mathrm{II}(1,\mathrm{A})}_{2,2,0,0}(k_1,k_2) \\
    &\equiv 
    4 \left(N^\mathrm{I}_{2}\right)^2 \int_{\hbk_1,\hbk_2,\bx_1,\bx_2} 
    e^{-i(\bk_1-\bk_2)\cdot(\bx_1-\bx_2)} 
    W^{\gamma}_{11}(\bx_1)W^{\rmg}_{11}(\bx_2)W^{\gamma}_{11}(\bx_2)W^{\rmg}_{11}(\bx_1) 
    \mathcal{L}^{m=2}_{2}(\hbk_1 \cdot \hbx_1) \mathcal{L}^{m=2}_{2}(\hbk_2 \cdot \hbx_2) \\
    &= 4 \left(N^\mathrm{I}_{2}\right)^2 \int_{\hbk_1,\hbk_2} 
    \left[ \int_{\bx_1} e^{-i(\bk_1-\bk_2)\cdot\bx_1} W^{\gamma}_{11}(\bx_1) W^{\rmg}_{11}(\bx_1) \mathcal{L}^{m=2}_{2}(\hbk_1 \cdot \hbx_1) \right] 
    \left[ \int_{\bx_2} e^{i(\bk_1-\bk_2)\cdot\bx_2} W^{\gamma}_{11}(\bx_2) W^{\rmg}_{11}(\bx_2) \mathcal{L}^{m=2}_{2}(\hbk_2 \cdot \hbx_2) \right] \\
    &\equiv 4 \left(N^\mathrm{I}_{2}\right)^2 \int_{\hbk_1,\hbk_2} 
    \mathcal{Q}^{\gamma{\rmg}}_2(\bk_1-\bk_2; \hbk_1) \mathcal{Q}^{\gamma{\rmg}*}_2(\bk_1-\bk_2; \hbk_2),
\end{align*}
with the associated Legendre polynomials, $\mathcal{L}^{m=2}_{2}(\mu)=3(1-\mu^2)$, decomposed into the sum of the products of Cartesian components to be computed by FFTs:
\begin{align*}
    \mathcal{Q}^{\gamma{\rmg}}_2(\bk_1-\bk_2; \hbk_1) 
    &= \int_{\bx} e^{-i(\bk_1-\bk_2)\cdot\bx} W^{\gamma}_{11}(\bx) W^{\rmg}_{11}(\bx)~ 3\left(1-(\hbk_1\cdot \hbx)^2\right) \\
    &= 3\underbrace{\int_{\bx} e^{-i(\bk_1-\bk_2)\cdot\bx} W^{\gamma}_{11}(\bx) W^{\rmg}_{11}(\bx)}_{1~\mathrm{FFT}}
    + 3\hk_1^i\hk_1^j \underbrace{\int_{\bx} e^{-i(\bk_1-\bk_2)\cdot\bx} W^{\gamma}_{11}(\bx) W^{\rmg}_{11}(\bx) \hx_i\hx_j}_{6~\mathrm{FFTs}}. 
\end{align*}
Similarly, we have
\begin{align*}
    &\mathcal{W}^{\mathrm{II}(1,\mathrm{B})}_{2,2,0,0}(k_1,k_2) 
    \equiv 
    \left(N^\mathrm{I}_{2}\right)^2 \int_{\hbk_1,\hbk_2,\bx_1,\bx_2} 
    e^{-i(\bk_1-\bk_2)\cdot(\bx_1-\bx_2)} W^{\gamma}_{22}(\bx_1)W^{\rmg}_{22}(\bx_2) \nonumber\\
    &\quad\quad\quad\quad\quad\quad\quad\quad\quad\times 
    \mathcal{L}^{m=2}_{2}(\hbk_1 \cdot \hbx_1) \mathcal{L}^{m=2}_{2}(\hbk_2 \cdot \hbx_1) 
    \bigg\{ e^{-2i\phi_{\hbk_1,\hbx_1}}e^{2i\phi_{\hbk_2,\hbx_1}} + e^{2i\phi_{\hbk_1,\hbx_1}}e^{-2i\phi_{\hbk_2,\hbx_1}} \bigg\} \\
    &= \left(N^\mathrm{I}_{2}\right)^2 \int_{\hbk_1,\hbk_2} 
    \left[ \int_{\bx_1} e^{-i(\bk_1-\bk_2)\cdot\bx_1} W^{\gamma}_{22}(\bx_1) 
    \mathcal{L}^{m=2}_{2}(\hbk_1 \cdot \hbx_1) \mathcal{L}^{m=2}_{2}(\hbk_2 \cdot \hbx_1) 
    \bigg\{ e^{-2i\phi_{\hbk_1,\hbx_1}}e^{2i\phi_{\hbk_2,\hbx_1}} + e^{2i\phi_{\hbk_1,\hbx_1}}e^{-2i\phi_{\hbk_2,\hbx_1}} \bigg\} \right] \\
    &\quad\quad\quad\quad\quad\quad\times 
    \left[ \int_{\bx_2} e^{i(\bk_1-\bk_2)\cdot\bx_2} W^{\rmg}_{22}(\bx_2) \right] \\
    &\equiv 
    \left(N^\mathrm{I}_{2}\right)^2 \int_{\hbk_1,\hbk_2} 
    \mathcal{Q}^{\gamma\gamma}_{22}(\bk_1-\bk_2; \hbk_1, \hbk_2) \mathcal{Q}_0^{\mathrm{gg}*}(\bk_1-\bk_2). 
\end{align*}
Hence we need 1 FFT computation for $\mathcal{Q}_0^\mathrm{gg}$ and 21 FFTs for $\mathcal{Q}^{\gamma\gamma}_{22}$ as
\begin{align*}
    \mathcal{Q}^{\gamma\gamma}_{22}(\bk_1-\bk_2; \hbk_1, \hbk_2) &
    = 
    \int_{\bx} e^{-i(\bk_1-\bk_2)\cdot\bx} W^{\gamma}_{22}(\bx) 
    ~3\left(1-(\hbk_1\cdot \hbx)^2\right) 3\left(1-(\hbk_2\cdot \hbx)^2\right)
    \bigg\{ e^{-2i\phi_{\hbk_1,\hbx}}e^{2i\phi_{\hbk_2,\hbx}} + e^{2i\phi_{\hbk_1,\hbx}}e^{-2i\phi_{\hbk_2,\hbx}} \bigg\} \\
    &= 36 \underbrace{\int_{\bx} e^{-i(\bk_1-\bk_2)\cdot\bx} W^{\gamma}_{22}(\bx) 
    \bigg\{ e^*_{ij}(\hbx)e_{kl}(\hbx) + e_{ij}(\hbx)e^*_{kl}(\hbx) \bigg\}}_{21~\mathrm{FFTs}} \hk^i_1\hk^j_1\hk^k_2\hk^l_2, 
\end{align*}
where we have used the definition of the phase factor: $e^{-2i\phi_{\hbk,\hbx}} = 2e^*_{ij}(\hbx) \hk_i\hk_j / (1-(\hbk \cdot \hbx)^2)$. 
Note that we need more arrays for the higher order multipoles ($\ell>0$) roughly scaling as $\ell^2$. 
However since the monopole moment (isotropic part) should have a dominant contribution and the higher order anisotropies of the window function should be sub-dominant as shown in Fig.~\ref{fig:window}, we ignore higher order moments than hexadecapole moment to save computational resources.

We next consider the double angular integration, $\int_{\hbk_1,\hbk_2} \cdots$, for each $(k_1,k_2)$-bin. 
We carry out the $\hbk_1$-integration taking the sub-sample average of $N_\mathrm{samp}$ points randomly drawn from wave vectors which belong to the $k_1$-shell ($N_\mathrm{samp} \leq N_\mathrm{mode}(k_1)$).
For the $\hbk_2$-integration, we refer to $\mathcal{Q}(\Delta \bk)$ within a sphere centered at the endpoint of each $\bk_1$ with a sufficiently large radius $|\Delta \bk| < k_\mathrm{sph}$ to take into account the smearing effect due to the survey window function, i.e. we set $k_\mathrm{sph} \gg 1/R_\mathrm{survey}$, and add them to the $(k_1,k_2)$-bin satisfying $k_2 = |\bk_1 - \Delta \bk|$.
In this work, we set $N_\mathrm{samp}=10000$ and $k_\mathrm{sph}=0.03~h\Mpc^{-1}$.

\subsection{Validation Tests for Covariance Matrices} 
\label{subsec:cov_validation}
We validate the analytic covariance for the estimated power spectra by comparing the evaluated covariance with that estimated from the mock data for the BOSS survey. 
Since the mathematical forms of window functions, $\mathcal{W}$, in both clustering and IA parts are fundamentally similar to each other, we first do the same validation test for the galaxy clustering part, $\bC^\mathrm{GG}_{\ell_1\ell_2}$, as done in Ref.~\cite{Wadekar&Scoccimarro2020:CovPT} to check our implementation of the numerical integrals, $\int_{\hbk_1,\hbk_2,\bx_1,\bx_2} \cdots$, in $\mathcal{W}$s. 
We prepare the mock covariance by using 2048 realizations of BOSS DR12 MultiDark-Patchy mock catalogs \cite{Kitaura+2016:PatchyMock} (hereafter Patchy mocks) and compare it to the analytic results (Eqs.~\ref{eq:cov_GG_tot}-\ref{eq:cov_W_GG_3}). 
We checked that the analytic covariance (Gaussian and shot noise terms) and the mock covariance show good agreement with each other on
scales up to $k\simeq 0.05~h\Mpc^{-1}$ that is the maximum wavenumber used for our analysis of the galaxy density power spectrum. 

\begin{figure*}
    \centering
    \includegraphics[width=1.0\columnwidth]{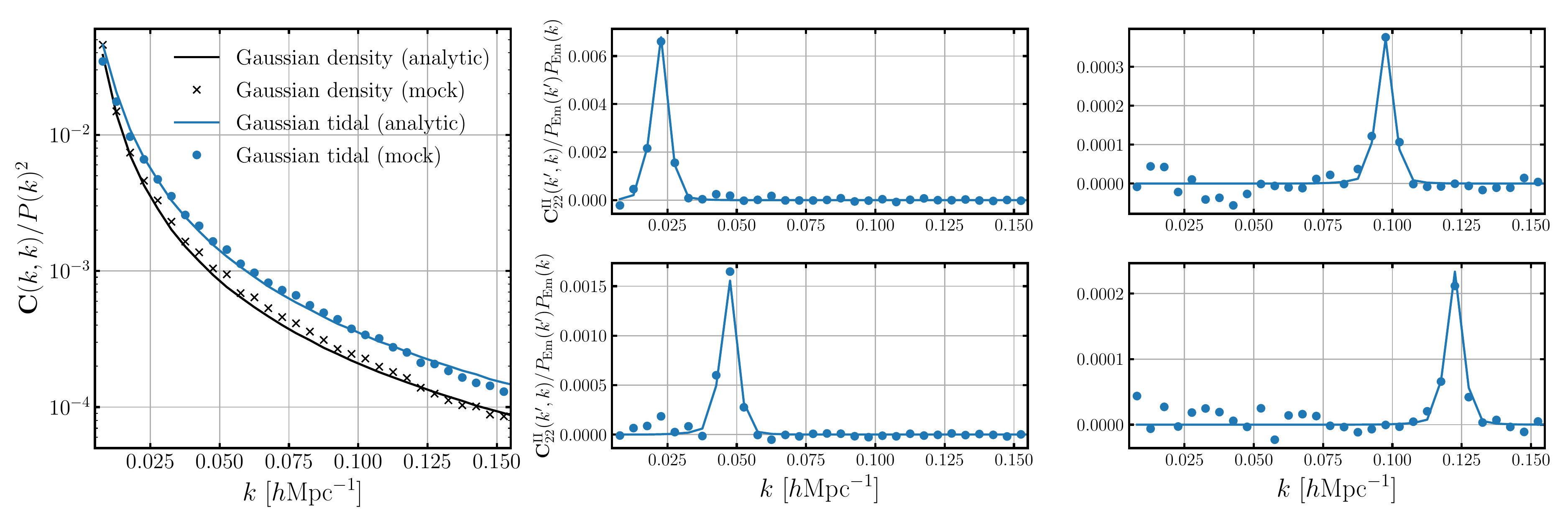}
    \caption{
    Comparison of the analytic covariance (solid line) with the mock covariance (dots) for the diagonal (left panel) and the off-diagonal terms (right), respectively. 
    The blue line and symbol correspond to the covariance of the shape power spectrum, $\bC^\mathrm{II(cont.)}_{22}$, and the black ones correspond to that of the density power spectrum, $\bC^\mathrm{GG(cont.)}_{00}$.
    }
    \label{fig:cov_validation}
\end{figure*}
On the other hand, as far as we know, there currently does not exist a suite of realistic and well physically motivated mock catalogs for galaxy IA unlike the galaxy clustering such as the Patchy mocks. 
Thus we do a validation test using the simulated tidal field as a hypothetical IA signal including the observational effects, i.e, the projection and survey window effects. 
We adopt the same Gaussian random fields and tidal fields as the mock data used in Ref.~\cite{Kurita&Takada2022:AnalysisIAPS}. 
We briefly describe the data here. 
We first generate each realization of the matter density field, $\delta(\bk)$ using the linear matter power spectrum $P(k)$ at redshift $z=0$, in a simulation box with comoving side length of $3~h^{-1}{\rm Gpc}$ with $512^3$ grids.
The Nyquist frequency $k_{\rm Ny} \simeq 0.5 ~h{\rm Mpc}^{-1}$. 
As for the input $P(k)$ we assume the flat $\Lambda$CDM cosmology, which is consistent with the {\it Planck} CMB data \citep{Planck2015_cosmo}: 
\begin{align*}
    \{ \Omega_\rmm, \omega_\rmb, \omega_\rmc, n_\rms, \ln(10^{10}A_\rms) \} = \{ 0.3156, 0.02225, 0.1198, 0.9645, 3.094\}. 
\end{align*}
The input power spectrum corresponds to $\sigma_8=0.834$, the rms value of present-day mass fluctuations within a sphere of radius $8\,h^{-1}\Mpc$.
We then compute the tidal field $T_{ij}(\bk) \equiv (\hk_i \hk_j - \delta_{ij}/3) \delta(\bk)$ in Fourier space and inverse-Fourier-transform the field to obtain $T_{ij}(\bx)$ on each grid in configuration space.
We repeated the above procedures to generate 1000 realizations of $T_{ij}$ using different random seeds. 
For the power spectrum measurements, we further define the projected tidal field, i.e. observed ellipticities, viewed by an observer in the simulation box for each grid as 
$\gamma^\mathrm{obs}(\bx) \equiv e_{ij}(\hbx) T_{ij}(\bx)$
with $e_{ij}$ being the polarization tensor. 
Note that the underlying (unwindowed) density and shape power spectra in this test are then given by 
\begin{align}
    P_\mathrm{mm}(\bk) &= P(k), \nonumber\\
    P_\mathrm{Em}(\bk) &= \frac{1}{2} (1-\mu_k^2) P(k). 
    \label{eq:cov_validation_theory}
\end{align} 
Employing a survey window that mimics the BOSS NGC footprint, we measure the IA power spectrum by using our LPP estimator defined by Eq.~\ref{eq:lpp_estimator_IG} and compute the mock covariance by using the measurements from 1000 simulation realizations. 
For the analytic covariance, on the other hand, we generate random particles whose distribution traces the assumed footprint and calculate the window functions, $\mathcal{W}$, with the above implementation.
By multiplying them and the theoretical power spectra (Eq.~\ref{eq:cov_validation_theory}) together, we obtain the analytic covariance, $\bC^\mathrm{II(cont.)}_{L_1L_2}$ (Eq.~\ref{eq:cov_II_cont_tot}). 
In Fig.~\ref{fig:cov_validation}, we show the comparison of the analytic covariance and the mock covariance. 
The two covariances agree well for both the diagonal terms and the off-diagonal terms due to window smearing. 
In particular, the diagonal terms of the covariance of the density and shape power spectra show a scale-dependent difference that is not just a constant multiple, due to the different effects of the window function on the scalar and tensor quantities.

\section{Weak Lensing Effects on IA Power Spectrum} 
\label{sec:weak_lensing}

We here derive the contamination of weak lensing effects in the measured IA power spectrum by using similar approximations employed in Refs.~\cite{Hui+2007:MagReal,Hui+2008:MagFourier,Schmidt+2008:WL3ptCF} to estimate the magnification bias on the measured galaxy power spectrum. 
We extend their previous results to the IA power spectrum and also newly propose a general method to include the survey window effects using an actual random catalog.

\subsection{Observables} 
\label{subsec:wl_observables}
The spatial fluctuation of an observed galaxy number density at a certain redshift is determined by the gravity of not only the surrounding large-scale structure but also the foreground large-scale structure due to the weak lensing effects. 
Hence the observed density fluctuation field including the leading-order weak lensing effect can be written as
\begin{align}
    \delta_\rmg^\mathrm{obs}(\bx) &= \delg(\bx) + \delta_\rmg^\mathrm{WL}(\bx).
    \label{eq:delta_obs}
\end{align}
The first term is the standard, intrinsic density field arising from the galaxy distribution in the large-scale structure at redshifts of galaxies in the sample. 
The second term is the magnification bias defined by 
$\delta_\rmg^\mathrm{WL}(\bx) \equiv 2(\alpha_\mathrm{mag}-1) \kappa^\mathrm{WL}(\bx)$ 
with $\alpha_\mathrm{mag} \equiv 5s/2$
where $s$ is the slope of the cumulative galaxy number counts for galaxies brighter than magnitude $m$:
\begin{align*}
    s \equiv \frac{\rmd {\ln N(<m)}}{\rmd m},
\end{align*}
and $\kappa^\mathrm{WL}$ is the weak lensing convergence field:
\begin{align}
    \kappa^\mathrm{WL}(\bx) 
    \equiv \frac{1}{2} \hat{\nabla}^2 \phi^\mathrm{WL}(\bx) 
    \equiv \frac{3}{2}\Om H^2_0 \int_0^\chi \rmd \chi' \frac{(\chi - \chi')\chi'}{\chi} \frac{1}{a(\chi')} \delm(\bx'),
\end{align}
where $\phi^\mathrm{WL}(\bx)$ is the lensing potential, $\chi$ is the comoving distance and $\hat{\nabla}^2$ is the angular part of the Laplacian. 
We have used the Poisson equation at the second equality.

Similarly, we define the observed galaxy shape field, inferred from the spatial pattern of observed galaxy ellipticities:
\begin{align}
    \gamma^\mathrm{obs}(\bx) &= \gamma^\mathrm{IA}(\bx) + \gamma^\mathrm{WL}(\bx),
    \label{eq:gamma_obs}
\end{align}
where the first term is the standard IA field arising from the large-scale structure at redshifts of galaxies in the sample, and the second term is the weak lensing shear:
\begin{align}
    \gamma^\mathrm{WL}(\bx) 
    \equiv \frac{1}{2} \eth^2 \phi^\mathrm{WL}(\bx) 
    = 
    \frac{3}{2}\Om H^2_0 \int_0^\chi \rmd \chi' \frac{(\chi - \chi')\chi'}{\chi} \frac{1}{a(\chi')} 2e_{ij}(\hbx) 
    \int_{\bk'} \hk'_i\hk'_j \delm(\bk') e^{i\bk'\cdot\bx'}, 
\end{align}
where $\eth^2 \equiv 2e_{ij}(\hbx) \hat{\nabla}_i\hat{\nabla}_j$ and $e_{ij}$ is the polarization tensor.
In the following, we use the abbreviated notations, $C\equiv 3\Om H^2_0/2$ and $K(\chi,\chi')\equiv (\chi - \chi')\chi'/(\chi a(\chi'))$.

\subsection{Two-point Statistics}
\label{subsec:wl_2pt_statistics}
The cross correlation between the observed galaxy density field and shape field can be decomposed into four terms:
\begin{align}
    \avrg{\delta_\rmg^\mathrm{obs} \gamma^\mathrm{obs}} = \avrg{\delg \gamma^\mathrm{IA}} + \avrg{\delg \gamma^\mathrm{WL}} + \avrg{\delta_\rmg^\mathrm{WL} \gamma^\mathrm{IA}} + \avrg{\delta_\rmg^\mathrm{WL} \gamma^\mathrm{WL}}.
\end{align}
The second and third terms arise due to the breakdown of the thin redshift shell approximation and the last term is the pure weak lensing auto-correlation.
We estimate the order of magnitude of each contribution including actual survey window effects. 

We first start with the auto correlation functions of weak lensing:
\begin{align*}
    \avrg{\delta_\rmg^\mathrm{WL}(\bx_1) \gamma^\mathrm{WL}(\bx_2)} 
    &= 2(\alpha_\mathrm{mag}-1) \avrg{\kappa^\mathrm{WL}(\bx_1) \gamma^\mathrm{WL}(\bx_2)} \\
    &= 2(\alpha_\mathrm{mag}-1) C^2
    \int_0^{\chi_1} \rmd \chi'_1 K(\chi_1,\chi'_1) \int_0^{\chi_2} \rmd \chi'_2 K(\chi_2,\chi'_2) 2e_{ij}(\hbx) 
    \int_{\bk'} \hk'_i\hk'_j P(k';\chi'_1,\chi'_2) e^{i\bk'\cdot (\bx'_1-\bx'_2)} \\
    &\simeq 2(\alpha_\mathrm{mag}-1) C^2 
    \int_0^{\mathrm{min}(\chi_1,\chi_2)} \rmd \chi' K(\chi_1,\chi')K(\chi_2,\chi') 
    \int_{\bk'_{\perp}} e^{2i\phi_{\hbk'_\perp}} P(k'_{\perp};\chi') e^{i\bk'_{\perp} \cdot (\bx'_{1,\perp}-\bx'_{2,\perp})}.
\end{align*}
We have used the Limber approximation in the second line where the subscript ``${}_\perp$'' denotes the components perpendicular to the line-of-sight (LOS) direction, 
and introduced the notation of the phase factor, $2e_{ij}(\hbx) \hk'_i\hk'_j \equiv (1-(\hbk' \cdot \hbx)^2) e^{2i\phi_{\hbk'_\perp}} \simeq e^{2i\phi_{\hbk'_\perp}}$. 
Also by approximating $\chi_1,\chi_2$ as the (constant) mean redshift $\bar{\chi}$, we obtain
\begin{align}
    \avrg{\delta_\rmg^\mathrm{WL}(\bx_1) \gamma^\mathrm{WL}(\bx_2)} 
    \simeq 2(\alpha_\mathrm{mag}-1) C^2 
    \int_0^{\bar{\chi}} \rmd \chi' K^2(\bar{\chi},\chi')
    \int_{\bk'_{\perp}} e^{2i\phi_{\hbk'_\perp}} P(k'_{\perp};\chi') e^{i\bk'_{\perp} \cdot (\bx'_{1,\perp}-\bx'_{2,\perp})}.
    \label{eq:wlwl_nowindow}
\end{align}
Next, we consider the survey window effects by multiplying the weight function, e.g. $\tilde{\delta}_\rmg^\mathrm{WL}(\bx) \equiv W(\bx) \delta_\rmg^\mathrm{WL}(\bx)$. 
Hereafter we assume the separable form for the window function, i.e. $W(\bx) \simeq W_\parallel(x_\parallel) W_\perp(\bx_\perp)$ where the subscript ``${}_\parallel$'' denotes the LOS component (see the next subsection for justifications of this approximation). 
By performing the Fourier transform of Eq.~(\ref{eq:wlwl_nowindow}), we define the coordinate-independent power spectrum with the window effects as
\begin{align}
    \tilde{P}_{\delta_\rmg^\mathrm{WL} \gamma^\mathrm{WL}}(\bk) \equiv e^{-2i\phi_{\hbk_\perp}} |W_\parallel(k_\parallel)|^2 
    2(\alpha_\mathrm{mag}-1) C^2 \int_0^{\bar{\chi}} \rmd \chi' K^2(\bar{\chi},\chi') 
    \left( \frac{\bar{\chi}}{\chi'} \right)^2
    \int_{\bk'_{\perp}} e^{2i\phi_{\hbk'_\perp}} P \left(k'_{\perp} \frac{\bar{\chi}}{\chi'} ;\chi'\right) 
    |W_\perp(\bk_\perp - \bk'_\perp)|^2,
\end{align}
and then, we finally obtain the multipole moments with respect to the associated Legendre polynomials of $m=2$:
\begin{align}
    \tilde{P}^{(L)}_{\delta_\rmg^\mathrm{WL} \gamma^\mathrm{WL}}(k) 
    &\equiv (2L+1)\frac{(L-2)!}{(L+2)!} \int_{\hbk} \tilde{P}_{\delta_\rmg^\mathrm{WL} \gamma^\mathrm{WL}}(\bk) \mathcal{L}_L^{m=2} (\mu) \\
    &=(2L+1)\frac{(L-2)!}{(L+2)!} \int_{-1}^1 \frac{\rmd \mu}{2} \mathcal{L}_L^{m=2} (\mu) |W_\parallel(k_\parallel)|^2 2(\alpha_\mathrm{mag}-1) C^2 \nonumber\\
    &\quad\times 
    2\pi \int_0^\infty r_\perp \rmd r_\perp Q_\perp(r_\perp) J_2(k_\perp r_\perp) \left[ \int_0^\infty \frac{k'_\perp \rmd k'_\perp}{2\pi} 
    \left\{ \int_0^{\bar{\chi}} \rmd \chi' K^2(\bar{\chi},\chi') 
    \left( \frac{\bar{\chi}}{\chi'} \right)^2 P\left(k'_{\perp} \frac{\bar{\chi}}{\chi'} ;\chi'\right) \right\} 
    J_2(k'_\perp r_\perp)\right] \\
    &\equiv(2L+1)\frac{(L-2)!}{(L+2)!} \int_{-1}^1 \frac{\rmd \mu}{2} \mathcal{L}_L^{m=2} (\mu) |W_\parallel(k_\parallel)|^2 2(\alpha_\mathrm{mag}-1) C^2 \nonumber\\
    &\quad\times 
    \mathcal{H}^\mathrm{2D}_2 \left[ Q_\perp(r_\perp) \left(\mathcal{H}^\mathrm{2D}_2\right)^{-1} \left[ 
    \int_0^{\bar{\chi}} \rmd \chi' K^2(\bar{\chi},\chi') \left( \frac{\bar{\chi}}{\chi'} \right)^2 P\left(k'_{\perp} \frac{\bar{\chi}}{\chi'} ;\chi'\right) \right](r_\perp)  \right] (k_\perp), 
    \label{eq:PSWL:delg_WL-gamma_WL}
\end{align}
where in the second line, we have defined the auto correlation function of the perpendicular components of the window function, $Q_\perp$, as
\begin{align}
    |W_\perp(\bk_\perp)|^2 \equiv \int_{\br_\perp} Q_\perp (\br_\perp) e^{i\bk_\perp \cdot \br_\perp},
\end{align}
and used the 2D plane-wave expansion with the Bessel function $J_n$:
\begin{align}
    e^{-i\bk_\perp \cdot \br_\perp} = \sum_{n=-\infty}^\infty (-i)^n J_n(k_\perp r_\perp) 
    e^{-in (\phi_{\hbk_\perp} - \phi_{\hbr_\perp})}.
\end{align}
In the third line, we reexpress the result in terms of the 2D Hankel and inverse Hankel transforms, $\mathcal{H}^\mathrm{2D}_\ell$ and $ \left(\mathcal{H}^\mathrm{2D}_{\ell}\right)^{-1}$, explicitly to numerically implement it with 1D FFT (FFTlog).

We next calculate the galaxy-weak lensing cross correlation with the same approximations as
\begin{align*}
    \avrg{\delg(\bx_1) \gamma^\mathrm{WL}(\bx_2)}_{\chi_1<\chi_2} 
    &= C
    \int_0^{\chi_2} \rmd \chi' K(\chi_2,\chi') 2e_{ij}(\hbx) 
    \int_{\bk'} \hk'_i\hk'_j P_\mathrm{gm}(k';\chi_1,\chi') e^{i\bk'\cdot (\bx_1-\bx'_2)} \\
    &\simeq C 
    K(\chi_2,\chi_1) 
    \int_{\bk'_{\perp}} e^{2i\phi_{\hbk'_\perp}} P_\mathrm{gm}(k'_{\perp};\chi_1) e^{i\bk'_{\perp} \cdot (\bx_{1,\perp}-\bx_{2,\perp})} \\
    &\simeq C 
    \frac{\chi_2 - \chi_1}{a(\bar{\chi})}
    \int_{\bk'_{\perp}} e^{2i\phi_{\hbk'_\perp}} P_\mathrm{gm}(k'_{\perp};\bar{\chi}) e^{i\bk'_{\perp} \cdot (\bx_{1,\perp}-\bx_{2,\perp})}.
\end{align*}
Performing the Fourier transform taking into account the window function, we obtain the power spectrum:
\begin{align}
    \tilde{P}_{\delg \gamma^\mathrm{WL}}(\bk) \equiv e^{-2i\phi_{\hbk_\perp}} 
    \frac{G_\parallel (k_\parallel)}{2}
    \frac{C}{a(\bar{\chi})} 
    \int_{\bk'_{\perp}} e^{2i\phi_{\hbk'_\perp}} P_\mathrm{gm} \left(k'_{\perp} ;\bar{\chi} \right) |W_\perp(\bk_\perp - \bk'_\perp)|^2,
\end{align}
where 
\begin{align*}
    G_\parallel (k_\parallel) 
    \equiv \int \rmd x_{1,\parallel}\int \rmd x_{2,\parallel} W_\parallel(x_{1,\parallel}) W_\parallel(x_{2,\parallel}) |x_{1,\parallel} - x_{2,\parallel}| e^{-ik_\parallel (x_{1,\parallel} - x_{2,\parallel})} 
    \equiv \int \rmd r_\parallel |r_\parallel| Q_\parallel (r_\parallel) e^{-ik_\parallel r_\parallel}.
\end{align*}
Thus, the multipole moments are
\begin{align}
    \tilde{P}^{(L)}_{\delg \gamma^\mathrm{WL}}(k) 
    &\equiv (2L+1)\frac{(L-2)!}{(L+2)!} \int_{\hbk} \tilde{P}_{\delg \gamma^\mathrm{WL}}(\bk) \mathcal{L}_L^{m=2} (\mu) \\
    &\equiv (2L+1)\frac{(L-2)!}{(L+2)!} \int_{-1}^1 \frac{\rmd \mu}{2} \mathcal{L}_L^{m=2} (\mu) \frac{G_\parallel (k_\parallel)}{2} \frac{C}{a(\bar{\chi})} 
    \mathcal{H}^\mathrm{2D}_2 \left[ Q_\perp(r_\perp) \left(\mathcal{H}^\mathrm{2D}_2\right)^{-1} \left[ P_\mathrm{gm}\left(k'_{\perp};\bar{\chi}\right) \right](r_\perp)  \right] (k_\perp). 
    \label{eq:PSWL:delg-gamma_WL}
\end{align}
After similar calculations, we obtain the magnification-IA cross power spectrum:
\begin{align}
    \tilde{P}^{(L)}_{\delta_\rmg^\mathrm{WL} \gamma^\mathrm{IA}}(k) 
    \equiv (2L+1)\frac{(L-2)!}{(L+2)!} \int_{-1}^1 \frac{\rmd \mu}{2} \mathcal{L}_L^{m=2} (\mu) \frac{G_\parallel (k_\parallel)}{2} \frac{2(\alpha_\mathrm{mag}-1)C}{a(\bar{\chi})} 
    \mathcal{H}^\mathrm{2D}_2 \left[ Q_\perp(r_\perp) \left(\mathcal{H}^\mathrm{2D}_2\right)^{-1} \left[ P_{\gamma^\mathrm{IA}\rmm}\left(k'_{\perp};\bar{\chi}\right) \right](r_\perp)  \right] (k_\perp). 
    \label{eq:PSWL:delg_WL-gamma_IA}
\end{align}
For the numerical evaluation, we assume $P_\mathrm{gm} = b_1 P^\mathrm{NL}$ and $P_{\gamma^\mathrm{IA}\rmm} = b_K P^\mathrm{NL}/2$ where $P^\mathrm{NL}$ is the nonlinear matter power spectrum.

\subsection{Window Functions for Weak Lensing Signals} 
\label{subsec:wl_window_function}
Here we address the window function for the weak lensing signal.
As shown in the previous subsection, we approximate the window function as the separable form parallel/perpendicular to the LOS direction.
In this case, the auto correlation of $W$ also becomes separable:
\begin{align*}
    Q(\br) \equiv \int_{\bx} W(\bx)W(\bx + \br) 
    \simeq \int_{\bx} W_\parallel(x_\parallel)W(\bx_\perp) W_\parallel(x_\parallel+r_\parallel)W(\bx_\perp+\br_\perp)
    \equiv Q_\parallel(r_\parallel) Q_\perp(\br_\perp).
\end{align*}
To see how good this approximation is, we first measure $Q_\parallel$ and $Q_\perp$ from the random catalog, next reconstruct its (3D) multipole moments by the angular integration:
\begin{align}
    Q^\mathrm{rec}_\ell(r) \equiv (2\ell+1) \int \frac{\rmd \Omega_{\hbr}}{4\pi} Q_\parallel(r_\parallel) Q_\perp(r_\perp) \mathcal{L}_\ell (\mu),
    \label{eq:def_qell_rec}
\end{align}
where $(r_\parallel, r_\perp) = (\mu r, \sqrt{1-\mu^2} r)$, and then compare $Q^\mathrm{rec}_\ell$ with the \textit{true} multipole moments, $Q_\ell$, which is defined as
\begin{align}
    Q_\ell(r) \equiv \int_{\bx} W(\bx)W(\bx + \br) \mathcal{L}_\ell (\mu).
    \label{eq:def_qell}
\end{align}

To obtain $Q_{\parallel,\perp}$, we follow the pair-counting approach \cite{Wilson+2017} which was originally proposed to obtain $Q_\ell$.
We slightly modify it to evaluate each parallel/perpendicular component as follows.
We first review the methodology for the calculation of $Q_\ell$. 
The number of pairs connecting the infinitesimal volume $\rmd V (\bx)$ with $\rmd V' (\bx')$ is written by 
\begin{align*}
    RR(\bx,\bx') = W(\bx) \rmd V \cdot W(\bx') \rmd V',
\end{align*}
where $W(\bx) \equiv \bar{n}(\bx) w(\bx)$ is the ``window'' function defined by the product of the mean number density $\bar{n}$ and weight $w$. 
The total number of pairs over the survey region with the separation vector $\br \equiv \bx' - \bx$ is given by the summation:
\begin{align}
    RR^\mathrm{tot}(\br) 
    \equiv \int RR(\bx,\bx') 
    = \rmd V' \int_{\bx} W(\bx)W(\bx + \br) 
    = \rmd V' Q(\br).
    \label{eq:RRtot}
\end{align}
By using $\rmd V' = r^3\Delta (\mathrm{ln}r) \rmd \Omega_{\hbr}$ and taking the angular average with the Legendre polynomials, we have
\begin{align*}
    RR^\mathrm{tot}_\ell(r) 
    \equiv (2\ell+1) \int RR^\mathrm{tot}(\br) \mathcal{L}_\ell (\mu)
    = 4\pi r^3\Delta (\mathrm{ln}r) Q_\ell(r),
\end{align*}
where we define $\mu$ for each pair as the cosine between the separation vector and the midpoint vector towards the pair, $\mu \equiv \hbr \cdot \hbd$ with $\bd \equiv (\bx + \bx')/2$. 
Thus we obtain $Q_\ell$ by normalizing the total weighted pair counts $\propto RR^\mathrm{tot}_\ell/r^3$.

Similarly, under the assumption of the separable form of the window function, 
we can factorize Eq.~(\ref{eq:RRtot}) by using 
$\rmd V' = r_\parallel \Delta (\mathrm{ln}r_\parallel) \cdot r^2_\perp \Delta (\mathrm{ln}r_\perp) \rmd \phi_{\hbr_\perp}$: 
\begin{align*}
    RR^\mathrm{tot}(r_\parallel, \br_\perp) 
    \simeq r_\parallel \Delta (\mathrm{ln}r_\parallel)Q_\parallel(r_\parallel) \cdot r^2_\perp \Delta (\mathrm{ln}r_\perp) \rmd \phi_{\hbr_\perp} Q_\perp(\br_\perp), 
\end{align*}
where $r_\parallel \equiv \br \cdot \hbd$ and $\br_\perp \equiv \br - r_\parallel \hbd$.
Therefore the parallel and perpendicular components of the window-auto correlation can be estimated by the pair counts summed with respect to $r_\parallel$ and $r_\perp$, respectively;
$Q_\parallel \propto RR^\mathrm{tot}/r_\parallel$ and $Q_\perp \propto RR^\mathrm{tot}/r^2_\perp$.

\begin{figure*}
    \centering
    \includegraphics[width=1.0\columnwidth]{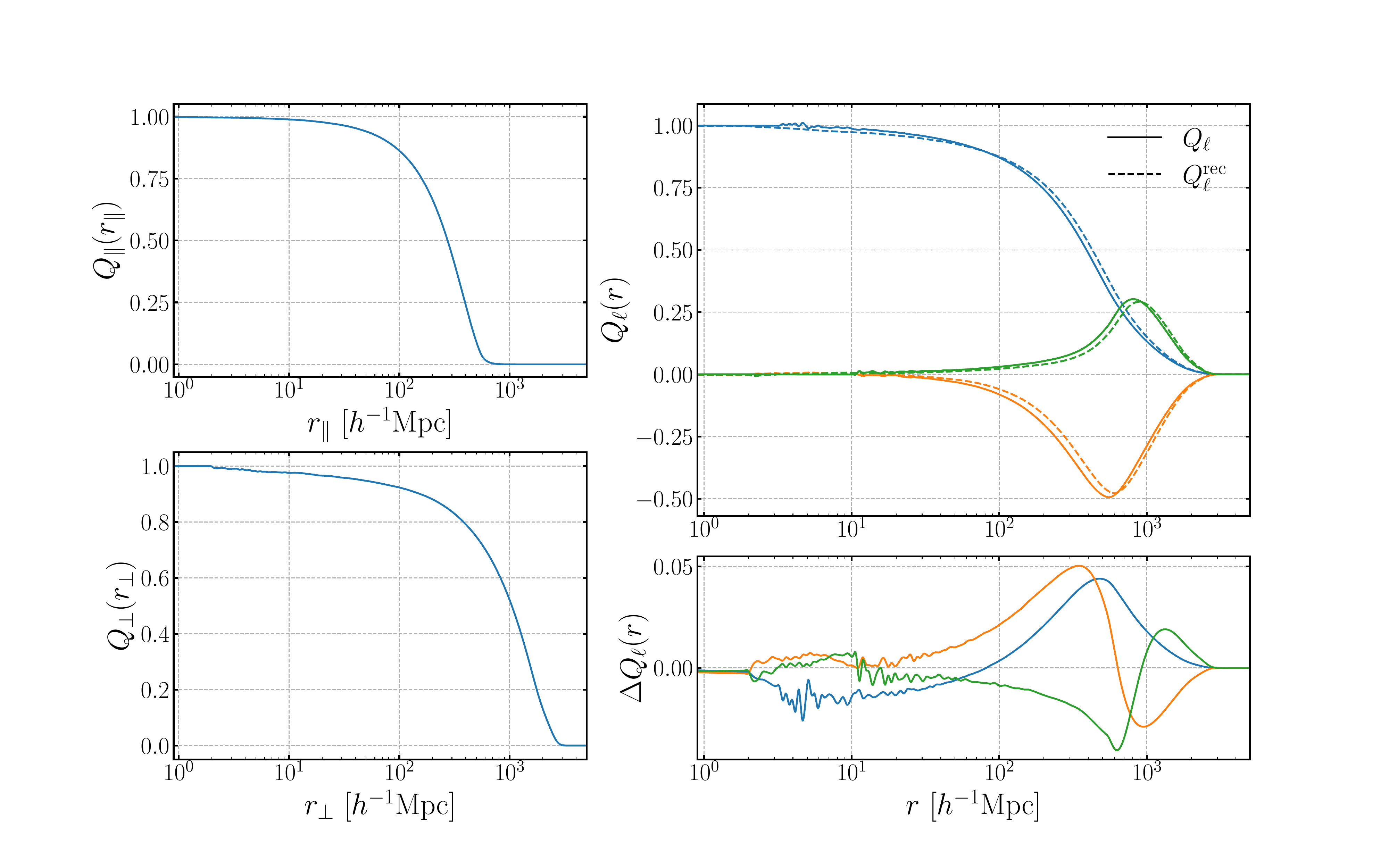}
    \caption{
    \textit{Left column:} The upper (lower) panel shows the LOS-parallel (-perpendicular) component of the window auto correlation function measured from the random catalog for NGC high-z sample ($0.5<z<0.75$).
    \textit{Right:} The comparison between the \textit{true} multipole moments defined by Eq.~(\ref{eq:def_qell}) (solid line), denoted as $Q_\ell$, and the \textit{reconstructed} moments from $Q_{\parallel,\perp}$ defined by Eq.~(\ref{eq:def_qell_rec}) (dashed line), denoted as $Q^\mathrm{rec}_\ell$. 
    The blue, orange and green curves correspond to the monopole, quadrupole and hexadecapole, respectively. 
    The lower panel shows the difference between them. 
    The jagged features at small scales are due to the resolution of the pair counting approach.
    }
    \label{fig:qell_lens}
\end{figure*}
The left panels in Fig.~\ref{fig:qell_lens} show the LOS-parallel and -perpendicular components of the window auto correlation, $Q_{\parallel}$ and $Q_{\perp}$, respectively. 
The dumping scales, $r_\parallel \sim 500~h^{-1}\Mpc$ and $r_\perp \sim 2000~h^{-1}\Mpc$, roughly correspond to the comoving range of galaxy distributions in each direction.
We show the comparison between the \textit{true} and \textit{reconstructed} multipole moments of the window auto correlation function in the right panel.
We find that the assumption of the LOS-parallel/perpendicular decomposition is not perfect, but the difference is less than $5\%$ for all scales.
Besides, since the order of amplitude of weak lensing signals is small enough compared to the statistical error as we will see in the next subsection, we stick to this assumption for the window function throughout this paper.

\subsection{Amplitude of Weak Lensing Signals} 
\label{subsec:wl_amplitude}
\begin{figure*}
    \centering
    \includegraphics[width=1.0\columnwidth]{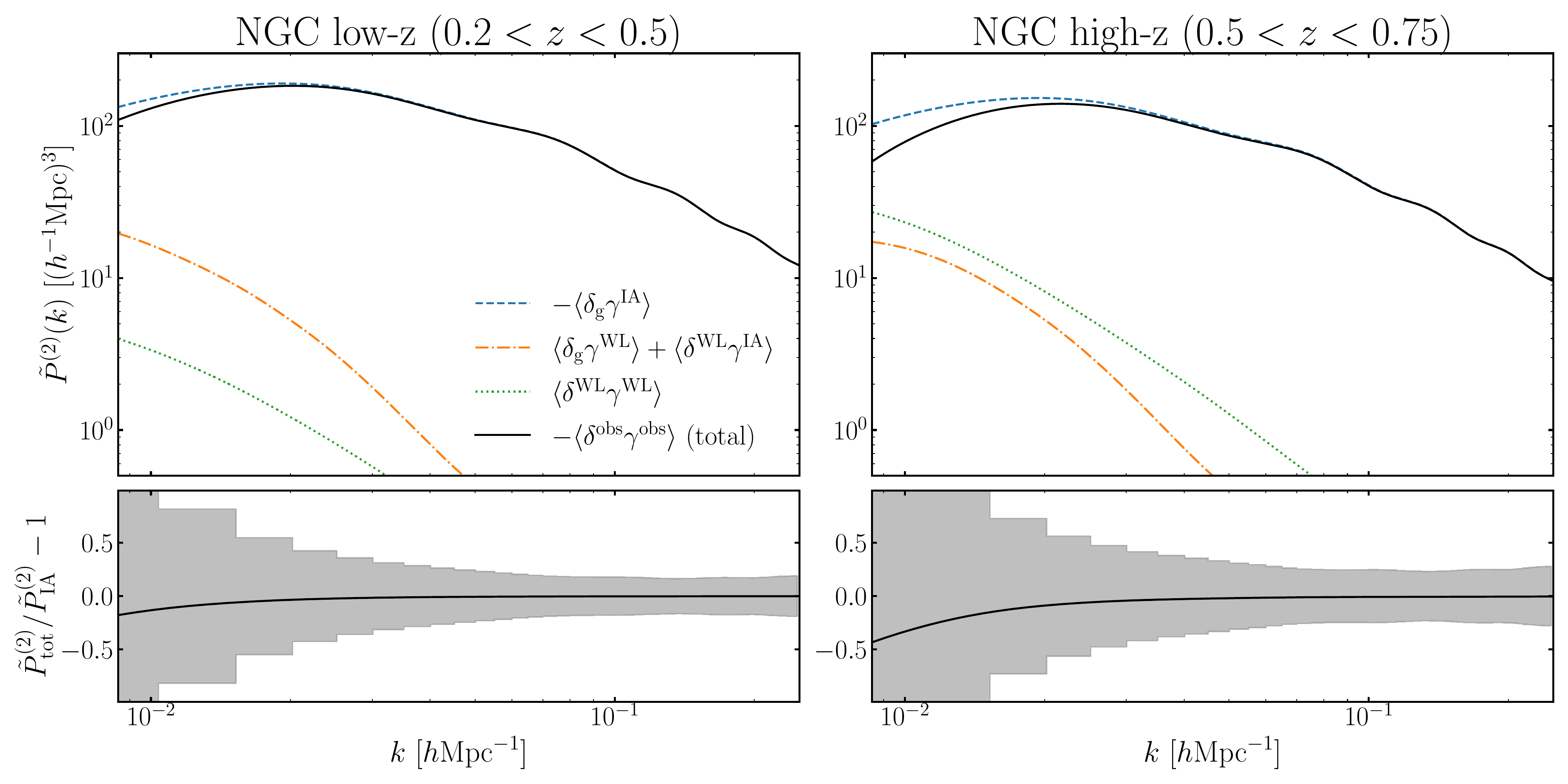}
    \caption{
    \textit{Upper panels:}
    The model predictions of of various power spectra, computed from our theoretical template of the linear-theory power spectra, for NGC low-z (\textit{left}) and high-z (\textit{right}) samples, respectively: 
    the ``intrinsic'' IA-density cross spectrum (blue, dashed line),
    the WL-density or WL-IA cross spectrum (orange, dot-dashed), and the WL auto spectrum (green, dotted). 
    Black, solid line show the total power that is the sum of the above power spectra, to be compared with the measured power spectrum. 
    Note that we plot the absolute values of power spectra because the signs of IA and WL signals are opposite. 
    \textit{Lower panels:}
    The ratio of the total power spectrum to the intrinsic IA power spectrum, where the latter does not include the WL contamination due to the foreground large-scale structure at different redshifts from those of galaxies in the sample. 
    The gray band corresponds to the statistical errors in each $k$ bin that are estimated from the diagonal elements of the covariance for each galaxy sample. 
    }
    \label{fig:wl_contribution}
\end{figure*}
Fig.~\ref{fig:wl_contribution} shows the density-shape cross power spectrum (multipole moment of $L=2$) including the survey window effect and the weak lensing contributions for each galaxy sample, NGC low-z/high-z. 
The effective redshifts are 0.38 and 0.61, respectively.
We assume $b_1=2,b_K=-0.045$ for the intrinsic alignment power spectrum (the linear alignment model), and for the magnification bias in the weak lensing signals, we use 
$\alpha^\mathrm{(low\mathchar`-z,high\mathchar`-z)}_\mathrm{mag} \equiv (1.93, 2.62)$ \citep{Joachimi+2021:KiDS1000_Methodology,Kramsta+2021:MagBias}. 
The amplitude of WL-cross power spectrum (orange, dot-dashed curve) is similar between the two samples because it is almost determined by the radial (finite) width of the galaxy distribution, $\Delta r_\parallel \sim 500~h^{-1}\Mpc$.
On the other hand, for WL-auto power spectrum (green, dotted), the high-z sample has greater amplitude than the low-z sample due to the higher weak lensing efficiency as expected. 
Since the intrinsic alignment and the weak lensing have opposite signs, i.e. radial and tangential distortions, the total power spectrum (black, solid) is smaller than the pure IA power spectrum (blue, dashed). 
We take this weak lensing contamination into account in the analysis as described in Section~\ref{subsec:model}.

\section{Further Tests for Cosmological Analysis} 
\label{sec:validation_test_for_analysis}

\subsection{Window Convolution with Primordial Non-Gaussianity} 
\label{subsec:validation_test_window_convolution}
The model of the IA power spectrum used in our analysis has been validated in Ref.~\cite{Kurita&Takada2022:AnalysisIAPS}.
However, the test was done in Ref.~\cite{Kurita&Takada2022:AnalysisIAPS} using a continuous tidal field, generated under the Gaussian initial condition, to simulate the IA signal. 
Since the observed galaxy density and shape fields are discrete and we aim at exploring the PNG (i.e. non-Gaussian initial condition) information from the observed power spectra, the previous tests would be considered insufficient. 
As we currently do not have a realistic mock signal of galaxy IA in the $k$-range we are interested in, $0.01\lesssim k \lesssim 0.2~\hMpci$, we conduct the following additional tests using dark matter halo samples obtained by $N$-body simulations under both Gaussian and non-Gaussian ($f_\mathrm{NL}^{s=2}=500$) initial conditions generated in Ref.~\cite{Akitsu+2021:IA_PNG}. 
We adopt $N_{\rm part}=2048^3$ particles and $4.096~h^{-1}{\rm Gpc}$ for the comoving simulation box size, which corresponds to the particle masses $m_{\rm p}\simeq 7.0\times 10^{11}~h^{-1}M_\odot$. 
We use halos identified by \texttt{Rockstar} \cite{Behroozi+2013:Rockstar} with their virial mass $M_\mathrm{vir} > 7\times10^{13}~h^{-1}M_\odot$.
Employing the same BOSS-like survey geometry as used in Ref.~\cite{Kurita&Takada2022:AnalysisIAPS} (also used in the validation test of the covariance matrix in Appendix~\ref{sec:derivation_covariance}), we measure the halo IA power spectrum including the survey window, the projection of halo shapes, and the redshift-space distortion effect for each line-of-sight direction. 
We chop out 32 different sub-regions, each of which mimics the BOSS survey window ($\sim 2~h^{-3}\mathrm{Gpc}^3$), from the entire simulation box ($\sim 69~h^{-3}\mathrm{Gpc}^3$) and use the signals measured from the sub-regions to estimate the mean signal and the error bars needed to compare the theoretical model with the measurements. 

\begin{figure*}
    \centering
    \includegraphics[width=1.0\columnwidth]{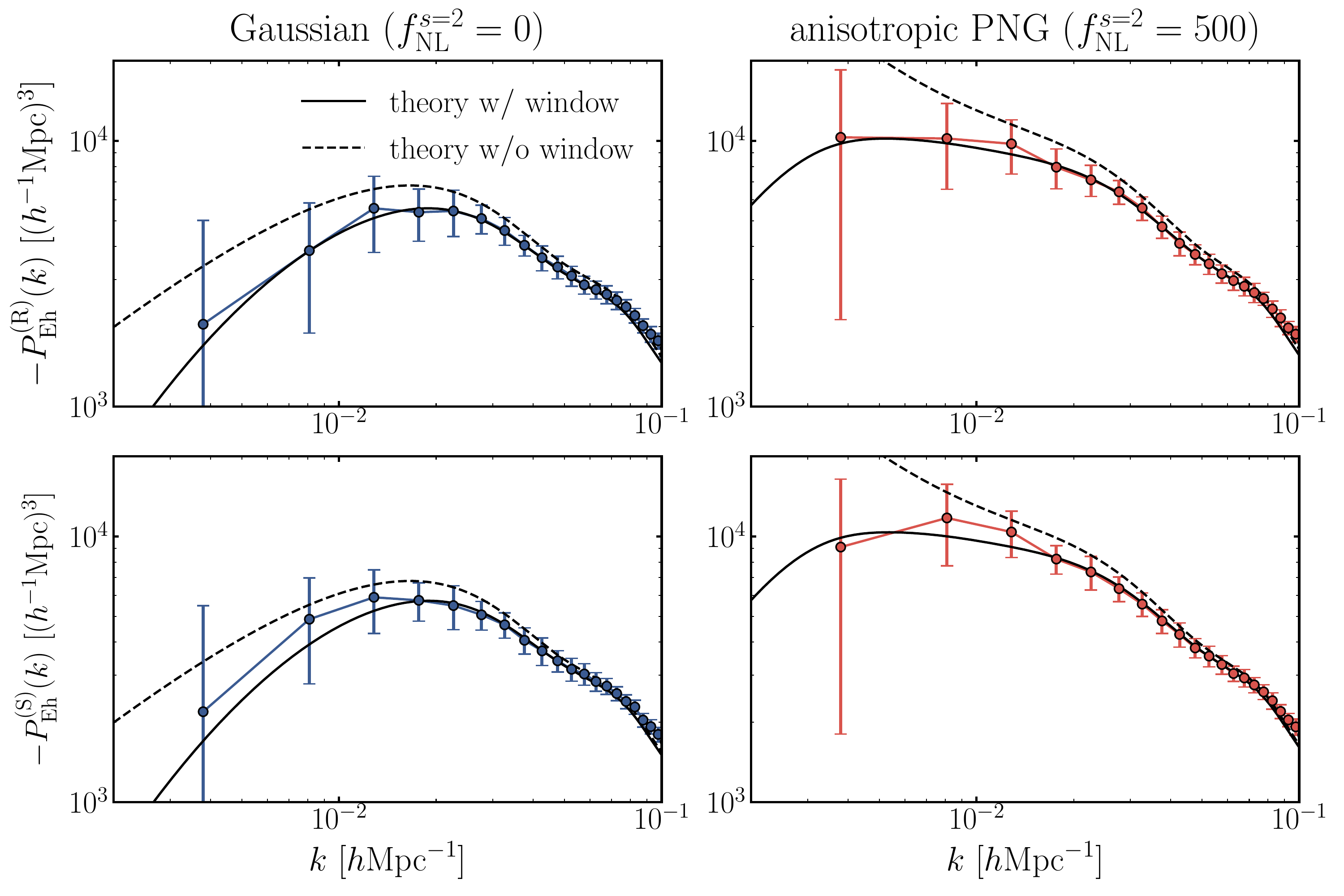}
    \caption{Comparison of the halo IA power spectra, measured by our estimator (points with error bars) from the mock data, with the theoretical predictions including the observational effects (solid lines) for Gaussian (\textit{left}) and anisotropic PNG (\textit{right}) initial conditions, respectively. 
    The upper panels are for the real-space power spectrum and the lower for the redshift-space power spectrum. 
    These are measured for the halo sample with $M_\mathrm{vir} > 7\times10^{13}~h^{-1}M_\odot$ in the mock data including the BOSS-like survey window (see text for details).
    Note that the data points are the mean signal of the 32 mock data realizations, while the error bars are computed from the standard deviations of the 32 measurements, which give an estimate of the statistical errors in the power spectrum measurement for the BOSS volume. 
    }
    \label{fig:png_validation}
\end{figure*}
Fig.~\ref{fig:png_validation} shows the comparison of the measured IA power spectrum and the linear (alignment) model prediction including observational effects. 
The figure gives validation of our theoretical model because the model predictions fairly well reproduces the simulation results, down to very small $k$ bins, for both the Gaussian and PNG initial conditions. 
Note that the deviation between the theory and measurement at $k \sim 0.1~\hMpci$ is due to the non-linearities of the evolution of IA and the RSD effect. 
We will discuss the impact of this nonilnear effect on the PNG parameter estimations in the next section.

\subsection{Determination of $k_\mathrm{max}$ for IA Power Spectrum} 
\label{subsec:validation_test_kmax}
We here study a proper choice of the maximum wavenumber $k_\mathrm{max}$ in the sense that our analysis using the theoretical template based on the linear model can recover the input PNG parameter in an unbiased manner.
In this work, we choose $k_\mathrm{max}$ so that the 1D systematic bias in $f_\mathrm{NL}^{s=2}$ parameter is smaller than the statistical error, 1$\sigma$.
For this purpose we need to prepare a realistic mock signal (data vector) that mimics the \textit{true} non-linearity of galaxy IA corresponding to the BOSS galaxy sample we use. 
However, again since we currently do not have a reliable mock of galaxy IA, we use the halo IA power spectrum generated in the previous section to approximate the non-linearity of the observed IA signal as follows. 
Since the density and IA power spectra of halos have greater amplitudes than do the BOSS galaxy power spectra, we make the following correction to make the mock catalog more realistic. 
We first estimate the linear density and shape bias parameters by fitting the linear power spectrum to the halo power spectrum up to $k=0.05~\hMpci$, where the linear model is valid. 
Then we rescale the halo power spectra by multiplying the constant factor so that the resulting halo power spectra have the linear bias parameters, $(b_1,b_K)=(2.0,-0.04)$, that are typical values for the BOSS galaxy samples we use. Nevertheless the halo power spectra have stronger nonlinearities in their clustering, IA and redshift-space distortion at the larger $k$ in the nonlinear regime, and therefore our validation tests to estimate the impact of these nonlinear effects can be considered as a conservative estimate. 

\begin{figure*}
    \centering
    \includegraphics[width=1.0\columnwidth]{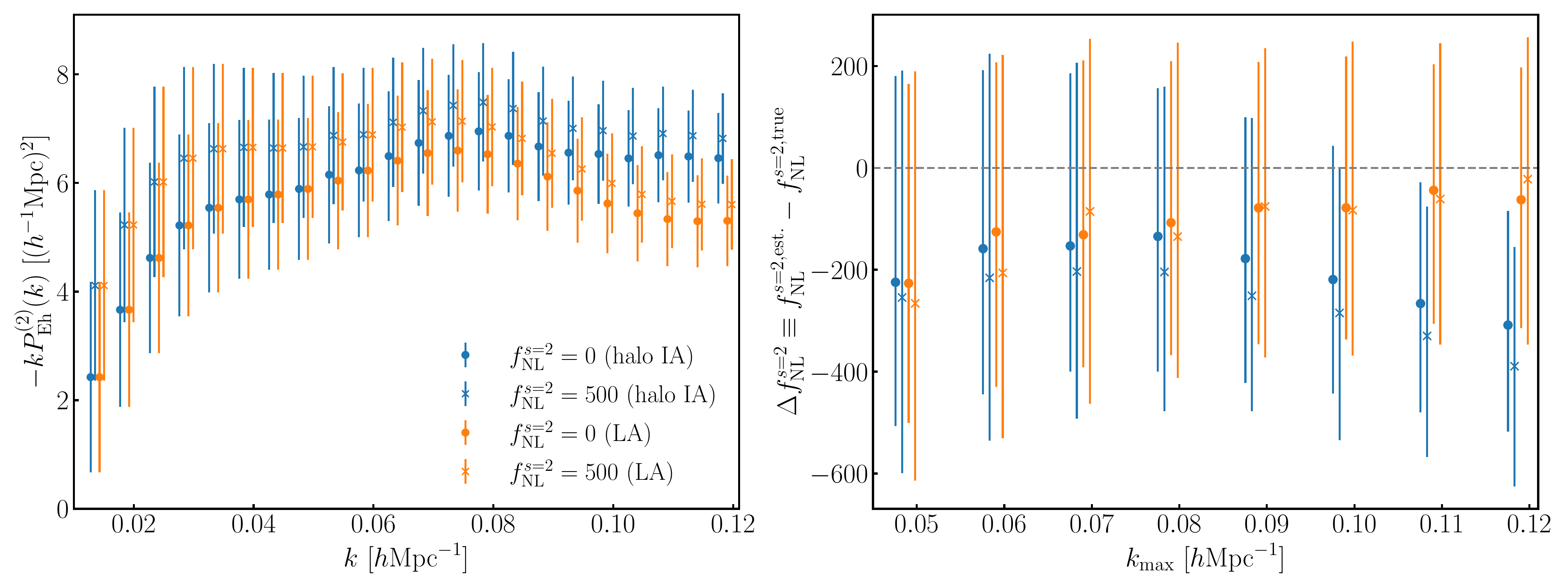}
    \caption{
    {\it Left panel}: Comparison of the mock data vector of the IA power spectrum in our test. 
    The orange points are the mock data made by our linear alignment model (``LA''), while the blue points are the mock data made by the halo power spectrum measured from the simulation (``halo IA''). 
    Here we adopted the linear shape bias parameter to mimic that for BOSS-like galaxies, while we ``rescaled'' the amplitude of the halo power spectrum to match with the linear-theory prediction for BOSS-like galaxies on linear scales (see text for details). 
    The circle symbols in the respective color are for the Gaussian initial conditions for the $\Lambda$CDM model, while the cross symbols are for the PNG model with $f^{s=2}_\mathrm{NL}=500$ (the other cosmological parameters are kept fixed to the fiducal values). 
    The error bars are from the diagonal components of the covariance for the BOSS NGC low-z galaxy sample.  
    {\it Right}: Results for the validation test of our analysis pipeline. 
    Shown is the difference between the estimated $f_\mathrm{NL}^{s=2}$ and its true value assumed in the mock data, where $f_\mathrm{NL}^{s=2}$ is estimated by comparing the model template of linear-theory power spectrum with the mock data in the range $k=[0.01~h{\rm Mpc}^{-1}, k_{\rm max}]$ as a function of $k_{\rm max}$ in the $x$-axis. 
    Each color and symbol corresponds to the respective result using the respective mock data in the left figure.
    }
    \label{fig:kmax_validation}
\end{figure*}

For comprehensiveness of our discussion we make two kinds of validation tests. 
For the first test, to generate the mock data vector $\bd$, we use the rescaled halo power spectra, as described above,  that are originally measured from halos in $N$-body simulations. 
Here we call this mock data vector as $\bd$ = ``halo IA''. 
For the second test we use the linear power spectra to make the mock data vector, where the model predictions are computed from the same model that is used in the theoretical template of the parameter inference ($\bd$ = ``LA''). 
Then we test whether our analysis pipeline can recover the input value of $f^{s=2}_\mathrm{NL}$. 
The latter test can quantify the impact of projection effect in a multi-dimensional parameter space in the Bayesian parameter inference, which refers to a bias that the input parameter value is not necessarily perfectly recovered if the posterior distribution in a full parameter space is non-Gaussian \citep{Nishimichi+2020:blinded_challenge,Kobayashi+2022:FullShape}. 

Fig.~\ref{fig:kmax_validation} shows the results. 
In the case of the ``halo IA'' (blue), our pipeline can recover the input $f^{s=2}_\mathrm{NL}$ only when using up to $k_{\rm max}\simeq 0.1~\hMpci$ to within $1\sigma$ error, and gives a parameter bias greater than $1\sigma$ for the larger $k_{\rm max}$ because of the stronger non-linearities that are not captured by the linear model. 
In the case of the ``LA'' (orange), although the pipeline can recover the input value of $f^{s=2}_\mathrm{NL}$ to within $1\sigma$ for all $k_{\rm max}$ of interest, the projection effect is larger for the smaller $k_{\rm max}$ due to the banana-shaped degeneracy between $b_K$ and $f^{s=2}_\mathrm{NL}$. 
We note that for both cases the estimated value of $f^{s=2}_\mathrm{NL}$ tends to be smaller than the true value, meaning that the analysis tends to underestimate the PNG amplitude. 
From these results we adopt $k_{\rm max}=0.1~\hMpci$ as our fiducial choice.

\bibliography{main}
\end{document}